\documentclass[useAMS,usenatbib]{mn2e}
\usepackage{txfonts}
\usepackage{graphicx}
\usepackage{pdflscape}
\usepackage{rotating}

\def\l30{$\textit{l}=30^{\circ}$}
\def\59l{$\textit{l}=59^{\circ}$}
\def\Tb{\textit{T}$_{b}$ }
\def\Ts{\textit{T}$_{s}$ }
\def\TL{\textit{T}$_{L}$ }
\def\Te{\textit{T}$_{e}$ }

\def\H2{H$_{2}$}
\def\HII{HII}
\def\neff{$n_{eff}$}
\def\Av{A$_{\mathrm{V}}$}
\def\um{$\mu$m}
\def\mum{$\mu$m}
\def\12CO{$^{12}$CO}
\def\CO13{$^{13}$CO}
\def\Ypah{Y$_{\mathrm{PAH}}$}
\def\Yvsg{Y$_{\mathrm{VSG}}$}
\def\Ybg{Y$_{\mathrm{BG}}$}

\title{The pros and cons of the inversion method approach to derive 3D dust emission properties in the ISM: the Hi-GAL field centred on (l,b)=(30$^{\circ}$,0$^{\circ}$).}

\author[A. Traficante, R. Paladini et al.]{A. Traficante$^{1,2,3}$\thanks{e-mail:
alessio.traficante@manchester.ac.uk}, R. Paladini$^{3}$, M. Compiegne$^{3,4}$,  M.\,I.\,R. Alves$^{5,1}$, L. Cambr\'esy$^{6}$, 
\newauthor S.\,J. Gibson$^{7}$, C.\,T. Tibbs$^{3}$, A. Noriega-Crespo$^{3}$, S. Molinari$^{8}$,  S.\,J. Carey$^{3}$, 
\newauthor J.\,G. Ingalls$^{3}$, P. Natoli$^{9,10,11}$, R.\,D.~Davies$^{1}$, R.\,J. Davis$^{1}$, C. Dickinson$^{1}$, G.\,A. Fuller$^{1}$\\
$^{1}$Jodrell Bank Centre for Astrophysics, School of Physics and Astronomy, The University of Manchester, Oxford Road, Manchester M13 9PL, UK\\
$^{2}$Dipartimento di Fisica, Universita' di Roma Tor Vergata, Italy\\
$^{3}$Infrared Processing Analysis Center, California Institute of Technology, Pasadena-CA. 91125, USA\\
$^{4}$Laboratoire d'Optique Atmospherique, UMR8518, CNRS-INSU, Universite' Lille 1, France\\
$^{5}$Institut d'Astrophysique Spatiale, CNRS (8617) Universite' Paris-Sud 11, Orsay 91405, France\\
$^{6}$Observatoire astronomique de Strasbourg, Universite' de Strasbourg, CNRS, UMR 7550, France\\
$^{7}$Department of Physics $\&$ Astronomy, Western Kentucky University, Bowling Green, KY 42101, USA\\
$^{8}$INAF, Istituto Fisica Spazio Interplanetario, I-00133 Rome, Italy\\
$^{9}$Dipartimento di Fisica e Scienze della Terra, Universit\`a di Ferrara e Sezione INFN Ferrara, Via Saragat 1, 44122 Ferrara, Italy\\
$^{10}$Agenzia Spaziale Italiana Science Data Center, c/o ESRIN, via Galileo Galilei, Frascati, Italy\\
$^{11}$INAF/IASF Bologna, Via Gobetti 101, Bologna, Italy}

\begin{document}

\date{}

\pagerange{\pageref{firstpage}--\pageref{lastpage}} \pubyear{2011}

\maketitle

\label{firstpage}

\begin{abstract}
\textit{Herschel} far-infrared (FIR) continuum data obtained as part of the Hi-GAL survey have been used, together with 
the GLIMPSE 8 \um\ and MIPSGAL 24 \um\ data, to attempt the first 3D-decomposition of dust emission 
associated with atomic, molecular and ionized gas at 15 arcmin angular resolution. Our initial test case is a 2$\times$2 square degrees 
region centred on \textit{(l,b)}=(30$^{\circ}$,0$^{\circ}$), a direction that encompasses the origin point of the Scutum-Crux Arm at the tip of the Galactic Bar. 
Coupling the IR maps with velocity maps specific for different gas phases (HI 21cm,  \12CO and \CO13, and Radio Recombination Lines, RRLs), 
we estimate the properties of dust blended with each of the gas components and at different Galactocentric distances along the Line of Sight (LOS). 

A statistical Pearson's coefficients analysis is used to study the correlation between the column densities estimated for each gas component and the intensity of 
the IR emission. This analysis provides evidence that the 2$\times$2 square degree field under consideration is characterized  by the presence of a gas component not accounted 
for by the standard tracers, possibly associated with warm \H2 and cold HI.

We demonstrate that the IR radiation in the range 8 $\mu$m $<$ $\lambda$ $<$ 500 $\mu$m is systematically dominated by emission originating within the Scutum-Crux Arm. 
By applying an inversion method, we recover the dust emissivities associated with atomic, molecular and ionized gas. 
Using the DustEM model, we fit the Spectral Energy Distributions (SEDs) for each gas phase, and 
find average dust temperatures of T$_{d,\ \mathrm{HI}}$=18.82$\pm$0.47 K, T$_{d,\ \mathrm{H}_{2}}$=18.84$\pm$1.06 K and T$_{d,\ \mathrm{HII}}$=22.56$\pm$0.64 K, respectively. 
We also obtain an indication for Polycyclic Aromatic Hydrocarbons (PAHs) depletion 
in the \emph{diffuse} ionized gas.

We demonstrate the importance of including the ionized component in 3D-decompositions of the total IR emission.

However, the main goal of this work is to discuss the impact of the missing column density associated with the \textit{dark gas} component on the accurate evaluation of 
the dust properties, and to shed light on the limitations of the inversion method approach when this is applied to a small section of 
the Galactic Plane and when the working resolution allows sufficient de-blending of the gas components along the LOS.

\end{abstract}


\section{Introduction}

A variety of space-borne experiments in the course of the last two decades (IRAS, Spitzer, \textit{Herschel}) have shown that the Galaxy is filled 
with relatively cold dust (15$\leq\mathrm{T}_{d}\leq30$ K, e.g. \citet{Ferriere01} distributed along each Line of Sight (LOS), and associated with the atomic, molecular and ionized gas phases. This finding has also been recently confirmed by the \textit{Planck} whole-sky observations \citep{Planck_A22,Planck_A23,Planck_A24,Planck_A25}.

Dust grains can absorb and re-emit a large fraction of the radiation provided by stellar sources. Depending on the size, they are stochastically heated  
(i.e. the smaller grains, which emit mainly in the Near/Mid-IR), or reach thermal equilibrium (i.e. the bigger grains, which emit predominantly in the Far-IR regime). 

Dust emission properties have been intensely studied at high Galactic latitudes, where each gas phase can be assumed relatively isolated from the others, and mixing 
along the LOS can be avoided \citep[e.g.][]{Boulanger88, Boulanger96,Lagache00}. 
On the contrary, in the Galactic Plane disentangling dust emission arising from different gas components and intrinsically different environments is a complicated problem which requires additional kinematic information on the emitting gas \citep[e.g.][]{Bloemen90}. 

The separation can be achieved with the so-called \textit{inversion method}, originally introduced by \citet{Bloemen86} to analyse the individual contribution of 
atomic and molecular gas to the integrated $\gamma$-ray emission. The model was later extended by \citet{Bloemen90} to include the FIR emission across 
the Galactic Plane observed with IRAS at 60 \um\ and 100 \um . 
So far, the application of the inversion method has been limited by the angular resolution of the available IR and ancillary data (equal or close to $1^{\circ}$) which does not allow to, e.g., spatially separate individual clouds \citep{Bloemen90, Giard94, Sodroski97, Paladini07, Planck_Marshall11}. These works have demonstrated that, on average,  most of the Galactic IR luminosity is 
associated with dust in the atomic gas component, which is primarily irradiated by the local radiation field. 
Dust associated with the molecular and ionized components, on the other hand, is mostly heated by O and B stars embedded in molecular clouds \citep{Sodroski97,Paladini07}. 
\citet{Planck_Marshall11} decomposed the emission from 12 $\mu$m to millimeter wavelengths using IRAS, COBE-DIRBE, WMAP and the recent \textit{Planck} data in the latitude range 
$\vert b\vert\leq10^{\circ}$ and found evidence of the existence of a significant amount of cold atomic and 
warm molecular gas not accounted for by the standard tracers. This gas is typically referred to as \textit{dark gas} \citep{Grenier05}. In particular, the authors of this work claim  
that the \textit{dark gas} column density is comparable - or greater - to the column density of molecular 
gas outside the Molecular Ring, i.e. a region of the Galaxy comprised between Galactocentric radius 4 kpc $<$ R $<$ 8 kpc, where 70 percent of the molecular gas resides \citep{Combes91}.

As well as using low-spatial resolution, early inversion works often did not include the ionized gas component, as 
historically its contribution to the overall IR emission was thought to be negligible  \citep{Bloemen90,Giard94}. A few studies 
\citep{Sodroski97, Paladini07, Planck_Marshall11} did take this phase of the gas into account, but used 
the only available data at the time for tracing ionized gas, that is radio continuum data. There are two problems with this approach. The first is that radio continuum emission 
is not uniquely associated with the interaction - and deceleration - of free electrons with ions in a plasma, the free-free or bremsstrahlung emission: at low frequencies (5 GHz or less), one has to estimate and subtract a possible contamination due to synchrotron emission \citep[e.g.][]{Paladini05}, while 
at relatively high frequencies ($>$ 10 GHz), spinning dust emission may become significant \citep[e.g.][]{Planck_Dickinson11, Planck_Marshall11}. The second, even more important, issue is the fact 
that radio continuum data are unable to provide the 3D-information on the location of the emitting gas which is required by inversion techniques.  

The work we describe in this paper is therefore motivated by the following considerations: 

\begin{enumerate}
\item the resolution and sensitivity of the newly available Spitzer and \textit{Herschel} data dramatically improve our view of 
the Galactic Plane at IR wavelengths: the combined GLIMPSE 
\citep{Benjamin03}, MIPSGAL \citep{Carey09} and Hi-GAL \citep{Molinari10_PASP} surveys have mapped 
the inner Galactic Plane in the wavelength range 8 $\mu$m $\leq\lambda\leq$ 500 $\mu$m with a resolution of 35 arcsec, or higher. These new 
data allow us to set more stringent constraints on dust properties and on their variations with Galactocentric radius;

\item the last couple of years have witnessed a tremendous improvement in the quality of available data on the ionized gas. 
In particular, hydrogen recombination lines (RRLs) have been observed for large portions of the Galactic Plane \citep{Staveley-Smith96,Alves11}. These data are sampled with a resolution of $\simeq15$ arcmin,  which allows us to work with a $\sim$4 times better resolution than the previous inversions. Even more important, they provide unprecedented information about the properties and distribution of the ionized gas component along the LOS;

\item in previous inversion works, the decomposition into Galactocentric bins has been done {\em{blindly}}, that is 
without taking into account local features present at a given location of the Galaxy or on specific angular scales. The higher resolution of 
the IR as well as of the ancillary data allow now a more targeted approach; 

\item last but not least, if on one side the higher angular resolution of the available data makes it possible to devote 
 more attention to the specific content of the Galactic region to {\em{invert}}, on the other hand it sheds light on the limitations 
 of the inversion technique itself. Hence, one of the goals of this work is to investigate and describe these limitations.

\end{enumerate}

In this first paper we concentrate on a $2^{\circ}\times2^{\circ}$ field centred on \textit{(l,b)}=(30$^{\circ}$.0,0$^{\circ}$.0) observed by the 
GLIMPSE, MIPSGAL and Hi-GAL surveys. This field was one of two observed during the Hi-GAL Science Demonstration Phase (SDP). Therefore, hereafter 
we will refer to it as SDPF1 ({\em{Science Demonstration Phase Field 1}}). 

The paper is organized as follows. A review of the content of SDPF1 is presented 
in Section \ref{sec:l30_field}. The IR and ancillary radio data used for the analysis are described in Section \ref{sec:infrared_ancillary}. 
The inversion model is discussed in Section \ref{sec:model_data}, as well as the Galactocentric 
region subdivision and the gas column density evaluation for each gas phase. We find evidence of untraced cold atomic and possibly warm molecular gas features extending up to several arcmins. These features do not allow a correct evaluation of the gas column densities, 
thereby preventing the inversion model from working properly. The regions where these features are dominant, however, can be predicted by analyzing 
the correlation between the gas column densities and the IR maps, as explained in Section \ref{sec:pearsons}. 
In Section \ref{sec:results_discussion} we present our results, and discuss the limitations 
of the inversion model in its current stage of development. We also investigate the importance of accounting for dust associated with the ionized gas, by demonstrating with 
a simple test the erroneous conclusions that one may reach by neglecting this component. In Section \ref{sec:conclusions} we describe our conclusions.

\section{The content of SDPF1}\label{sec:l30_field}

According to the \citet{Russeil03} model of the Galactic spiral structure, SDPF1 intercepts the Sagittarius arm twice, and 
both the Norma-Cygnus arm and the Perseum arm once. Furthermore, the LOS also cuts through the Scutum-Crux arm at the tangent point which, adopting 
a Sun-Galactic centre distance of 8.5 kpc, is found at a Galactocentric distance R=4.25 kpc. Both the source content and local Interstellar Medium (ISM) of SDPF1 
have been extensively investigated (see Figure \ref{fig:pmw_l30}). 

The mini-starbust complex W43, located near l=30$^{\circ}.8$ \citep{Bally10}, is the brightest source in the field and one of the brightest FIR sources in the entire Galaxy. 
It is located at d$\simeq$5.5 kpc, 
corresponding to a Galactocentric distance R$\simeq$ 4.65 kpc. This giant HII region contains a cluster of OB and Wolf-Rayet stars and its FIR continuum luminosity is $\sim 3.5\times 10^{6}$ L$_{\sun}$ \citep{Smith78}. 

An additional 29 HII regions are located in SDPF1 \citep[][]{Anderson09}, most of which are evolved \citep{Paladini12}. 
Toward the centre of the field is located the well-studied ultra-compact HII region G29.944-0.04 \citep{Quireza06}.  The N49 HII bubble, 
centred on \textit{(l,b)}=(28.83$^{\circ}$,-0.23$^{\circ}$), is a known case of triggering scenario \citep{Zavagno10}. Two separate HII regions forming the RCW175 group and located towards the edge of the field at $(l,b)=(29.07,-0.68)$ 
exhibit spinning dust emission (Tibbs et al. 2012). 

About 50 molecular clouds sit in the proximity of W43, and a giant cloud with a diameter of $\sim 20$ pc surrounds W43 itself \citep[][and references therein]{Bally10}. 
In total, $\sim$75 molecular clouds can be found in SDPF1 \citep[][see Figure \ref{fig:pmw_l30}]{Rathborne09}, together with $\sim 340$ Infrared Dark Clouds (IRDC) from 
the \citet{Peretto09} catalogue. 

A cold layer of atomic hydrogen \citep{Gibson04} has also been observed in SDPF1. As discussed in Section \ref{sec:missing_column}, 
this layer is seen in absorption in the HI 21cm line. It is composed of a combination of 
HI Self-Absorption features \citep[HISA, e.g.][]{Gibson10}, which require a warmer background provided by HI itself, and HI Continuum Absorption 
features \citep[HICA,][]{Strasser04}, which instead require a continuum background. The most relevant features are seen in correspondence of W43.

A detailed investigation of the properties of the ISM in SDPF1 is presented in \citet{Bernard10}. By comparing 
the Hi-GAL maps with a 3D-extinction map, these authors find that dust temperature is higher when the LOS intercepts the spiral arms. The relation 
between dust emissivity spectral index $\beta$ and dust temperature T$_{d}$ is instead analysed by \citet{Paradis10}, by combining Hi-GAL and IRAS data. 
Evidence for a T$_{d}-\beta$ anti-correlation is reported in their work, although the authors caution against possible spurious effects due to temperature mixing 
along the LOS.

\begin{figure}
\centering
\includegraphics[width=8cm]{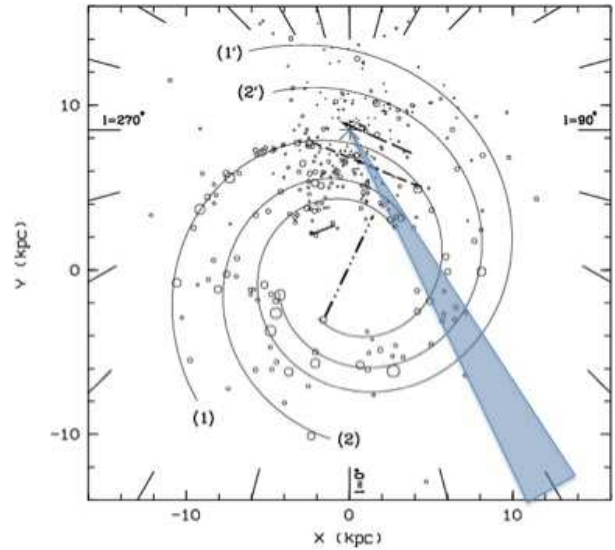}
\caption{
Superposition of the \l30 LOS with the \citet{Russeil03} Galactic model. The circles represent the star forming complexes used by the author to estimate the Galactic spiral arm pattern.  
The best-fit model has four arms. 1: Sagittarius-Carina arm; 2: Scutum-Crux arm; 1 arcmin: Norma-Cygnus arm; 2 arcmin: Perseus arm. A star denotes the position of the Sun. 
The \l30 LOS is tangent to the Scutum-Crux arm. The short dashed line, which passes through the \l30 LOS, is the expected departure 
from a logarithmic spiral arm observed for the Sagittarius-Carina arm. More information about the model can be found in \citet{Russeil03}.}
\label{fig:Galactic_model}
\end{figure}

\begin{figure*}
\centering
\includegraphics[width=12cm]{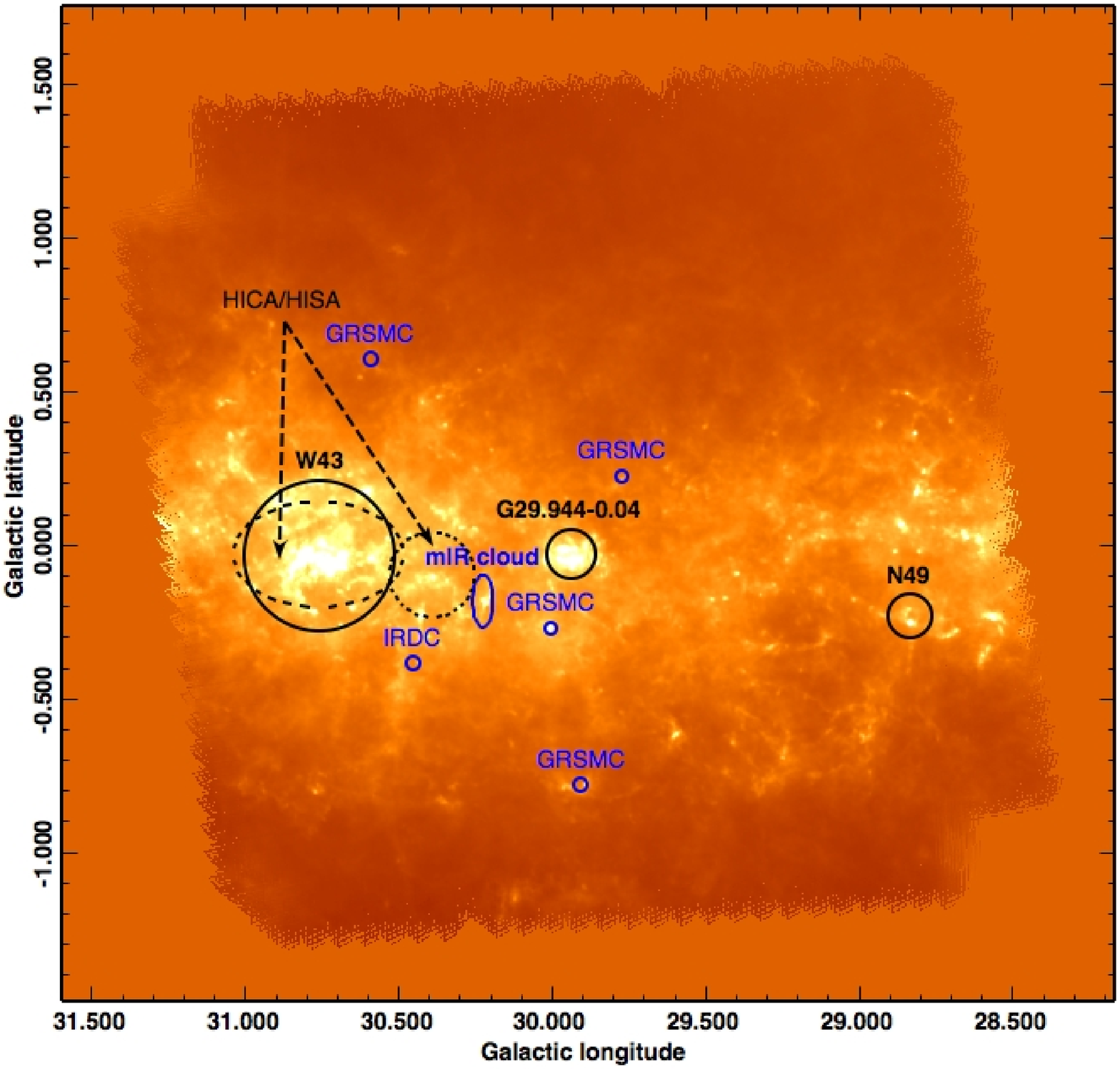}
\caption{Hi-GAL 500 $\mu$m data for SDPF1. The black circles represent W43, G29.944-0.04 and N49. The blue circles denote a sub-samples of the $\sim$75 molecular clouds found in SDPF1, as well as one of the 
most prominent IRDCs catalogued by \citet{Peretto09} and an example of a mid-IR bright and mid-IR dark cloud studied by \citet{Battersby11} in these  2$\times$2 square degrees. Finally, the black-dotted circles indicate HI absorption features,
a combination of HICA and HISA (see Section \ref{sec:l30_field} for details).}
\label{fig:pmw_l30}
\end{figure*}

\section{IR and Ancillary radio data}\label{sec:infrared_ancillary}
In this Section we describe the input IR data and the ancillary radio data used to trace the different gas phases.

\subsection{IR data}\label{sec:IR_data}

We consider IR imaging data at 7 different wavelengths covering the range 8 $\mu$m $\leq\lambda\leq$ 500 $\mu$m. 
From the GLIMPSE and MIPSGAL surveys we use the 8 $\mu$m and 24 $\mu$m data respectively. The map resolution is 
2 arcsec for GLIMPSE 8 $\mu$m and 6 arcsec for MIPSGAL 24 $\mu$m . Details of the data reduction can be found in \citet{Benjamin03} 
and in \citet{Carey09} for GLIMPSE and MIPSGAL respectively. 
At longer wavelengths, we use data from the Hi-GAL survey, which has mapped the whole Galactic Plane in five bands in the range 
70 $\mu$m $\leq\lambda\leq$ 500 $\mu$m. The first Hi-GAL survey covers almost 280 square degrees of the inner Galaxy, with a $2^{\circ}$ wide strip centred on the Galactic Plane in the longitude range $\vert l\vert\leq 70^{\circ}$. The scanning strategy of the survey is organized in tiles of 2$\times$2 square degrees. During the \textit{Herschel} SDP two Hi-GAL 
tiles were observed, the SDPF1 field studied in this work and another, which also covers 2$\times$2 square degrees, centred on \textit{(l,b)} = (59$^{\circ}$, 0$^{\circ}$). 
The data have been taken with both the PACS \citep{Poglitsch10} and SPIRE 
\citep{Griffin10} instruments in parallel mode and are reduced with the ROMAGAL pipeline \citep{Traficante11}. PACS has observed the sky at 70 $\mu$m and 160 $\mu$m, while 
SPIRE at 250 $\mu$m, 350 $\mu$m and 500 $\mu$m. The Hi-GAL spatial resolution is $\simeq10$ arcsec and $\simeq13.5$ arcsec for PACS and 18, 24 and 34.5 arcsec 
for SPIRE \citep{Traficante11}. The  zero-level offsets in the Hi-GAL maps are evaluated as described in \citet{Bernard10}.

The seven maps are point-source subtracted in order to avoid contamination at high spatial frequencies. Point sources were identified and removed by using two software tools, one tailored for Spitzer and the other for \textit{Herschel}. For the Spitzer data,  ie. IRAC 8 $\mu$m and MIPS 24 $\mu$m, we used the Spitzer Science Center
APEX package \citep{Makovoz05} under the MOPEX tools{\footnote{http://irsa.ipac.caltech.edu/data/SPITZER/docs/dataanalysistools/tools/mopex/}}. 
For the \textit{Herschel} data, for which we had to build our own Point Spread Functions from the data itself, we used ``Starfinder"{\footnote{http://www.bo.astro.it/StarFinder/paper6.htm}} \citep{Diolaiti00}. 
In both cases the removal is carried out in such a way that the residuals from the point source subtraction match as close as possible the noise properties of the original images.

The seven IR maps are shown in Figure \ref{fig:IR_maps}.

\subsection{Ancillary data}\label{sec:ancillary_data}

\subsubsection{Atomic hydrogen}\label{sec:atomic_column}

The atomic gas phase (HI) is traced with the 21 cm line, which can be easily detected across the Galaxy \citep{Stil06}. 
We used the VLA Galactic Plane Survey (VGPS) data, covering the longitude region 
$18^{\circ}\leq l\leq67^{\circ}$ for $\vert b\vert\leq1.3^{\circ}$ up to $\vert b\vert\leq 2.3^{\circ}$ \citep{Stil06}. 
These data are part of the International Galactic Plane Survey (IGPS) which includes the VGPS, 
the Canadian Galactic Plane Survey (CGPS) and the Southern Galactic Plane Survey (SGPS) and covers $\sim$90 percent of the 21 cm line 
Galactic disk emission. The VGPS brightness temperature (\Tb) maps have an angular resolution of 1 arcmin with a velocity resolution of $1.56$ km s$^{-1}$. The survey covers the velocity range $-113\leq \mathrm{V_{LSR}}\leq 167$ km s$^{-1}$. The \textit{r.m.s.} noise per channel is $1.8$ K on average, depending on the location and velocity \citep{Stil06}.

Since the VGPS 21 cm line data are continuum subtracted, a few pixels in correspondence of bright HII regions have very low (even negative) values due to strong HI continuum emission which arises from the HII regions. In order to identify and flag these pixels, we have used the VGPS continuum data \citep{Stil06}. We masked the pixels in all velocity channels of the 21 cm data cube with continuum temperature T$_{c}\ge 50$ K. This threshold allows us to mask pixels in correspondence of bright HII regions but without including the cold diffuse regions, which also appear as absorption features in the HI 21 cm data (see Section \ref{sec:missing_column}). In total, less than 1 percent of the pixels have been masked, the majority of them in correspondence of W43. The same pixels flagged in the 21 cm data cube were also flagged in the data cubes for the other gas tracers.

\subsubsection{Molecular hydrogen}\label{sec:H2_column}

The molecular gas (\H2) has no observable transitions under the conditions typical of molecular clouds, but it can be indirectly traced 
with carbon monoxide (CO) emission. \H2 is primarily traced by measuring the $J=1\rightarrow0$ rotational 
transition line of the most abundant CO isotope in the Galaxy, \12CO. However, SDPF1 is characterized by the presence of many molecular clouds, IRDCs and HI 
absorption features (see Section \ref{sec:l30_field}). These potentially cold, dense regions are better traced with the $J=1\rightarrow0$ line emitted from \CO13 
isotope (see Section \ref{sec:molecular_column}).

The $J=1\rightarrow0$ lines of $^{12}$CO and $^{13}$CO used in this work were observed with the Massachusetts-Stony Brook Galactic 
plane CO survey \citep[UMSB,][]{Sanders86} and the Galactic Ring Survey \citep[GRS,][]{Jackson06} respectively. 
Both surveys were carried out with the Boston-FCRAO 14m telescope. The UMSB survey covers the range $8^{\circ}\leq l\leq 90^{\circ}$, 
$\vert b\vert\leq 1.05^{\circ}$. The spatial resolution is 44 arcsec, even though the sky was 
sampled with a 3 arcmin step. The pixel size of the map is 3 arcmin, which subsamples the telescope beam and does not preserve the spatial information below this value. In order to alleviate the effect of the under sampling, we first regridded the data to 44 arcsec, and then assigned each pixel value in the initial map to a pixel in the new map. We then interpolated consecutive pixel values using a cubic interpolation routine, and finally smoothed and re-binned the map to the original 3 arcmin spatial resolution. This approach assures a better continuity in surface brightness across the map. However, we notice that the arc minute structures present in the data are smeared out at our final working resolution of 14.8 arcmin (see Section 4.1), and the results of the analysis do not vary significantly if we do not apply the cubic interpolation step described above. 

The spectral resolution is 1 km s$^{-1}$ with a \textit{r.m.s.} sensitivity of 0.4 K per velocity channel.  

The GRS has mapped the first quadrant of the Milky Way for $18^{\circ}\leq l\leq 55.7^{\circ}$, $\vert b\vert\leq 1^{\circ}$. 
The sensitivity is $\leq$0.4 K per velocity channel, the spectral resolution is 0.212 km s$^{-1}$ and the angular resolution is 46 arcsec sampled with a 22 arcsec grid. 

Since we use both the CO isotopes to evaluate molecular hydrogen column density (see Section \ref{sec:molecular_column}) we rebinned the \12CO and \CO13 data on the same grid and, using a gaussian kernel, we convolved both datasets to 9 arcmin. The convolved maps have a pixel size of 3 arcmin, which is the same spatial resolution of the \12CO map and allows us to properly sample the kernel.

\subsubsection{Ionized hydrogen}

The Warm Ionized Medium (WIM) is generally traced using either free-free continuum emission or optical (e.g. H$\alpha$) and radio (RRL) transition lines. 
In the Introduction, we have already discussed the limitations of using free-free emission in the context of inversion works. Both the H$\alpha$ line and RRLs 
are part of the cascade of recombination transitions. However, the H$\alpha$ line is emitted when a transition occurs between relatively low quantuum levels, i.e. from 
$n$ = 3 to 2, while RRLs correspond to transitions between high quantuum levels, with $n$ typically $>$ 40. With respect to H$\alpha$, RRLs have the advantage 
of not being affected by dust absorption \citep[e.g.][]{Dickinson03}, thus providing a {\em{clean}} view of the ionized gas across the Galaxy. 

In this work, to trace \HII\ and evaluate the corresponding column density we make use primarily of RRL data. In particular, we use RRLs corresponding 
to three different transitions, i.e. 
H166$\alpha$, H167$\alpha$ and H168$\alpha$. These have been observed by the HI Parkes All-Sky Survey \citep[HIPASS,][]{Staveley-Smith96} and the associated Zone of Avoidance (ZOA), aimed at detecting galaxies in the local Universe. The integration time of the ZOA survey, 2100 s per beam, is five times higher than that of the HIPASS. The data from the two surveys have been combined and the three RRLs have been stacked to achieve a final \textit{r.m.s.} sensitivity of 3 mK/beam/channel. The beam is 14.8 arcmin, the pixel dimension is set to 4 arcmin and the spectral resolution is 20.0 km s$^{-1}$ \citep{Alves11}. Details of the data reduction are given in \citet{Alves10}.

We complement the information provided by the RRL data, especially in terms of electron density of the observed ionized gas, with free-free radio continuum data. 
The free-free data for our region are available from the observations made at 5 GHz (i.e. 6 cm) with the Parkes 64 m telescope \citep{Haynes78}, digitised by the Max Planck Institute in Bonn{\footnote{http://www3.mpifr-bonn.mpg.de/survey.html}}.
The survey covers the range $-170^{\circ}\leq l\leq40^{\circ}$, $\vert b\vert\leq1.5^{\circ}$ at an angular resolution of 4.1 arcmin and a sensitivity of 
around 10 mJy. We apply baseline/offset removal (of 0.1 K and 1 K respectively) and estimate/subtract the synchrotron contribution according to 
Figure 7 in \citet{Paladini05}.

\section{The analysis method}\label{sec:model_data}

In the following we describe the inversion method used to disentangle the dust properties in each gas phase and along different 
Galactocentric positions. The model requires a subdivision of the Galaxy into Galactocentric bins (hereafter referred to as \textit{rings}) and evaluation of the gas column density for each phase (Section \ref{sec:dataset}).

\subsection{The 3D-inversion model}\label{sec:model}

We follow the prescription of the inversion model of \citet{Bloemen90} by assuming that dust emissivities vary with  
Galactocentric distance, thus remaining constant within fixed Galactocentric rings.  
This working hypothesis, although not entirely true, has yet its foundations in the real-case scenario. For instance, \citet{Dunham11} find, by 
comparing the emission at 1.1mm with that for different transitions ((1,1), (2,2) and (3,3)) of NH$_{3}$, that the properties of 
cold clumps seem to vary as a function almost exclusively of Galactic radius. 
In our inversion model, we further assume a single dust temperature and gas-to-dust mass ratio, hence 
a constant emissivity value per H atom, for each gas phase and Galactocentric ring. We emphasise that this hyphothesis implies that dust emits (or absorbs) coherently within each gas phase and Galactocentric ring, thus ignoring local effects which may alter the emissivities along specific LOS, and this is an intrinsic limitation of the model. Under these assumptions, however, we can express the IR emission at a wavelength $\lambda$ in a pixel $j$, $I^{j}(\lambda)$, as a linear combination 
of gas column densities and dust emissivities associated with each gas phase in each ring $i$

\begin{centering}
\begin{eqnarray}\label{eq:decomposition}
I^{j}(\lambda) & = & \sum_{i=1}^{n}\epsilon_{\mathrm{H_{I}}}(\lambda,R_{i})N_{\mathrm{H_{I}}}^{j}(R_{i}) + \epsilon_{\mathrm{H_{2}}}(\lambda,R_{i})N_{\mathrm{H_{2}}}^{j}(R_{i}) + \nonumber \\  
               &   &   + \epsilon_{\mathrm{HII}}(\lambda,R_{i})N_{\mathrm{HII}}^{j}(R_{i})
\end{eqnarray}
\end{centering}

In the expression above, $\epsilon_{\alpha}$ are the dust emissivities (in units of MJy sr$^{-1}$ / $10^{20}$ atoms) for the three gas phases ($\alpha=(\mathrm{H_{I}}, \mathrm{H_{2}}, \mathrm{HII})$) 
in each ring $R_{i}$, and $N_{\alpha}^{j}$ are the corresponding column densities in pixel $j$. The summation is taken over the $n$ rings adopted 
for the decomposition. Once the gas column densities have been evaluated for each ring, the model allows us to disentangle the fraction of the integrated IR emission which is generated within 
each ring and gas phase, and to estimate the corresponding dust emissivities by solving Equation \ref{eq:decomposition} with a least-squares fit analysis.

The column density for each gas component and ring is convolved with a gaussian kernel at the lowest available resolution, 14.8 arcmin (the RRL data resolution), and the maps rebinned with a pixel scale of 4 arcmin. We note that this spatial resolution is four times  higher than in 
previous inversion works ($\simeq1^{\circ}$). Due to the reduced size of the region under investigation (4 square degrees), the number of available pixels 
that we can use to estimate the dust emissivities is only $\sim 700$ \citep[i.e. a factor $\sim10^{2}$--$10^{3}$ lower with respect to, e.g.,][]{Paladini07,Planck_Marshall11}. 
Since the error bars associated with the emissivities in our model are estimated with a bootstrap procedure, a small number of samples induces an increase of the 
random sampling errors arising from the bootstrap method itself \citep{Efron79}. In Section \ref{sec:no_HII} we will show that, if we do not include in the inversion matrix the RRL data, we can afford working with a smaller pixel size, although at the expense of not being able to correctly account for the contribution of 
dust emission associated with the ionized medium.

Noteworthy, the IDL{\footnote{Interactive Data Language}} code which evaluates the gas column densities and solves Equation \ref{eq:decomposition} is based on the code of \citet{Paladini07}. The code 
has been fully re-designed and now only takes $\simeq 4$ minutes 
to complete the inversion of a 2$^{\circ}\times$2$^{\circ}$ Hi-GAL tile, 
rather than $\simeq$ 100 minutes required by the old version. The test has been performed on an Intel Core 2 Duo at 2.2 GHz. This code optimization turned out to be critical for this work, as it allowed to test 
a large number of model configurations.

\subsection{Selection of Galactocentric rings}\label{sec:ring_selection}

\begin{figure*}
\centering
\includegraphics[width=12cm]{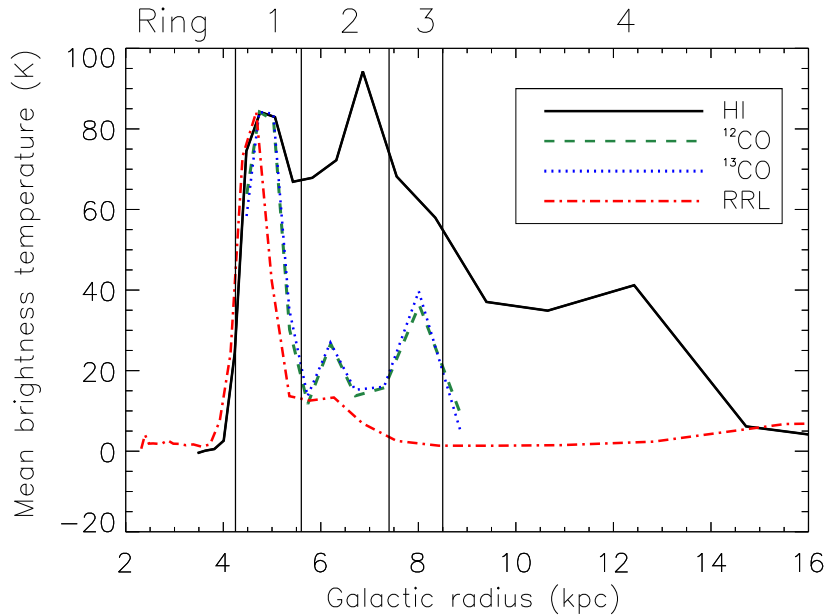}
\caption{Mean brightness temperature, as a function of Galactocentric radius, for HI (solid line), \12CO (dashed line), \CO13 (dotted line) and RRLs (dash-dotted line), respectively. 
Each datacube has been rebinned to the same velocity resolution, namely 20.0 km s$^{-1}$. For each gas phase and velocity channel, a median brightness temperature has been computed. 
All datasets are normalized to the amplitude of the first HI peak. The boundaries of the Galactocentric rings are R=[4.25, 5.6, 7.4, 8.5,16.0] kpc.}
\label{fig:radius_vs_intensity}
\end{figure*}

Figure \ref{fig:radius_vs_intensity} presents the median brightness temperature \Tb values for each gas tracer as a function of the Galactocentric radius. Radial velocities are converted into 
Galactocentric distances by applying the \citet{Fich89} rotation curve and by assuming that the circular motion is the dominant component of velocity. With this assumption, the emission at a particular radial velocity increment translates identically to emission at a single Galactocentric bin. We note also that since the Galactocentric distance is a function of the radial velocities and of the longitude (see the \citet{Fich89} model), some velocity channels can have part of the emission belonging to one ring and part of it that belongs to the subsequent ring. The longitude value that determines the separation is in correspondence of the Galactocentric distance equal to the ring edges.

In order to allow comparison among different data sets, all data cubes have been rebinned with the RRL velocity resolution, i.e. $~20.0$ km s$^{-1}$. 
In addition, the mean profiles for each gas phase have been normalized to the peak in the first Galactocentric ring of the HI profile. 

From Figure \ref{fig:radius_vs_intensity}, we see that HI has two prominent peaks, one at R$\simeq$ 5 kpc and the other at R$\simeq$7 kpc. There is no clear correlation between HI emission and spiral arms pattern in our Galaxy \citep[e.g.][]{Gibson05}, however the first peak is likely due to emission from 
the Scutum-Crux arm, being located at the tangent point (see Figure \ref{fig:Galactic_model}). HI data can be used to trace the Galactic edge up
to distances of $\sim20-25$ kpc from the Galactic centre, depending on the LOS \citep[e.g.,][]{Weaver74}. In our case, we truncate the 
HI data to R=16 kpc since, beyond this value, the S/N drops significantly.

The molecular gas peaks at approximately R$\simeq$4.5 kpc and R$\simeq$8 kpc. The first peak falls in correspondence of the the Scutum-Crux arm. 
The second peak in CO data is either produced by \H2 located in the local Sagittarius-Carina arm or, alternatively, in the interarm region between the Sagittarius-Carina arm  
(at the second passage through the \l30 LOS) and the Perseus arm. It is commonly accepted that interarm regions, despite being characterized by an average
gas density $\sim3.6$ lower than the spiral arms \citep{Ferriere01}, are not devoid of molecular gas. We truncate both the \12CO and \CO13 data at R$\simeq$8.5 kpc since the \H2 column density is known to drop off rapidly beyond R=$R_{\sun}$ \citep[e.g.][]{Combes91,Ferriere01}.

RRL data also peaks around R$\simeq4.5-5$ kpc, in the Scutum-Crux arm region. This arm contains 80 percent (i.e. 23 out of a total of 29) of the known HII regions in SDPF1  \citep{Anderson09}, 
including the W43 complex and the bright source G29.944-0.04. This trend mimics the overall radial distribution of ionized medium across the Galaxy \citep{Ferriere01}. 
The RRL signal decreases exponentially in the outer Galaxy, being consistent with pure noise at R$\simeq$8.5 kpc. 

To perform the 3D-inversion, we must subdivide the Galaxy into Galactocentric rings by minimizing their cross-correlations. In previous analyses, this step has been performed 
following a mathematical approach, i.e. using the eigenvalues of the cross-correlation matrix. In our case, we use the physical properties of the gas tracers to guide our decomposition. 
In Figure \ref{fig:radius_vs_intensity}, four distinct regions of emission can be clearly identified. The first region (Ring 1, 4.25 kpc $\leq$R$\leq$ 5.6 kpc) is characterized by a peak of emission 
in all tracers (HI, \12CO and \CO13, RRLs) and defines the edges of the Scutum-Crux arm. The second region (Ring 2, 5.6 kpc $\leq$R$\leq$ 7.4 kpc) presents a prominent feature in HI. The third region (Ring 3, 7.4 kpc $\leq$R$\leq$ 8.5 kpc) 
contains significant CO emission and cold HI features (HICA and HISA, see Section \ref{sec:missing_column}), while HII emission is entirely accounted for by a diffuse component. 
Finally, region four (Ring 4, 8.5 kpc $\leq$R$\leq$ 16 kpc) corresponds to the outer Galaxy and is visible in HI emission only.

\subsection{Column density evaluation}\label{sec:dataset}

In the following we describe how we evaluate the column densities of each gas phase using the tracers described in Section \ref{sec:ancillary_data}.

\subsubsection{Atomic hydrogen column density}\label{sec:HI_data}

Due to the blending in SDPF1 of cold and warm HI, the 21 cm line brightness temperature, \Tb, in each ring cannot be converted into HI column density, 
N(HI), with a fixed spin temperature, \Ts, following the standard approach of previous inversion works 
\citep[see e.g.][]{Paladini07,Planck_Marshall11}. In fact, since in the cold neutral medium (CNM) 10 K $\leq T_{s}\leq100$ K, and in the warm neutral medium (WNM) 
$T_{s}$ ranges from $\sim500$ K to $T_{s}\sim5000$ K \citep[e.g.][]{Strasser04,Heiles03}, the assumption of a constant $T_{s}$ is too simplistic. 
However, the HI column density can be evaluated by assuming a single emission component in the \textit{optically thin} limit \citep{Dickey90}:

\begin{equation}\label{eq:hi_column_opt_thin}
\mathrm{N(HI)}_{\tau << 1} \simeq C_{0} \int{T_{b}(\mathrm{v}) d\mathrm{v}} \qquad 10^{20} \ \mathrm{H\ atoms / cm}^{2}
\end{equation}   
where $C_{0}=1.823\times 10^{18}$ cm$^{2} \mathrm{/ (K\  km\ s^{-1})}$.  \Tb$(\mathrm{v})$ is the observed 21 cm line brightness temperature at the velocity position $\mathrm{v}$. The integral is taken over the ensemble of the velocity channels corresponding to each ring. The integrated \Tb map in Ring 3 shows an offset around \textit{l}=29.5$^{\circ}$ and \textit{l}=30.8$^{\circ}$ in correspondence of the longitudes in which some velocity channels have their emission split among Ring 3 and Ring 2. These offsets are present also in Ring 2. However, the steps amount to only a few K and induce a step in column density of $\simeq 1\times 10^{20}\  \mathrm{H\ atoms / cm}^{2}$ from the two regions, lower than the mean and the \textit{r.m.s.} value of the map ($\simeq 20\times 10^{20}\  \mathrm{H\ atoms / cm}^{2}$ and $\simeq 2.5\times 10^{20}\  \mathrm{H\ atoms / cm}^{2}$ respectively) and in Ring 2 they are not visible (see Figure \ref{fig:HI_column_density}). 

The HI column density evaluated under these hypothesis could underestimate the ``true" column density by a factor of 1.3 -- 1.5, depending on the location \citep{Strasser04}. As a simple test, we tried to rescale the HI column density in each Ring by a fixed fraction (the HI column density in each Ring has been multiplied by a factor 1.3). The estimated emissivities do not sensibly change from the results showed in Section \ref{sec:results_discussion}, most likely because the regions in which the assumption of having one component in the optically thin limit fails have their own spatial structure along the LOS. 

Since in previous inversion studies the HI column density was evaluated by assuming a fixed spin temperature, we compared the HI column density and dust emissivities for the three phases estimated with the two methods: the optically thin limit and the assumption of a spin temperature. Setting a unique value for the spin temperature is not straightforward. However, considering that \Ts has to be higher then the observed \Tb, we can fix \Ts at the 
highest \Tb value measured along the LOS \citep[e.g.][]{Johannesson10}. In SDPF1, \Tb peaks at \Tb=160 K. We then adopt this value to compute N(HI) for all pixels. In doing so, we need to remember that this operation can lead to overestimate \Ts in regions far from the peak. The total HI column density estimated with this approach is $\simeq30$ percent higher than the value obtained in the optically thin limit, with peaks of $\sim80$ percent for pixels in correspondence of very bright HI emission. Interestingly, such a column density increase is accompanied by a change in morphology of HI emission in the rings. We also run a series of tests adopting a spin temperature in the range $160\leq T_{s}\leq250$ K, following \citet[][]{Planck_Marshall11}. All these tests show that dust emissivities do not change significantly over this temperature range. Median emissivity differences (for all rings) between the isothermal and optically thin models in HI, H2, and HII are, respectively, 55 , 2, and 8 percent for \Ts = 160 K and 29, 3, and 6 percent for \Ts = 250 K. In addition, the dust temperature derived from the emissivities using both methods are the same, as is the convergence of the code in different rings (see Section \ref{sec:results_discussion}), which are the main results of this work.

Both models are not defining accurately the temperature distribution along the LOS, but they are useful to describe a limiting case (the optically thin assumption) or a weighted average (the spin temperature assumption) for obtaining column densities. At the same time, assuming the optically thin condition does not require the evaluation of the optical depth or the spin temperature along the LOS. It represents a well-defined lower limit to the total HI column density, despite the fact that it cannot correctly reproduce  the amount of gas along the LOS in cold, dense regions. We therefore decided to apply Equation \ref{eq:hi_column_opt_thin} without any further assumptions.

\subsubsection{Molecular hydrogen column density}\label{sec:molecular_column}

The \H2 column density is evaluated in the Local Thermal Equilibrium (LTE) approximation by assuming a constant  $[\textrm{CO}/\textrm{H}_{2}]$ ratio and by using both the $^{12}$CO and $^{13}$CO isotopes. 

In the denser and colder part of the molecular clouds,  \12CO becomes optically thick while $^{13}$CO can still be considered optically thin. On the other hand, 
in the external regions of the clouds \12CO is optically thin but $^{13}$CO emission is too weak to detect. 
Thus, we use $^{12}$CO where only this isotope is observed, and both $^{13}$CO and $^{12}$CO otherwise. We emphasise that this approach, although it allows us to better recover the molecular hydrogen column density, is nevertheless affected by some limitations. Noteworthy, CO lines can be self-absorbed, and \H2 cannot be traced in optically thick regions \citep{Pineda08}.

For each velocity channel, we searched for pixels where both $^{12}$CO and $^{13}$CO are detected following these steps:

\begin{enumerate}
\item[1.] we compute the median value over all the pixels;
\item[2.] we compute the sigma ($\sigma$) with respect to the median;
\item[3.] we isolate the pixels with a value greater than the median value plus 2.5$\sigma$ in both $^{12}$CO and $^{13}$CO maps;
\item[4.] we flag the pixels isolated in both maps as common regions.
\end{enumerate} 

The sigma evaluated with respect to the median, rather than the mean,  assures more stability to fluctuations induced by the signal, especially in channels where bright star forming regions 
strongly contribute to the overall emission. The 2.5$\sigma$ threshold can, in some cases, lead to an inclusion of noisy outliers. A value higher than 2.5$\sigma$, 
however, results in missing pixels where $^{13}$CO emission is clearly seen, while a lower 
value induces the identification of too many outliers. With a threshold of 2.5$\sigma$ we identify an average of 5 percent common pixels in each velocity channel, concentrated in the densest regions.
Assuming only residual white noise, the outliers are randomly distributed across the map, so the probability that the same outliers are falsely detected in both the \12CO and \CO13 
maps is $<0.1$ percent.

In pixels where only \12CO is observed, \Tb is converted into molecular hydrogen column density N(\H2) using the optically thin hypothesis.  Then, for 
a given ring and in each pixel, N(\H2) is proportional to $\int{T_{b}(\mathrm{v})}d\mathrm{v}$, where the integration is taken over the velocity range corresponding to that ring. We 
set the conversion factor $X_{\mathrm{CO}}$, that is the ratio of the molecular hydrogen column density (in units $10^{20}$ H\ atoms cm$^{-2}$) to the velocity-integrated CO intensity (K km s$^{-1}$), equal to 1.8 $\times\ 10^{20}$ cm$^{-2}$ (K km s$^{-1}$)$^{-1}$, in agreement with the recommended value suggested by \citet{Bolatto13}. 
Recent studies suggest that $X_{\mathrm{CO}}$ may vary with metallicity, in particular increasing with 
Galactocentric distance by a factor 5 -- 10, with an estimated value at R$\simeq 1.5$ kpc of $X_{\mathrm{CO}}=0.3$ \citep[e.g.,][]{Strong04,Abdo10}. If this is true, our working hypothesis of $X_{\mathrm{CO}}$ = 1.8 has the effect of producing an overestimate of the effective molecular gas column densities.

When $^{13}$CO is also observed, in order to evaluate the column density, $\textrm{N}^{13}(\textrm{CO})$, we need to know the excitation 
temperature $T_{ex}$ and the \CO13 optical depth $\tau_{13}$ at each velocity $v$ \citep[][]{Duval10,Pineda10}. 
The details of this analysis are given in Appendix \ref{sec:appendix}. N($^{13}$CO) is then converted into N(H$_{2}$) applying \citep{Stahler05}

\begin{equation}
\mathrm{N({H_{2}})}=7.5\times 10^{5}\ \mathrm{N(^{13}CO)} \qquad (10^{20} \mathrm{atoms}\ \mathrm{cm}^{-2})
\end{equation}

A total of $\sim$27 percent pixels of the column density maps are observed in both the \12CO and \CO13 maps. For $\sim$ 20 percent of these common pixels, the estimated column density is $\geq$20 percent higher than the one calculated with the $^{12}$CO data only, confirming the necessity of combining the information, when this is available, from the two isotopes.

\subsubsection{Ionized hydrogen column density}\label{sec:ionized_column}

At low Galactic latitudes, the Warm Ionized Medium (WIM) is characterized by a thin layer which consists of both individual HII regions and diffuse emission \citep{Paladini05}. 
RRLs allow us to trace simultaneously these two components. 

The RRL brightness temperature peaks in the Scutum-Crux ring where the majority of the HII regions in SDPF1, in particular W43,  
are located (Figure \ref{fig:radius_vs_intensity}) and it is significantly low in the other rings. The ionized hydrogen column density, N(\HII), is proportional to the RRLs brightness temperature, as we are going to show in the 
following, therefore the HII column density also peaks in Ring 1. 

The HII column density can be defined as the mean electron density $\langle n_{e}\rangle$ integrated along the LOS:

\begin{equation}
\mathrm{N(HII)} = \int {\langle n_{e}\rangle} ds
\end{equation}

The quantity above can be estimated from the Emission Measure (EM). If we assume along each LOS a standard effective electron density, $n_{eff}$:

\begin{equation}\label{eq:EM_Ne}
\textrm{EM}=\int n_{e}^{2}ds=n_{eff}\times\int \langle n_{e}\rangle ds=n_{eff}\times \mathrm{N(HII)}
\end{equation}

EM is evaluated in turn from the integral of the RRL line temperature $T_{L}$ \citep{Alves10}:

\begin{equation}\label{eq:TL_EM}
\int T_{L}d\nu=1.92\times10^{3}T_{e}^{-1.5}\ \textrm{EM}
\end{equation}

where $T_{e}$ is the mean electron temperature and $d\nu$ is the frequency interval in kHz . The continuum emission brightness temperature \Tb is also proportional to EM \citep{Mezger67}:

\begin{equation}\label{eq:Tb_EM}
T_{b} = 8.235 \times 10^{-2} a(T_{e})T_{e}^{-0.35}\nu_{\mathrm{GHz}}^{-2.1}(1+0.08)\ \mathrm{EM}
\end{equation}

where the factor $(1+0.08)$ takes into account the additional contribution to \Tb from helium, and $\nu_{\mathrm{GHz}}$ is the frequency expressed in GHz. Combining this Equation with Equation \ref{eq:TL_EM} we obtain:

\begin{equation}\label{eq:TL_Tb}
\frac{\int{T_{L}}dV}{T_{b}} = 6.985\times10^{3}\frac{1}{a(T_{e})}\frac{1}{n\mathrm{(He)}/[1+n\mathrm{(H)}]}T_{e}^{-1.15}\nu_{\mathrm{GHz}}^{1.1}
\end{equation}

In the expression above, $a$(\Te) is a slowly varying function of \Te and $dV$ is expressed in km s$^{-1}$. The factor $n(He)/[1+n(H)]$ is equivalent to the factor (1+0.08) in Equation \ref{eq:Tb_EM}.

In order to derive N(\HII) from Equation \ref{eq:EM_Ne} to \ref{eq:Tb_EM}, we need to know both \Te and \neff\ for each of the physical regimes we are considering, that is HII regions and 
the diffuse medium. At the RRL data angular resolution (14.8 arcmin), each of the 29 HII regions in SDPF1 falls  
within a single pixel, except for the W43 complex. Therefore, apart from W43 which we treat as a special case, we assume that the HII emission in the field originates exclusively in the diffuse emission, thus neglecting the contribution from individual HII regions.

\begin{itemize}
\item\textit{Mean electron temperature:} For HII regions, $T_{e}$ is known to increase with Galactocentric radius \citep[e.g.,][]{Shaver83, Paladini04}, varying in the 
range 4000 to 8000 K. \citet{Alves11} have recently shown that the diffuse ionized gas along the Galactic Plane has a similar electron temperature of that of HII regions. Therefore, for simplicity, we assume a constant $T_{e}$ for both W43 and the diffuse HII component and, 
following \citet{Alves10}, we take $T_{e}$ = 5500 K.\\ 

\item\textit{Effective electron density:}
In general \neff\ strongly depends on the physical conditions of a given environment. Previous inversion works have often used values close or equal to 10 cm$^{-3}$ \citep[e.g.][]{Paladini07,Planck_Marshall11}. 
This value is in agreement with measurements in compact HII regions. In the diffuse ionized medium \neff\ is tipically lower, i.e. $0.03<n_{eff}<0.08$ cm$^{-3}$ \citep[e.g.,][]{Haffner09}, especially at intermediate and high Galactic latitudes. 

To estimate $n_{eff}$ for both the diffuse medium in SDPF1 and W43, we use the Parkes 5 GHz data. To this end, we note that the observed free-free emission is proportional to EM and follows Equation \ref{eq:Tb_EM}. If we assume that $\langle n_{e}\rangle=n_{eff}$, it follows that

\begin{equation}\label{eq:EM_free_free}
n_{eff} = \left(\frac{T_{b}} {8.235} \frac{10^{2}}{a(T_{e})} T_{e}^{0.35}\nu_{\mathrm{GHz}}^{2.1}\frac{1}{(1+0.08)}\frac{1}{d}\right)^{1/2}  
\end{equation}
where $d$ is the linear size of the parcel of emitting gas. We emphasize that Equation \ref{eq:EM_free_free} holds true only if the electron density does not vary significantly 
along the LOS. To ensure that this condition is satisfied, we solve Equation  \ref{eq:EM_free_free} independently for W43 and the diffuse medium:

\begin{itemize}

\item[1.]  \textit{W43 complex}: In this case, the assumption that the emission is dominated by the star forming complex with minor 
contributions from the foreground/background emission is motivated by the fact that \TL in correspondence of the HII region is 
several orders of magnitudes higher with respect to its surroundings.  In the RRL data cube, W43 is at V$_{LSR}$=107 km s$^{-1}$, which corresponds to a Galactocentric radius of
$R\simeq$ 4.65 kpc. The complex is known to be on the near side of the Galaxy \citep{Wilson70}, so its solar distance is D$\simeq$ 5.5 kpc, in agreement with \citet{Bally10}. 
We measure an angular size of $\simeq29.1$ arcmin  which, at a distance of 5.5 kpc, is equivalent to an effective size d$\simeq$47.8 pc. Assuming a spherical geometry, we 
adopt  this value to evaluate \neff\ from Equation \ref{eq:EM_free_free}. For \Tb = $T_{b, W43peak}$, we obtain $n_{eff, W43peak}=43.8\ \textrm{cm}^{-3}$.

\item[2.] \textit{diffuse emission}: For the diffuse component, the definition of the edges of the emitting region is less straightforward. In Section 4.2, we have 
seen that the emission from HII detected through RRLs drops dramatically beyond $R\simeq$ 8.5 kpc. At the far side of the Galaxy the $R\simeq$ 8.5 kpc circle intersects \l30 at a distance of D$\simeq$ 14.7 kpc from the Sun. We take this as the outer boundary of the emitting region. To derive $n_{eff}$, we then first mask 
the pixels across the W43 complex, and then compute the median of the values obtained by applying Equation \ref{eq:EM_free_free}. The resulting 
effective electron density is  $n_{eff, diffuse}=0.39\ \textrm{cm}^{-3}$.

\end{itemize}

At this stage, in order to obtain a continuous solution, we scale $n_{eff, W43peak}$ to $n_{eff, diffuse}$. For this purpose, we use a 2d-Gaussian profile with a full-width-half-maximum equal to the measured angular size of W43. Having estimated $n_{eff}$, we derive N(\HII) by combining Equations \ref{eq:EM_Ne}, \ref{eq:TL_EM} and  \ref{eq:TL_Tb}. 

\end{itemize}

\subsubsection{Column density distribution}\label{sec:total_density}

We compare the contribution to the total gas column density provided by each ring and 
obtain that most of the gas ($\sim$ 65 percent) is located in Ring 1. Ring 2 accounts for another $\sim$ 20 percent, while the 
remaining $\sim$15 percent is distributed between Ring 3 and 4. 

The column density maps for each gas phase and Ring are in Figures \ref{fig:HI_column_density}, \ref{fig:H2_column_density} and \ref{fig:HII_column_density}.

\section{Statistical correlation analysis}\label{sec:pearsons}
The inversion model is based on the hypothesis that the total IR emission at a given wavelength can be decomposed into dust emission associated with different gas phases and Galactocentric rings. This hypothesis implies that the integrated IR dust emission at each given wavelength linearly correlates with the dust emission associated with each gas component and Galactocentric ring and the degree of correlation depends on the contribution of dust associated with each gas component in each ring to the total IR emission. A low degree of correlation can be the natural result of either a gas phase being less abundant (hence with a lower column density) or intrinsically less emissive with respect to other phases.

Two caveats of this approach have to be kept in mind: 

\begin{table*}

\begin{center}
\begin{tabular}{c|c|c|c|c|c|c|c|c|c|c|c|c|c}
\hline
\hline
\multicolumn{14}{c}{\textit{Pearson's coefficient}} \\
& & & & & & & & & & & & &\\
\hline
\textit{Band} & \multicolumn{4}{c}{\textit{Ring 1}} & \multicolumn{4}{c}{\textit{Ring 2}} & \multicolumn{4}{c}{\textit{Ring 3}} & \textit{Ring 4} \\
($\mu$m)  &  \multicolumn{4}{c}{(4.25-5.6 kpc)} & \multicolumn{4}{c}{(5.6-7.4 kpc)} & \multicolumn{4}{c}{(7.4-8.5 kpc)} & (8.5-16.0 kpc) \\
\\
\textit{gas phase} & & HI & \H2 & \HII & & HI & \H2 & \HII & & HI & \H2 & \HII & HI\\

\hline
\\
8 & & 0.51 & 0.74 & 0.82 & & 0.38 & 0.52 & 0.41 & & -0.44 & 0.28 & -0.01 & -0.10\\
\\               
24 & & 0.27 & 0.52 & 0.65 & & 0.21 & 0.34 & 0.28 & & -0.44 & 0.14 & -0.04 & -0.06\\
\\                  
70 & & 0.11 & 0.51 & 0.67 & & 0.19 & 0.31 & 0.40 & & -0.53 & 0.10 & 0.02 & -0.02\\
\\                  
160 & & 0.42 & 0.79 & 0.86 & & 0.39 & 0.55 & 0.53 & & -0.46 & 0.21 & 0.07 & -0.08\\
\\
250 & & 0.47 & 0.85 & 0.87 & & 0.45 & 0.61 & 0.56 & & -0.41 & 0.21 & 0.15 & -0.04\\
\\
350 & & 0.48 & 0.86 & 0.87 & & 0.47 & 0.62 & 0.57 & & -0.41 & 0.22 & 0.14 & -0.06\\
\\
500 & & 0.50 & 0.86 & 0.86 & & 0.48 & 0.62 & 0.56 & & -0.40 & 0.21 & 0.15 & -0.05\\
\hline 
\end{tabular} 
\end{center}
\caption{Pearson's correlation coefficients for the input IR maps and the column density maps for each gas phase and ring. In Ring 1 (which includes the Scutum-Crux arm), there is 
a high degree of correlation between the IR maps - at all wavelengths - and the column density maps of HI, \H2 and \HII. A lower degree of correlation, even a signature of anticorrelation, is instead revealed by the 
Pearson's coefficients between the column densities in Ring 2, 3 and 4 and the input IR maps.} 
\label{tab:scatterplot_columns_IR}
\end{table*}

\begin{itemize}
 \item[1.] in the short IR bands (e.g. 8 \mum), the emission is proportional to the product N$_{H}$ $\times$ G$_{0}$, where N$_{H}$ is the total hydrogen column density and G$_{0}$ 
  represents a scaling of the \citet{Mathis83} radiation field in the solar neighbourhood \citep[see e.g.][]{Compiegne10}. This is also true at 24 $\mu$m and, partly, at 70 $\mu$m, given that these bands 
  are contributed to by emission from very small grains (VSGs) which, as PAHs, are stochastically heated by the local radiation field. Conversely, the emission at wavelength longer than 70 \mum\ is produced by big grains (BGs) which are in thermal equilibrium with the radiation field. The G$_{0}$ intensity defines the BG equilibrium temperature, while the total hydrogen column density determines the absolute level of the BG emission. Hence, if the radiation field (therefore the BG equilibrium temperature) varies smoothly across the region, the intensity variations in the FIR emission are dominated by the column density variations across the field \citep{Compiegne10}. As a result, the net effect is an increasing degree of correlation going from shorter to longer wavelengths.  

\item[2.] A second limitation of the inversion approach lies in the fact that, in order to solve Equation \ref{eq:decomposition}, the column densities associated to each gas phase for each LOS have to be accurately recovered. However, this might not occur, either when part of the gas in a given phase is not traceable by standard methods (i.e. warm \H2), or when a fraction of the gas along the LOS absorbs rather than emitting (i.e. the cold HI). In the first case, the morphology of the column density maps is artificially altered, leading to an ``excess'' of IR emission and to a consequent lower degree of correlation between the column density maps and the input IR maps. In the second case, if the optically thick regions are in correspondence of strong IR emission features, the column density maps, estimated in the optically thin limit, and the IR maps will show an anti-correlation.

\end{itemize}

Such a correlation can be investigated in terms of the Pearson's coefficients \citep[e.g.,][]{Edwards76}. Given two vectors \textit{X} and \textit{Y}, the Pearson's 
coefficient $\rho$ is defined as the ratio between the covariance $cov(X,Y)=\sum_{i=1}^{n}(X_{i}-\bar{X})(Y_{i}-\bar{Y})$ and the product of their standard 
deviations $\sigma_{X}$ and $\sigma_{Y}$

\begin{equation}
\rho(X,Y)=\frac{cov(X,Y)}{\sigma_{X}\sigma_{Y}}
\end{equation}

$\rho$ is defined in the range $\vert \rho \vert\leq 1$ and $\rho \simeq1$ indicates strong correlation, 
while $\rho \simeq$ -1 suggests a strong anti-correlation.

\begin{figure*}
\centering
\includegraphics[width=8cm]{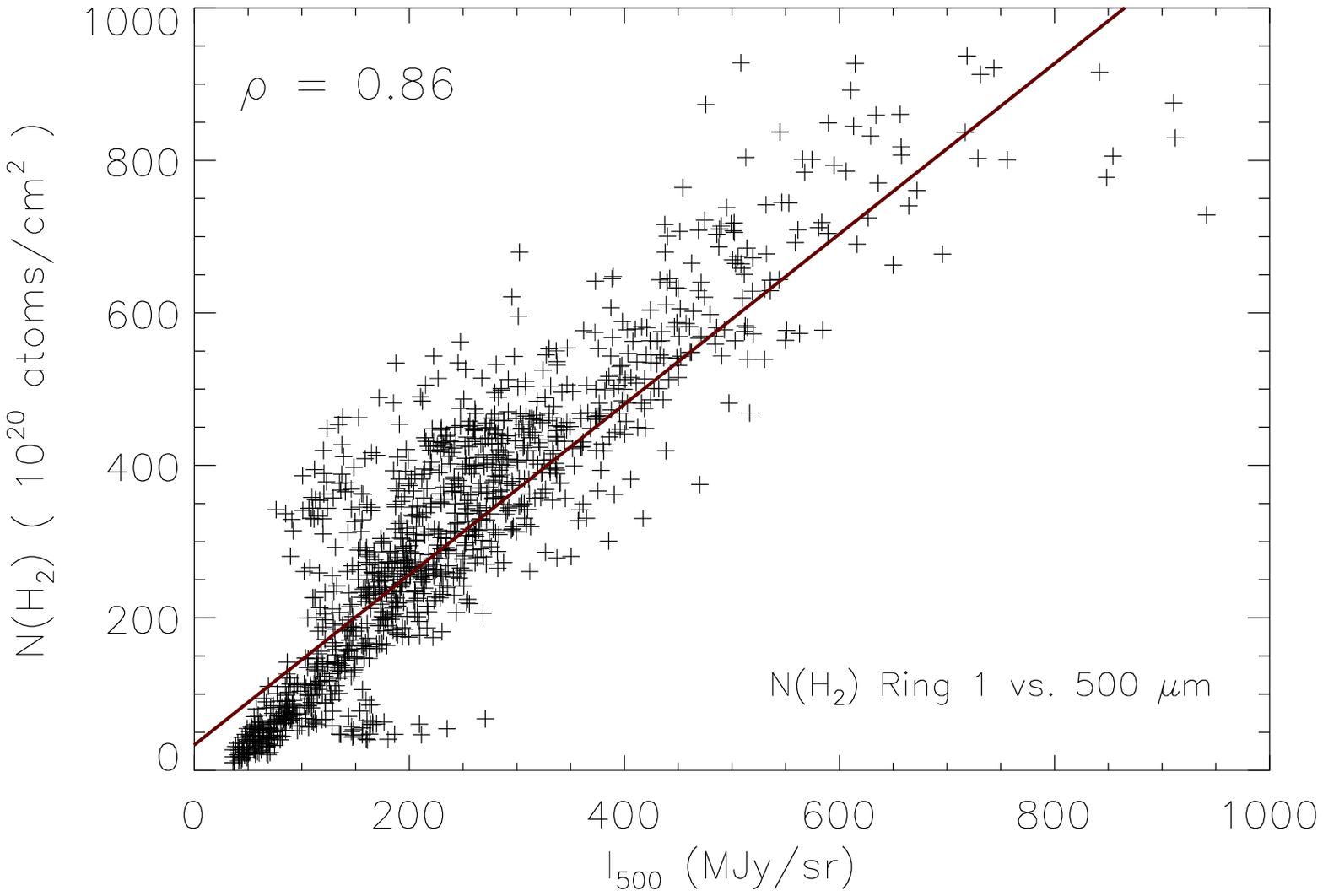}
\includegraphics[width=8cm]{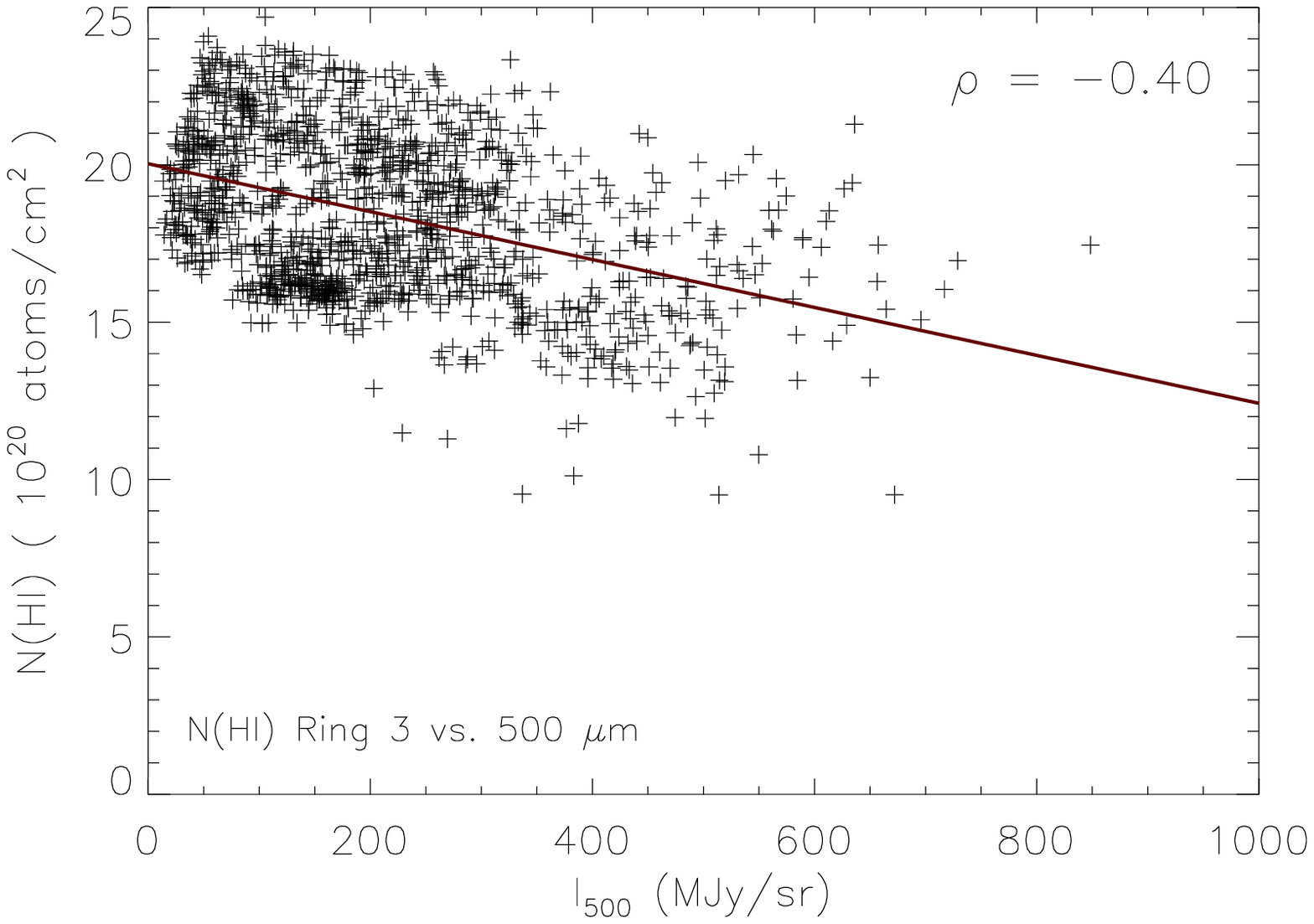}\\
\caption{Correlation plots between Hi-GAL 500 $\mu$m map and N(\H2) Ring 1 map (left image) and Hi-GAL 500 $\mu$m map and N(HI) Ring 3 map (right). The Pearson's coefficients are $\rho=0.86$ and $\rho=-0.40$ respectively, the lowest and the highest between this IR map and the column density maps in this region.}
\label{fig:scatterplot}
\end{figure*}

Table \ref{tab:scatterplot_columns_IR} provides a summary of our computed Pearson's coefficients ($\rho$). Figure \ref{fig:scatterplot} shows two examples of correlation plots, obtained from comparison of the 500 $\mu$m emission with the molecular and atomic column densities. These plots correspond to the highest (N(\H2), Ring 1) and lowest (N(HI), Ring 3) degree of correlation among all the considered cases for the 500 $\mu$m emission.

The correlations in Ring 1 are the highest, 
at all wavelengths and for each gas phase. The correlation is significant (up to 0.87) for both the \H2 and HII column densities, while it  
is noticeably lower for the atomic phase. Averaging across wavelengths, we obtain: $\rho_{HI,Ring1}$ = 0.39, $\rho_{H{_{2}},Ring1}$ = 0.73 and 
$\rho_{H{II},Ring1}$ = 0.80. 

In general, the aromatic infrared bands (AIB, 8 \mum\ and 24 \mum) and the 70 $\mu$m band display a lower degree of correlation with respect to longer wavelengths, hence corroborating the hypothesis 
that they trace the intensity of the radiation field as well as the total hydrogen column density.

For Ring 2, we have that the average correlations are: $\rho_{HI,Ring2}$ = 0.37, $\rho_{H{_{2}},Ring2}$ =  0.51 and
$\rho_{HII,Ring2}$ = 0.48. These values are lower than for Ring 1, especially for \H2 and \HII, indicating that Ring 2 either contributes less than the Scutum-Crux region to the total IR emission or that part of the gas is not entirely seen in emission with the standard tracers.

In Ring 3 the IR maps are poorly correlated with the molecular and ionized gas column densities and partially anticorrelated with the atomic gas column density. 
This is possibly due to the presence of gas not entirely traced by the standard tracers, as supported by observations of a cold layer of HI and strong individual absorption features located at 7.4 kpc$\leq$R$\leq$8.5 kpc 
(see Section \ref{sec:missing_column}).   

A weak anticorrelation is also present in Ring 4. In this case, the negative Pearson's coefficients can be explained through a combination of both 
absorption features and intrinsic weak emission.

\begin{table*}\label{tab:scatterplot_HI_CO_HII}

\begin{center}
\begin{tabular}{c|c|c|c}
\hline
\hline
\multicolumn{4}{c}{\textit{Pearson's coefficient}} \\
& & &\\
\cline{1-4}
\textit{Gas}  &  \textit{Ring 1}& \textit{Ring 2}& \textit{Ring 3} \\
\textit{phases}  & (4.25-5.6 kpc)  & (5.6-7.4 kpc)  & (7.4-8.5 kpc) \\
\hline
\\
HI - \H2  & 0.34 & 0.61 & -0.21\\
\hline               
HI - \HII  & 0.24 & 0.38 & 0.14\\
\hline                  
\H2 - \HII & 0.77 & 0.64 & -0.09\\
\hline 
\end{tabular} 
\end{center}
\caption{Pearson's coefficients for the correlation between different gas phases in each ring. In Ring 1 and 2, the molecular component is well correlated with the ionized component, 
while the atomic component, which is more ubiquitous, is less correlated with both \H2 and \HII. The low/negative 
coefficients in Ring 3 are likely the consequence of an underestimate of the total column density.}
\label{tab:scatterplot_HI_CO_HII}
\end{table*}

The Pearson's coefficients also measure the correlation among the various gas phases in each ring. These coefficients are shown  
in Table \ref{tab:scatterplot_HI_CO_HII}.  The correlation between atomic and molecular gas in Ring 1 is not very strong ($\rho_{HI,H_{2},Ring1}$ = 0.34), likely a consequence of the ubiquitous distribution of 
HI with respect to \H2. Conversely, molecular and ionized hydrogen appear to be strongly correlated ($\rho_{H_{2},HII,Ring1}$ = 0.77), as expected in star forming regions. The correlation between HI and HII ($\rho_{HI,HII,Ring1}$ = 0.24) likely reflects the fact that 
both phases have a diffuse component.

In Ring 2, the Pearson's coefficients follow the same trend as in Ring 1, with a high correlation between molecular and ionized components ($\rho_{H_{2},HII,Ring2}$ = 0.64), confirming the tight spatial 
correlation among these two phases, and a partial correlation among HI and HII. 

In Ring 3, there is evidence of anticorrelation between HI and \H2 components, indicating the presence of pixels with detected \H2 in correspondence of 
HI absorption features. In Section \ref{sec:missing_column}, we investigate the possible reasons behind this effect. The ionized gas in this region 
is dominated by the diffuse component, as there are no cataloged HII regions at 7.4 kpc$\leq$R$\leq$8.5 kpc towards this LOS, and it appears not to 
be associated with either HI nor \H2, as illustrated by the poor correlations (and, even, anticorrelation) with these phases.

\section{Discussion and results}\label{sec:results_discussion}

In the following, we solve Equation \ref{eq:decomposition} to recover the emissivities for dust associated with the different gas phases in each ring. 
The code converges and retrieves positive and reliable  emissivity coefficients only in Ring 1 (Section \ref{sec:dustem}) which, as we show in Section \ref{sec:ring1_dominate} and 
Appendix \ref{app:model_test}, dominates the IR emission at all wavelengths. For Ring 2, 3 and 4, where more than 50 percent of the Pearson's coefficients 
across all the wavelengths are lower than 0.5, the code either does not converge or returns negative emissivities (Section \ref{sec:missing_column}). 
Whereas these results are difficult to interpret at first, they take a more clear meaning 
when we investigate the presence of matter not entirely accounted for by the gas tracers (Section \ref{sec:missing_column} and \ref{sec:extinction}). Finally, we analyze the consequences of excluding the ionized gas component from the decomposition, a procedure 
often adopted in past inversion works (Section \ref{sec:no_HII}).

\subsection{Ring 1: fitting the emissivities with DustEM}\label{sec:dustem}

\begin{figure*}
\centering
\includegraphics[width=10cm]{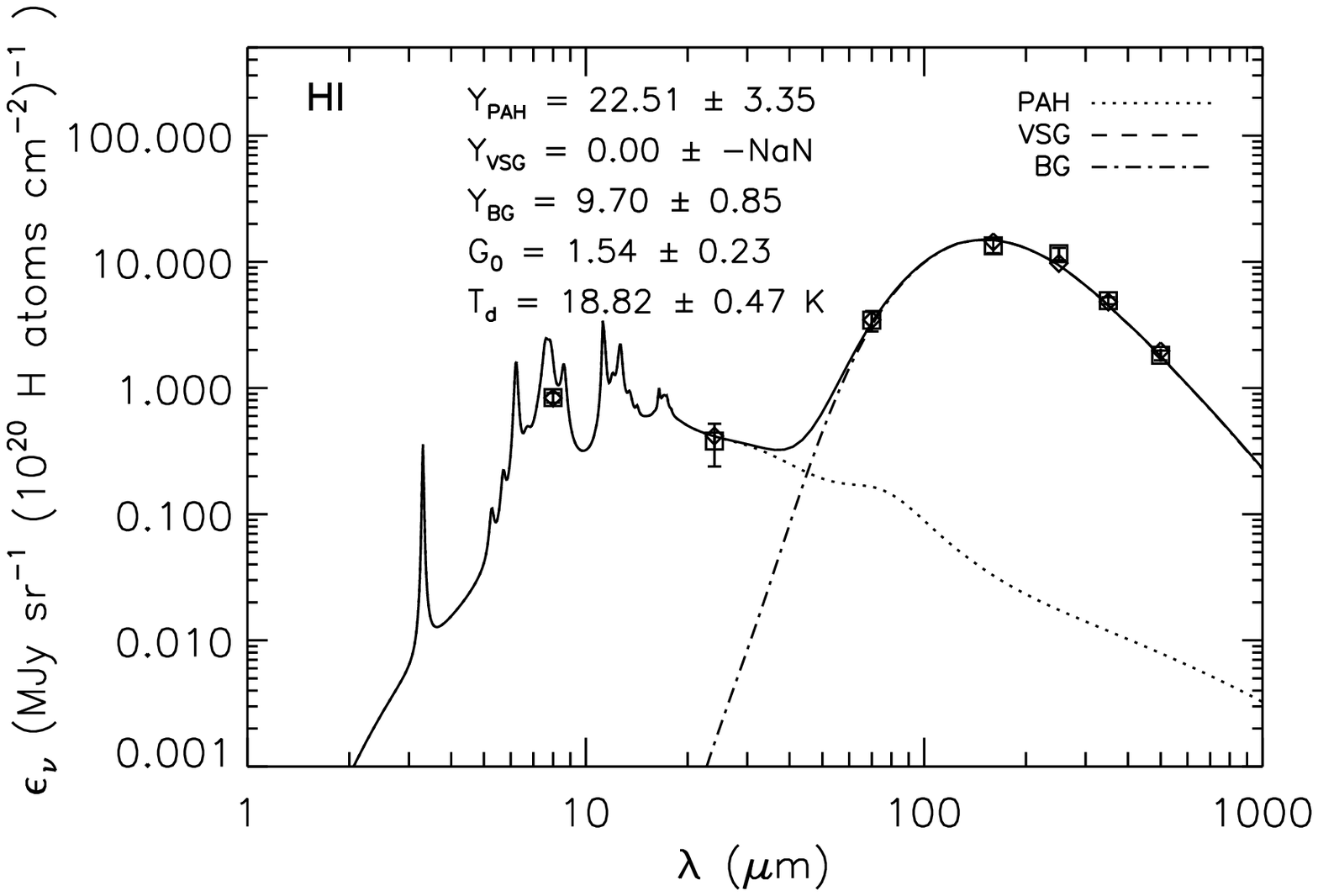}\\
\includegraphics[width=10cm]{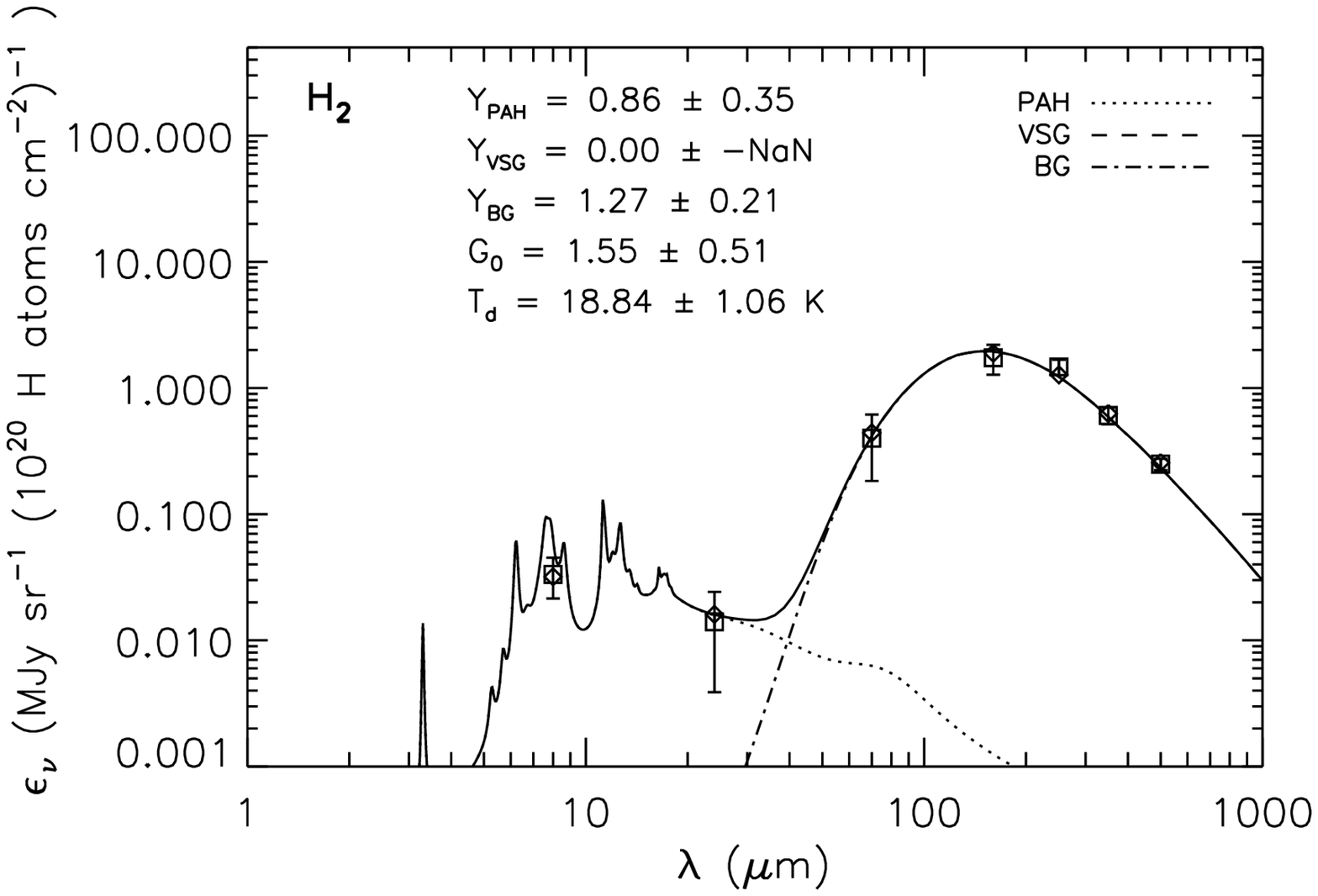}\\
\includegraphics[width=10cm]{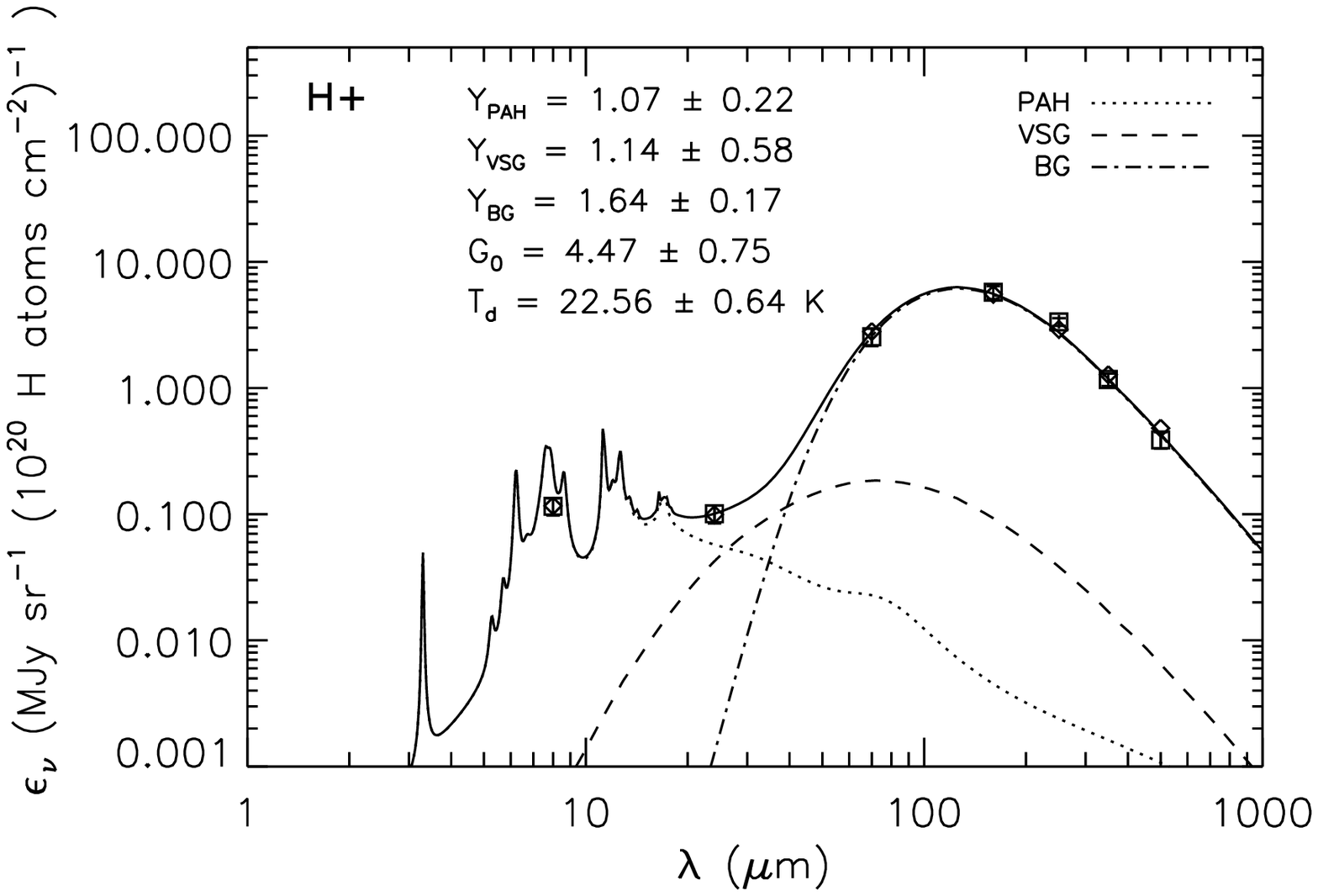}

\caption{DustEM fit of the emissivities evaluated with the inversion model for dust associated with the HI (top panel), \H2 (medium panel) 
and HII (lower panel) gas components. The PAHs, VSGs and BGs contributions to each SED are plotted with dotted, 
dashed and dash-dotted line, respectively. G$_{0}$ and Y$_{\mathrm{PAHs,VSGs,BGs}}$ are defined in the text. The results are  for Ring 1 only.}
\label{fig:dustem}
\end{figure*}

The emissivities for dust associated with the different gas phases in Ring 1 are shown in Table \ref{tab:emissivities}. We fit 
these with the DustEM model \citep{Compiegne11} which incorporates three populations of dust grains: PAHs, hydrogenated small amorphous carbons (VSGs),
and a combination of large amorphous carbons and amorphous silicates (BGs). The DustEM fits are shown in Figure \ref{fig:dustem}.  
We note that the dust properties in SDPF1 are evaluated with respect to a reference 
SED obtained for the diffuse ISM at high ($|b| >$ 15$^{\circ}$) Galactic latitudes (hereafter referred to as DHGL).

\begin{table*}
\begin{center}
\begin{tabular}{c|c|c|c}
\hline
\hline
\textit{Band} & & $\epsilon_{\nu}$ & \\
($\mu$m) & & [ MJy sr$^{-1}$ (10$^{20}$ H atoms cm$^{-2}$)$^{-1}$ ] & \\
\hline
& HI & \H2 & \HII \\
\hline
\\
8 & 0.839 $\pm$ 0.073 & 0.033 $\pm$ 0.012 & 0.116 $\pm$  0.018\\
\\
24 & 0.380 $\pm$ 0.141 & 0.014 $\pm$ 0.010 & 0.101 $\pm$ 0.016\\
\\
70 & 3.443 $\pm$ 0.629 & 0.400 $\pm$ 0.217 & 2.546 $\pm$ 0.401\\
\\
160 &  13.387 $\pm$ 1.687 & 1.736 $\pm$  0.463 &  5.743 $\pm$ 0.670 \\
\\
250 & 11.611 $\pm$  1.308 & 1.470 $\pm$ 0.204 & 3.343 $\pm$ 0.235\\
\\
350 & 4.922 $\pm$ 0.312 & 0.606  $\pm$ 0.092 & 1.168  $\pm$ 0.132\\
\\
500 &  1.820 $\pm$ 0.162 & 0.249 $\pm$ 0.026 & 0.388 $\pm$ 0.058\\
\hline 
\end{tabular} 
\end{center}
\caption{Emissivities, and corresponding uncertainties, of dust associated with the three gas phases in Ring 1.}
\label{tab:emissivities}
\end{table*}

From DustEM we estimate the intensity of the radiation field associated with each phase of the gas. We obtain, for the ratio between the local radiation field and the 
\citet{Mathis83} value for the solar neighborhood, G$_{0}$: G$_{0}$(HI)=1.54$\pm$0.23, G$_{0}$(\H2)=1.55$\pm$0.51 and G$_{0}$(\HII)=4.47$\pm$0.75. 
In the ionized phase, the radiation field is remarkably higher than in the other phases and consistent with a star formation scenario. 

Dust temperatures are evaluated by DustEM separately for each of the three gas phases by applying \citep{Bernard08}

\begin{equation}
T_{d} = 17.5 \times \mathrm{G}_{0} ^ {1/(4+\beta)}
\end{equation}
with the spectral emissivity index $\beta=1.9$, in agreement with the result of \citet{Paradis10} for this field. We find: 
$T_{d,\ \mathrm{HI}}=18.82\pm0.47$ K, $T_{d,\ \mathrm{H_{2}}}=18.84\pm1.06$ K and $T_{d,\ \mathrm{HII}}=22.56\pm0.64$ K. As expected, dust in the ionized gas phase is warmer 
compared to dust in the atomic and molecular phases since, in the proximity of star forming complexes, the radiation causing ionization of interstellar hydrogen also induces heating of dust particles. 

In DustEM, under the assumption of a constant gas-to-dust mass ratio in each gas phase, dust abundances are expressed relative to 
H atoms \citep{Compiegne10}, $[\mathrm{M}_{PAH}/\mathrm{M}_{H}]$, $[\mathrm{M}_{VSG}/\mathrm{M}_{H}]$ and $[\mathrm{M}_{BG}/\mathrm{M}_{H}]$. 
In addition, the code retrieves abundances normalized to the values in the DHGL \citep[][]{Compiegne10}. We refer to these normalized 
abundances as: \Ypah, \Yvsg\ and \Ybg. From the fits of the emissivities in Ring 1, we obtain: \Ypah(HI)=$22.51\pm$3.35, \Ypah(H$_{2}$)=0.86$\pm$0.35, \Ypah(\HII)=1.07$\pm$0.22 
and \Ybg(HI)=$9.70\pm$0.85, \Ybg(H$_{2}$)=1.27$\pm$0.21, \Ybg(\HII)=1.64$\pm$0.17. Surprisingly, both in the atomic and molecular phase, DustEM is able to reproduce 
the emissivity values without invoking a VSG contribution. Only in the ionized phase we have \Yvsg\ different from zero and equal to \Yvsg(\HII)=1.14$\pm$0.58. 
Noteworthy, the apparent lack of VSGs in the atomic and molecular gas phases could be ascribed to the limitation of the 3D-inversion 
model in accounting for the dependence of the AIB emission on the intensity of the radiation 
field, as discussed in the previous section. Furthermore, for very low values of \Ybg, DustEM is not able to distinguish between a PAH and a VSG contribution, 
and typically tends to increase \Ypah\ at the expense of \Yvsg. 

Most importantly, the result of the fit reveals a significant decrease of the relative abundance of PAHs in the molecular and ionized phase with respect 
to the neutral phase. In fact, while [\Ypah(HI)/\Ybg(HI)] = 2.3, [\Ypah(\H2)/\Ybg(\H2)] = 0.67 and [\Ypah(\HII)/\Ybg(\HII)] = 0.65, suggesting that 
PAHs are somehow depleted in these two environments. The destruction of PAHs in the ionized gas has been investigated from a 
theoretical standpoint by \citet{Draine11} and observational evidence of these predictions are reported by, e.g \citet{Povich07} and \citet{Paradis11}. For PAH 
depletion in the  
molecular gas component, although no theoretical prescription is readily available to interpret this result, we speculate, along the lines of \citet{Paradis11}, that this effect could be 
attributed to both the interstellar and local radiation field, i.e. by penetrating the cloud and causing partial destruction of the aromatic molecules.

\subsection{Ring 2-3-4: missing column density and \textit{dark gas}}\label{sec:missing_column}

In Ring 2, 3 and 4 the code either retrieves only partial positive solutions or does not converge. The emissivities and the associated errors for these rings 
are in Table \ref{tab:emissivities_ring234}. Although the  results in the Table do not have a direct physical interpretation (i.e. negative emissivities), they are an indication of the limitations of the basic assumptions of our model. From Equation \ref{eq:decomposition} and under the hypothesis of a single emissivity value for each gas component, the integrated IR emission is expected, by construction, to positively correlate with the 
gas column densities (see Section \ref{sec:pearsons}). In this framework, the negative emissivities obtained in Ring 2-3-4 suggests that in these regions 
the model fails to adequately reproduce the details of emission associated to each gas phase. In the following, we investigate whether the failure of the model can be attributed 
to a mis-match between the column densities retrieved by the tracers described in Section \ref{sec:dataset} and the actual total amount of gas located in these rings.

\begin{table*}
\begin{center}
\begin{tabular}{c|c|c|c|c|c|c|c}
\hline
\hline
\textit{Band} & \multicolumn{7}{c}{$\epsilon_{\nu}$}\\
($\mu$m) & \multicolumn{7}{c}{[ MJy sr$^{-1}$ (10$^{20}$ H atoms cm$^{-2}$)$^{-1}$ ]}\\
\hline
&  \multicolumn{3}{c}{HI} & \multicolumn{2}{c}{\H2} &  \multicolumn{2}{c}{HII} \\
& \textit{Ring 2} & \textit{Ring 3} & \textit{Ring 4} & \textit{Ring 2} & \textit{Ring 3} & \textit{Ring 2} & \textit{Ring 3} \\
\hline
\\
8 & 0.038 $\pm$ 0.166 & -1.766 $\pm$ 0.507 & 0.035 $\pm$ 0.198 & 0.086 $\pm$ 0.059 & 0.076 $\pm$ 0.043 & -0.059 $\pm$ 0.090 & -0.366 $\pm$ 0.134\\
\\
24 & 0.194 $\pm$ 0.112 & -0.945 $\pm$ 0.416 & 0.147 $\pm$ 0.157 & -0.003 $\pm$ 0.048 & 0.055 $\pm$ 0.057 & 0.013 $\pm$ 0.076 & -0.484 $\pm$ 0.156 \\ 
\\
70 & 2.219 $\pm$ 2.240 & -7.109 $\pm$ 2.517 & -3.600 $\pm$ 0.892 & -0.577 $\pm$ 0.326 & 1.686 $\pm$ 0.676 & 1.849 $\pm$0.343 & 0.439 $\pm$ 1.841 \\ 
\\
160 & 4.611 $\pm$ 7.196 & -14.980 $\pm$ 10.86 & -16.330 $\pm$ 2.316 & 2.189 $\pm$ 0.909 & 0.332 $\pm$ 2.241 & 3.318 $\pm$ 1.644 & 4.331 $\pm$ 5.278 \\ 
\\
250 & 2.374 $\pm$ 3.456 & -23.792 $\pm$ 6.040 & -6.929 $\pm$ 2.683 & 2.298 $\pm$ 0.624 & -3.470 $\pm$ 1.806 & 0.038 $\pm$ 1.276 & 3.154 $\pm$1.276 \\
\\
350 & 0.970 $\pm$ 1.409 & -6.615 $\pm$ 3.321 & -2.587 $\pm$ 0.604 & 0.698 $\pm$ 0.450 & -1.346 $\pm$ 0.746 & 0.202 $\pm$ 0.447 & 1.681 $\pm$ 0.739 \\
\\
500 & 0.198 $\pm$ 0.391 & -3.122 $\pm$ 1.627 & -0.555 $\pm$ 0.203 & 0.268 $\pm$ 0.174 & -0.570 $\pm$ 0.401 & -0.090 $\pm$ 0.153 & 0.701 $\pm$ 0.377 \\ 
\hline 
\end{tabular} 
\end{center}
\caption{Emissivities, and corresponding uncertainties, of dust associated with the three gas phases in Ring 2-3-4.}
\label{tab:emissivities_ring234}
\end{table*}

\citet{Grenier05} have shown, by comparing HI/CO data with $gamma$-ray emission, the existence in the solar neighbourhood of the so-called \textit{dark gas}, 
a mixture of cold HI, which is optically thick in the 21cm line, and warm \H2 which cannot be observed with the standard tracers. They claim that, although it is not clear which component (cold HI or warm \H2) dominates, 
the contribution from dark \H2 must be considerable, up to 50 percent of the total column density in regions where 
there is no CO detection. An important dark gas contribution to the overall IR emission is also reported in \citet{Planck_Bernard11}. In this inversion analysis, applied to the entire Galactic Plane at 1$^{\circ}$ resolution, the 
authors find that \textit{dark gas} is mostly distributed around major molecular cloud complexes.

Theoretical predictions from \citet{Wolfire10} indicate that a significant amount of warm \H2 is located in the exterior of photodissociation regions, where the transition of atomic into molecular gas occurs. Recent \textit{Herschel}-HIFI \texttt{CII} observations by \citet{Pineda13} have allowed to slightly revise the early estimate by \citet{Grenier05} relative to the total amount of dark \H2 in the Galaxy. These authors estimate that warm \H2 is likely to account for $\simeq30$ percent of the total molecular gas, and find that the fraction of dark \H2 increases with Galactocentric distance. Therefore, with the standard CO tracers, we have an intrinsic limitation in estimating all the \H2 along the LOS.

In light of the considerations above, we have re-analyzed the content of Ring 2-3-4 and checked if the negative emissivities derived for these rings can be due to the presence of untraced gas, specifically warm \H2 and cold HI. In Ring 2, we note (Figure \ref{fig:radius_vs_intensity}) that both CO isotopes are characterized by a pronounced peak of emission, indicating the presence of molecular clouds which, following the results from \citet{Planck_Marshall11}, might be associated with warm \H2. The same peak of emission in \12CO and \CO13 is also observed in correspondence of Ring 3. However, in this case, as mentioned in Section \ref{sec:l30_field} and Section \ref{sec:HI_data}, a cold HI layer is also found, as reported by \citet{Gibson04}. The existence of cold (i.e. optically thick) HI could explain the anti-correlation and the lack of  convergence of the code (Section \ref{sec:pearsons}). From the inspection of the VGPS data cube for SDPF1, we have identified two additional prominent features (either HISA or HICA) in the same range of Galactocentric radii:

\begin{itemize}
\item[1] H30.74-0.05 at  $11\leq\mathrm{V_{LSR}}\leq15$ km s$^{-1}$
\item[2] H30.39-0.24 at $8\leq\mathrm{V_{LSR}}\leq14.5$ km s$^{-1}$
\end{itemize}

\begin{figure}
\centering

\includegraphics[width=8cm]{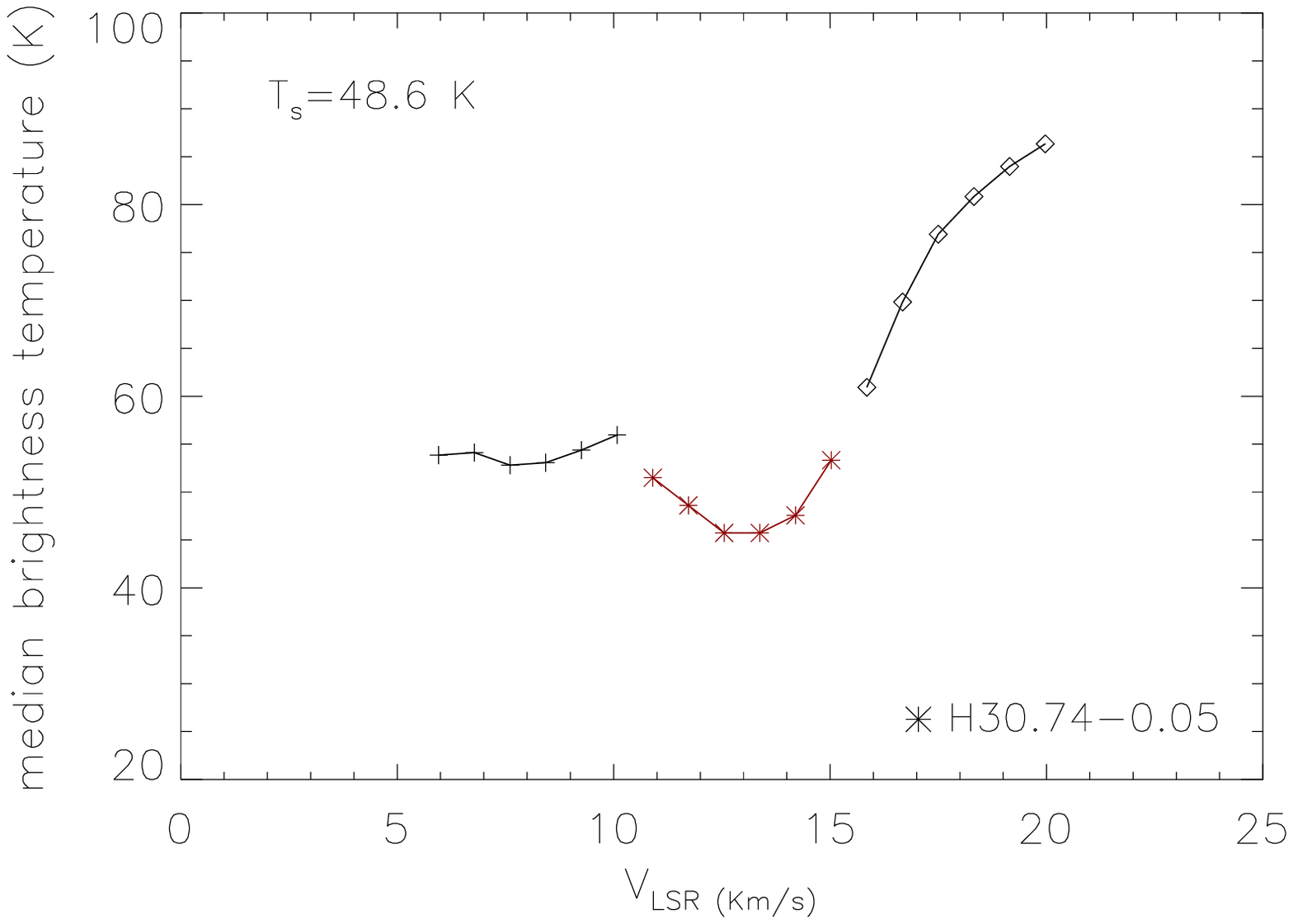}\\
\includegraphics[width=8cm]{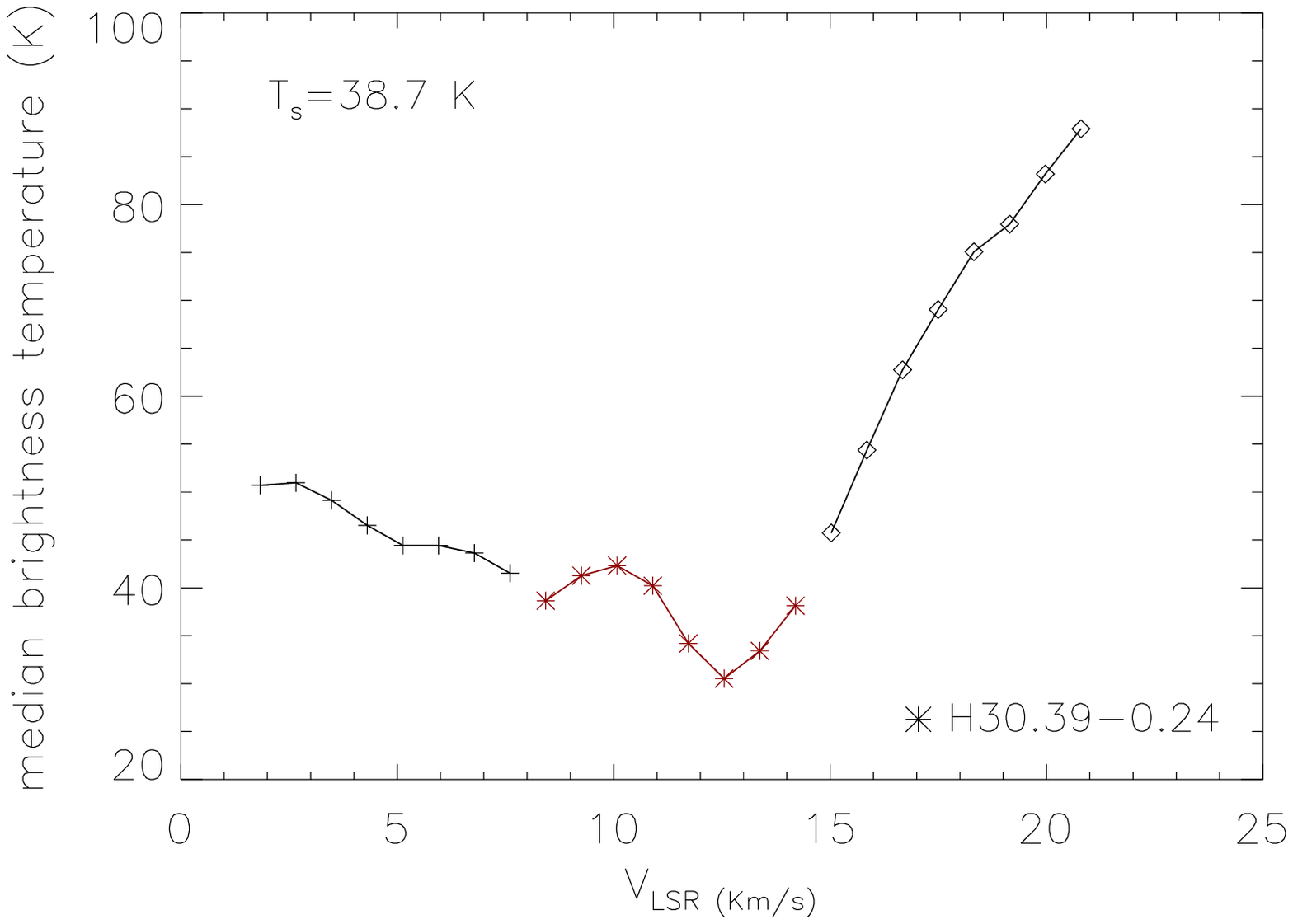}

\caption{Maximum of the median brightness temperature (red line) in the surrounding of H30.74-0.05 and H30.39-0.24. 
Shown is the mean of these values across each feature. For comparison, the black line denotes the maximum of the median brightness temperatures measured in the same region at the nearby Galactocentric positions.}

\label{fig:HISA_profiles}
\end{figure}

For these cold regions, we have made an attempt to derive an indicative upper limit of their spin temperatures by measuring the maximum of the median \Tb values evaluated in the pixels in proximity of the regions of absorption. We recall that, as discussed in Section \ref{sec:HI_data}, the spin temperature of cold HI is noticeably lower than in the WNM, and that 
\Ts is meaningful only in the case of warm HI in optically thin conditions. For this reason, it cannot be used to derive HI column densities in regions populated by different HI components \citep[e.g.,][]{Strasser04}. Figure \ref{fig:HISA_profiles} 
shows the brightness temperature profiles for the cold features H30.74-0.05 and H30.39-0.24, for which we obtain a mean value in each feature of 
$\mathrm{T_{b}}$=48.6 K and $\mathrm{T_{b}}$=38.7 K, respectively. 

Interestingly, HISA features have been found mixed with molecular clouds. For instance, in the Perseus star forming region, a significant fraction of cold HI is undergoing the transition to the molecular phase, and the HISA features appear to be non-gravitationally bound regions of molecular material not detected in CO \citep{Klaassen05}.

In summary, we speculate that in Ring 2 the derivation of dust emissivities possibly fails due to a significant amount of warm \H2 not properly accounted for by the \12CO and \CO13 data, while in Ring 3 the lack of convergence could be ascribed to both (untraced) warm \H2 and (badly traced) cold HI. Regarding Ring 4, our current hypothesis is that, after subtracting the contributions from the other rings and given the small amount of HI (only a few percent, see Section \ref{sec:total_density}), the code has little to no leverage to return meaningful values.

\subsubsection{Solving for the distance ambiguity of cold HI features}
\begin{figure*}
\centering
\includegraphics[width=8.5cm]{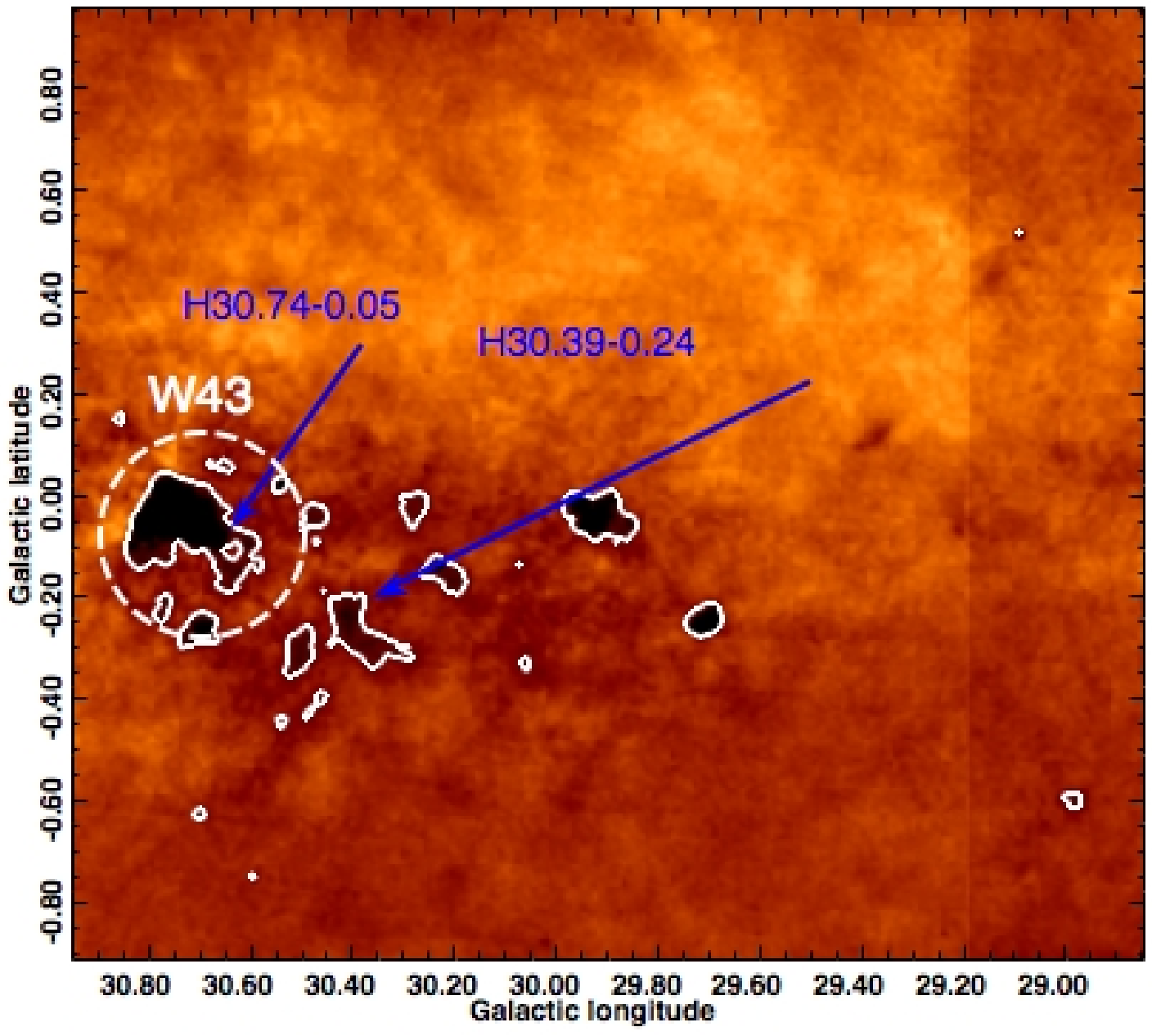} \qquad
\includegraphics[width=8.5cm]{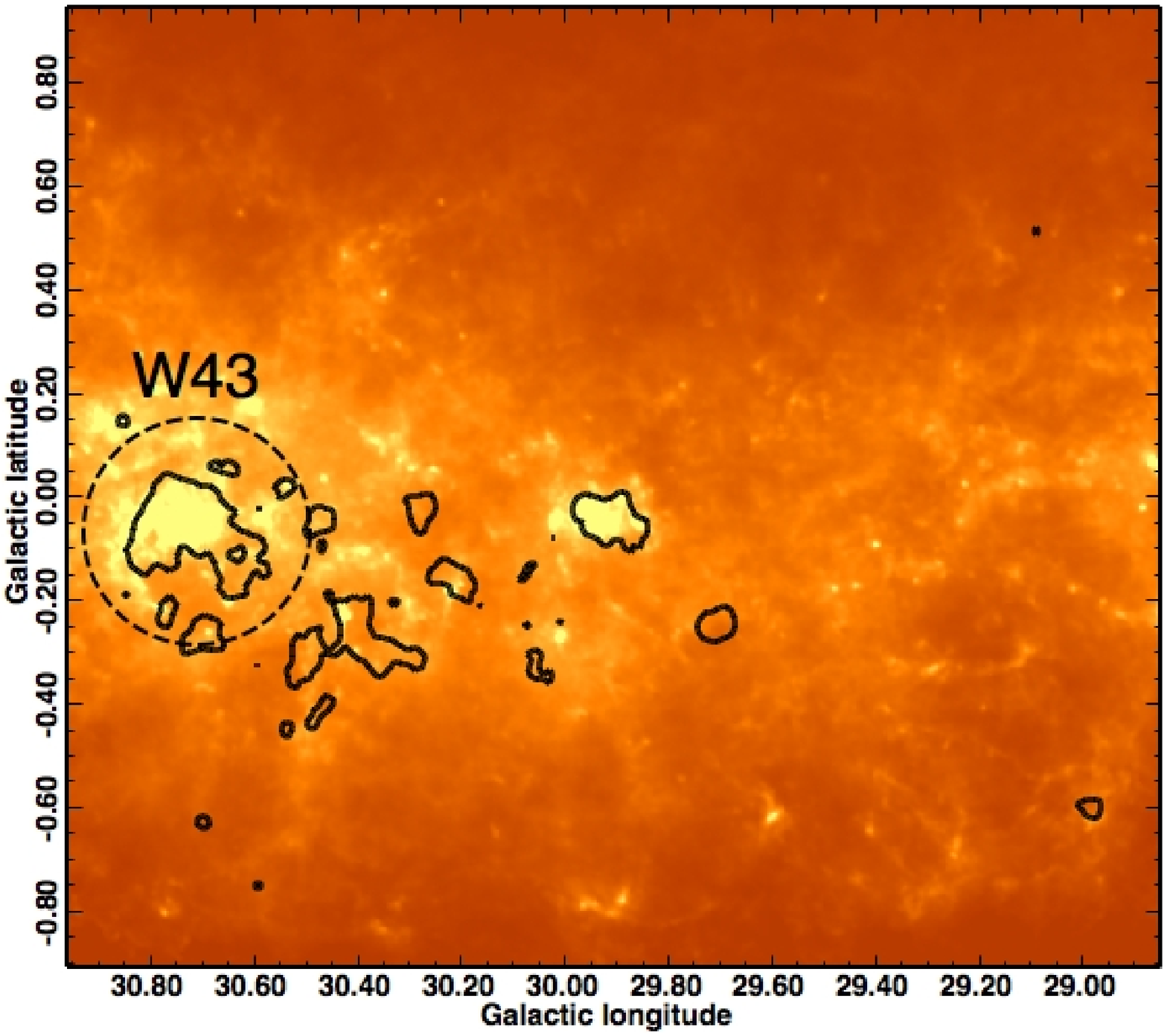}

\caption{HI brightness temperature map integrated over Ring 3 (left panel) and Hi-GAL 500 $\mu$m data (right panel). 
The white contours in the HI map denote HICA and HISA features. 
The same regions at 500 $\mu$m (the black contours) overlap with either the W43 complex or with other HII regions in the field. In the HI map, some of the dark pixels within the cold features are pixels where \Tb is negative due to strong absorption. The majority are found associated with the brightest HII regions, especially W43 and G29.944-0.04.}  
\label{fig:PSW_HISA}
\end{figure*}

As a by-product of our study, we can solve for the distance ambiguity of some of these cold features by comparing the HI data with the maps 
for the other tracers as well as with the input IR maps. In Figure \ref{fig:PSW_HISA} we show the absorption features in silhouette against the warm background provided by the HII regions in SDPF1  
compared to the VGPS 21-cm map integrated along Ring 3. H30.74-0.05 is dominated by absorption of the 
continuum background provided by W43, and it is therefore primarily an HICA feature. Since W43 is at the tangent point of the Scutum-Crux arm, 
we can solve the distance ambiguity for this cold structure and locate it at 0.7 kpc $\leq\mathrm{d}\leq$ 1.1 kpc, 
i.e. on the Sagittarius-Carina arm (see Figure \ref{fig:Galactic_model}). In addition, the 29 HII regions in SDPF1 are distributed between Ring 1 and 2 
\citep{Anderson09}, thus all the absorption features, including H30.39-0.24, observed in Ring 3 in correspondence of these 
HII regions are also likely associated with the Sagittarius-Carina arm.

\subsection{The dominant contribution of Ring 1 to the integrated intensity maps}\label{sec:ring1_dominate}

\begin{figure*}
\centering
\includegraphics[width=8cm]{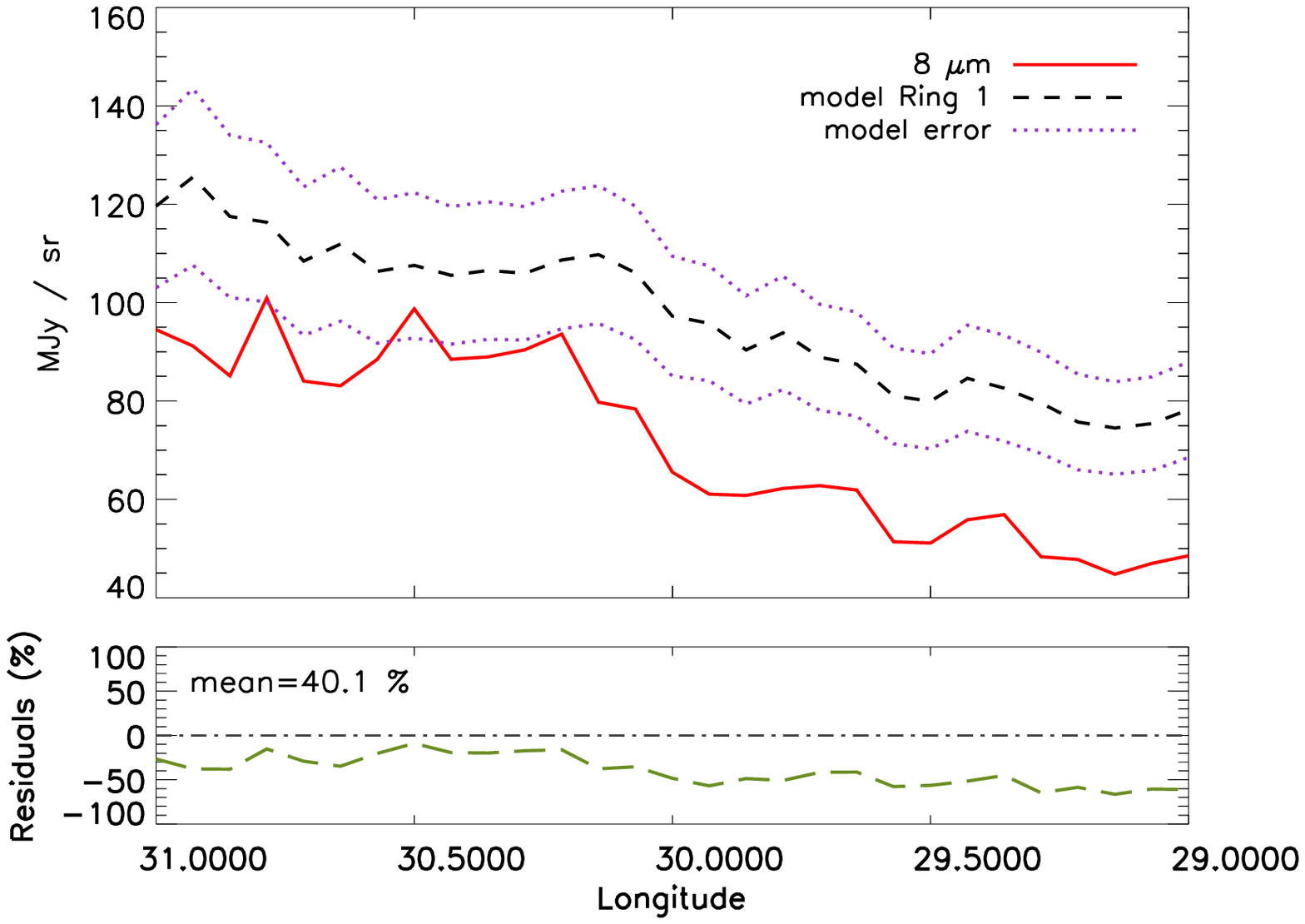} \qquad 
\includegraphics[width=8cm]{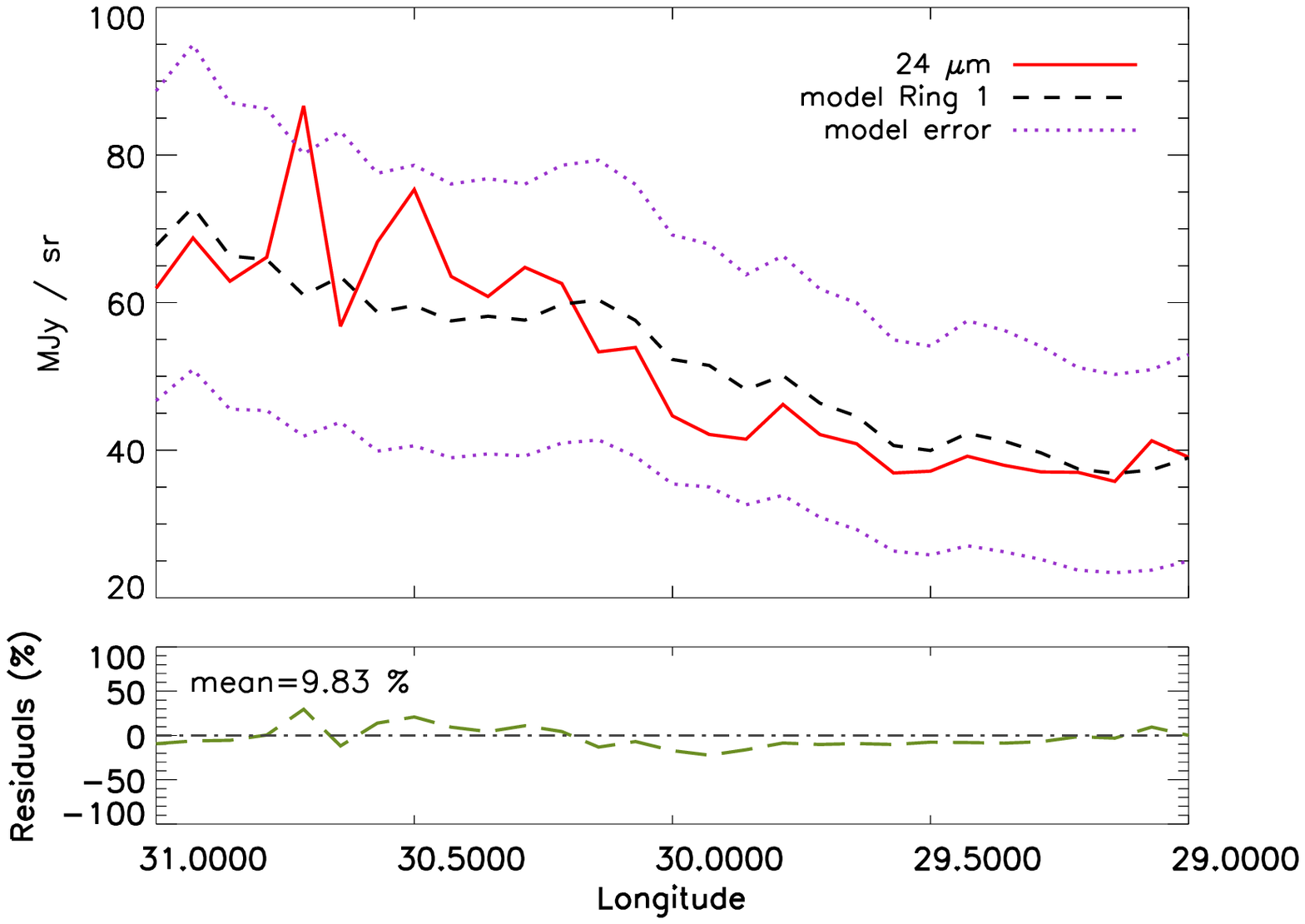}\\
\includegraphics[width=8cm]{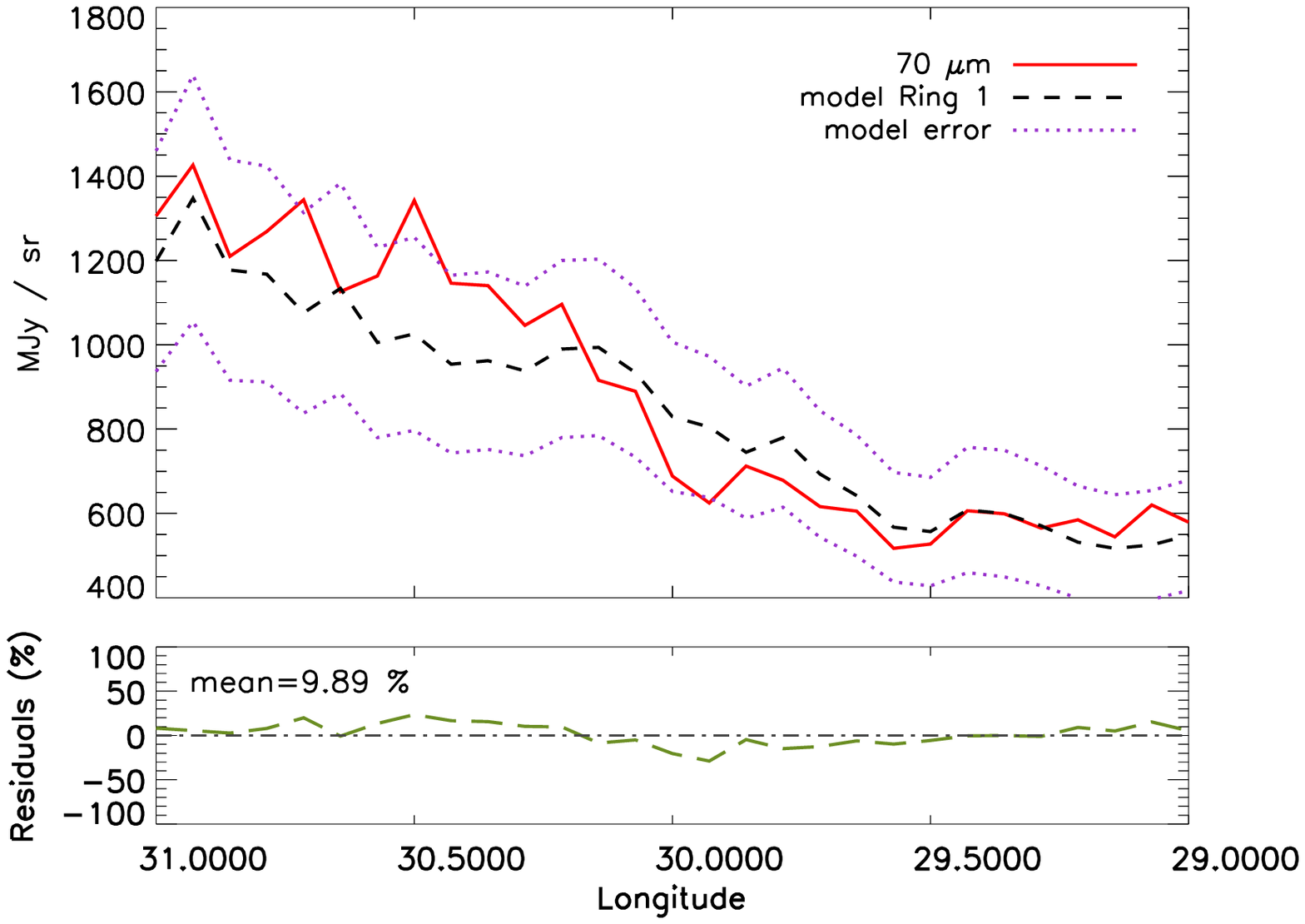}  \qquad
\includegraphics[width=8cm]{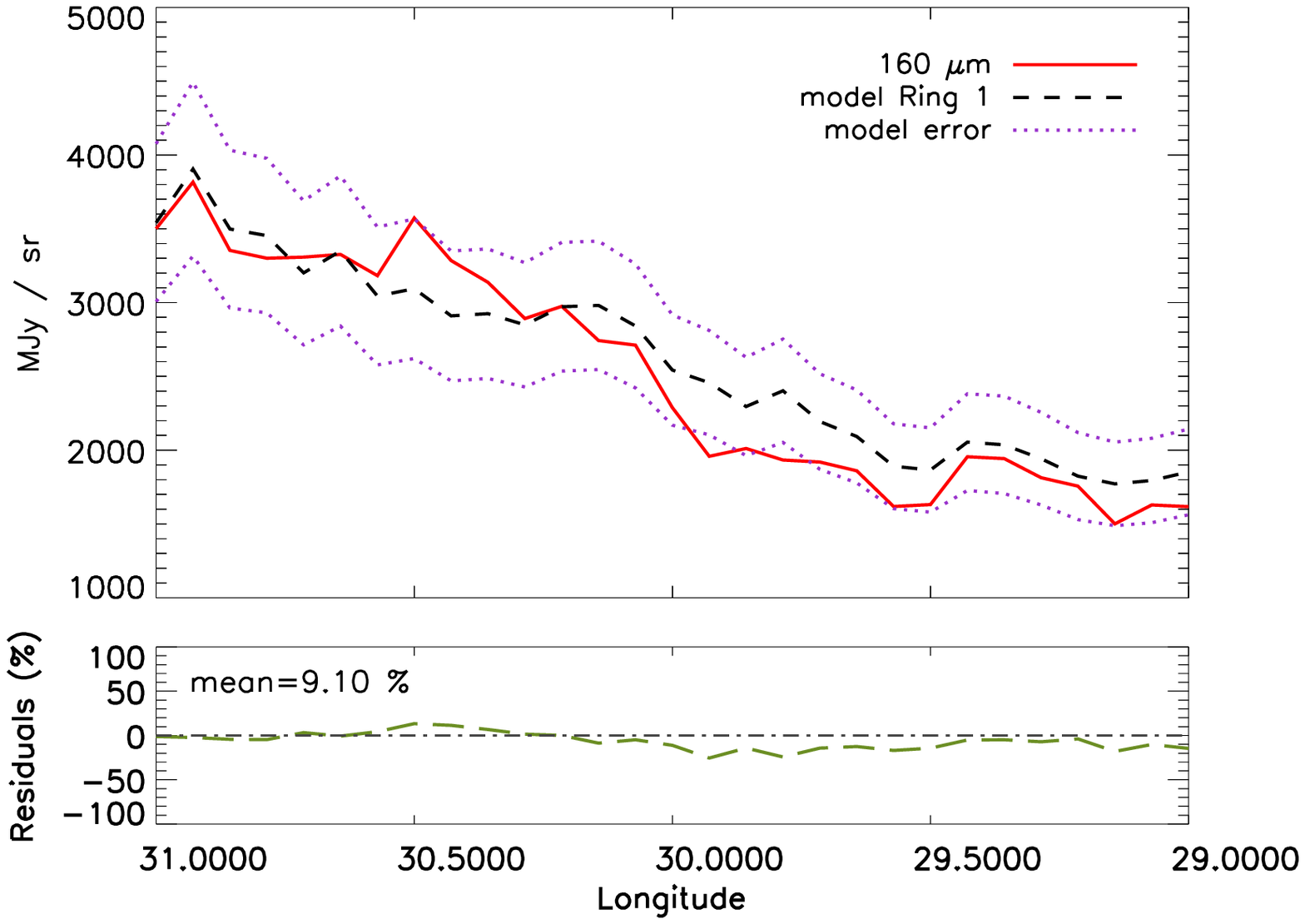} \\
\includegraphics[width=8cm]{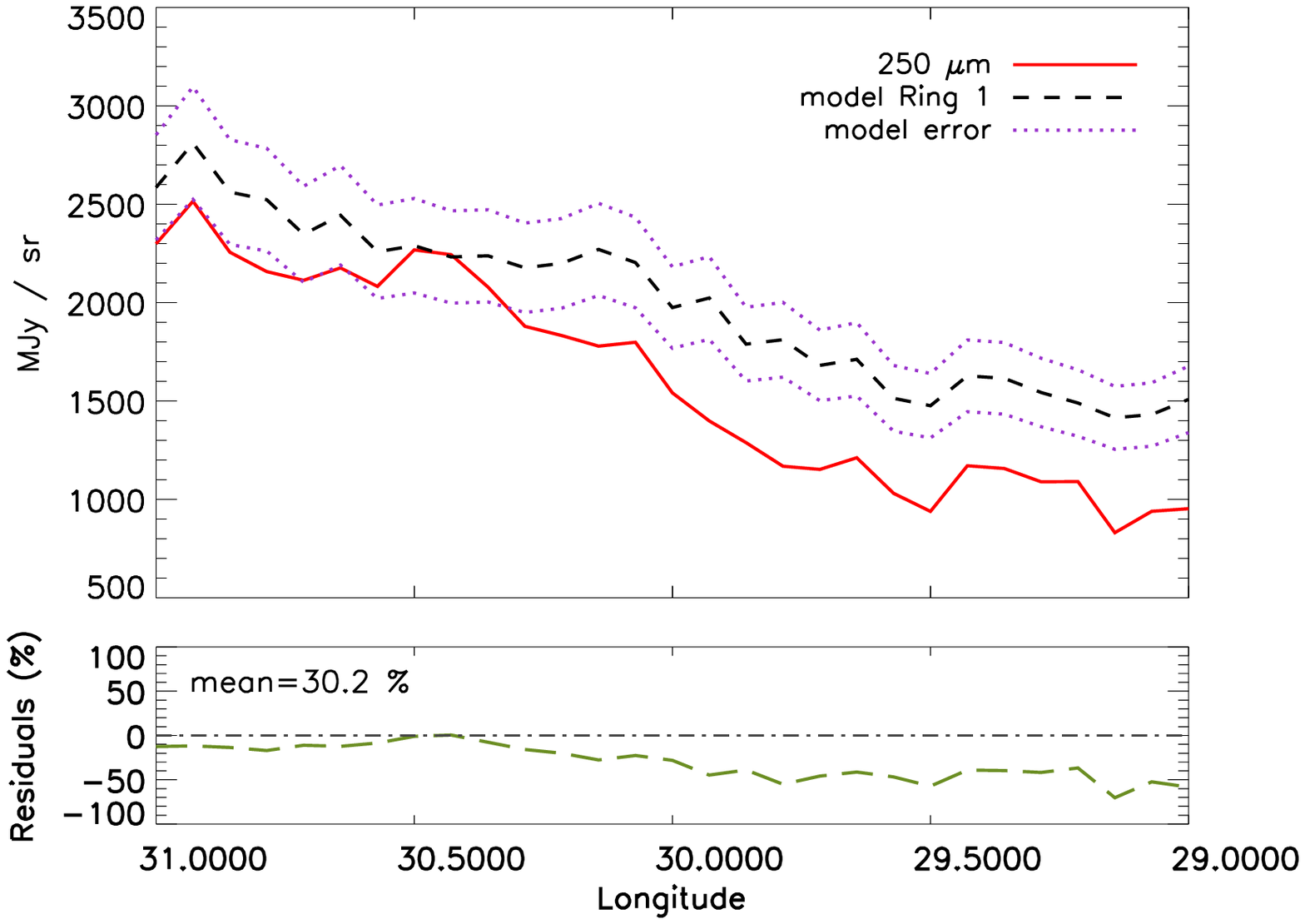} \qquad
\includegraphics[width=8cm]{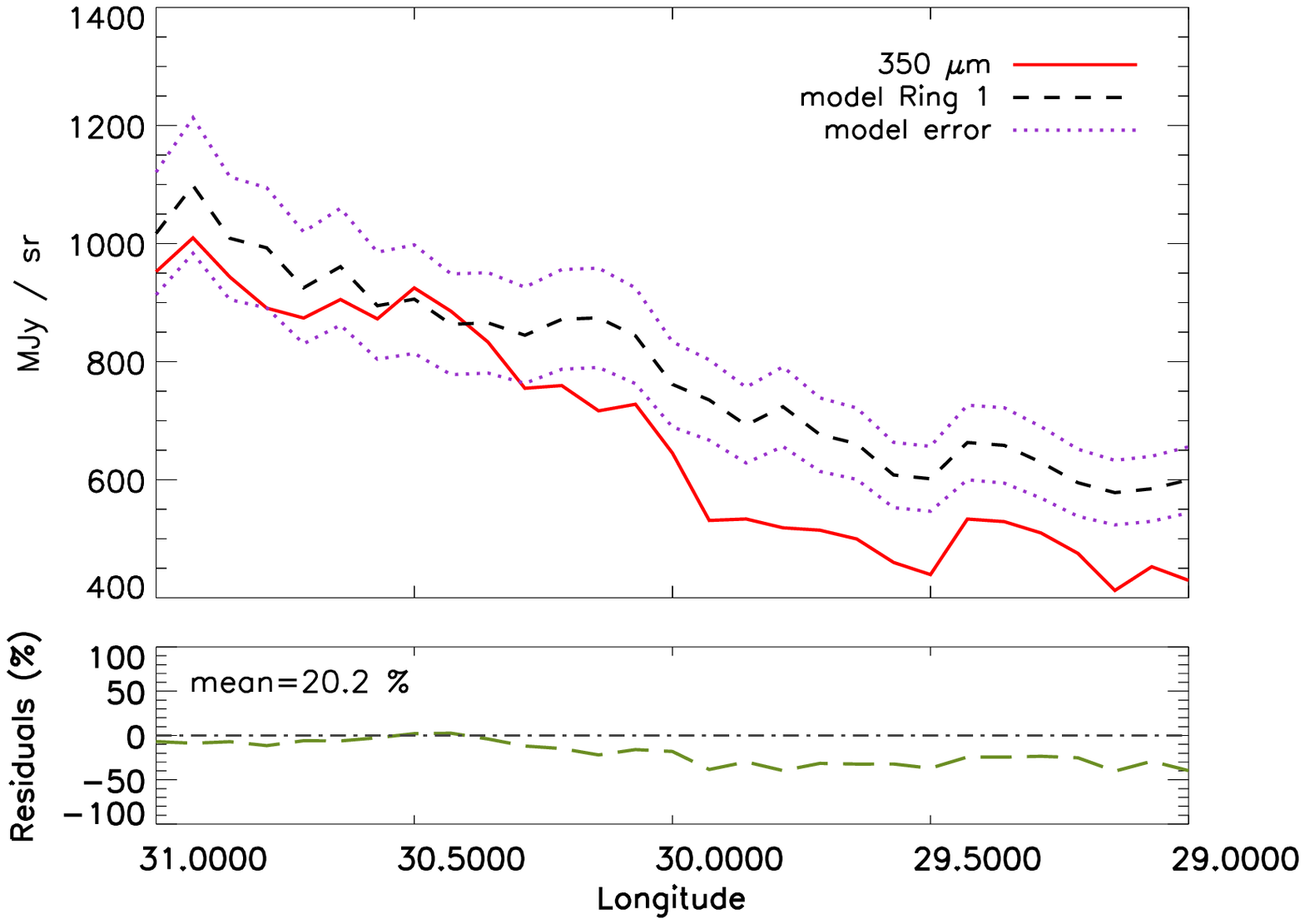} \\
\includegraphics[width=8cm]{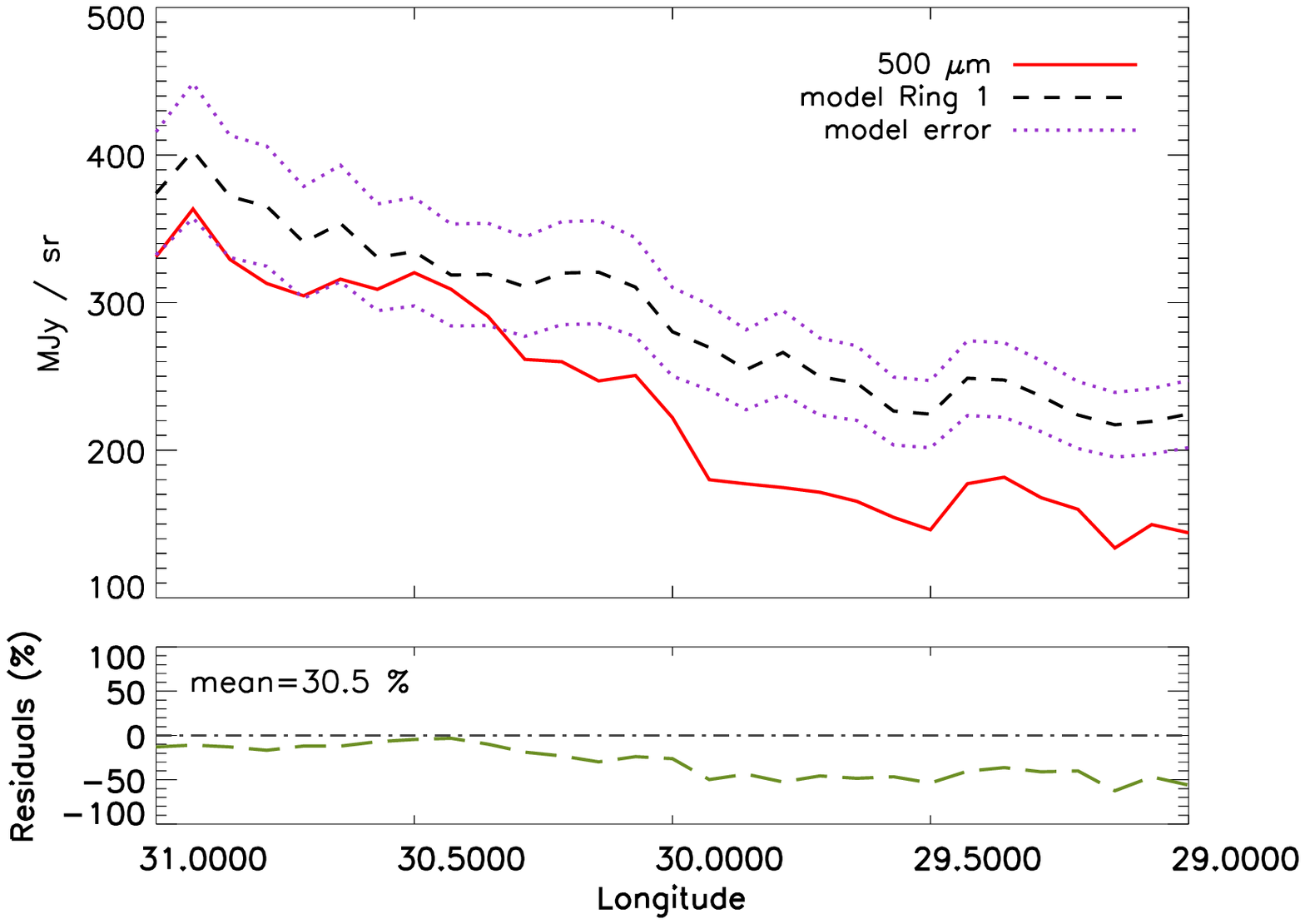} 
\caption{Comparison of the intensity longitude profiles for the input maps (red line) and for the Ring 1 inversion model 
(black dotted line). Overlaid are the $\pm$ 1$-\sigma$ boostrap errors (magenta dotted line). The separate panels at the bottom 
of the longitude profiles show the residuals (in percentage) obtained by subtracting the inversion model for Ring 1 from the 
input profiles (dashed green line).}
\label{fig:ring1_profile_8_160}
\end{figure*}

In Section 4.3.4 we have seen that roughly 65 percent of the total gas column density of SDPF1 is located 
in Ring 1. In this Section, we want to show that dust emission associated with this ring accounts indeed for the bulk of dust 
emission in the input maps. To this end, we compare the longitude profiles of the output model with those of the input maps, and in doing so 
we consider the model contribution from Ring 1 only. The longitude 
intensity profiles are generated by averaging, for a given Galactic longitude, all latitude pixel values. These profiles, 
as well as the residuals obtained by subtracting the model from the input maps are shown in Figure \ref{fig:ring1_profile_8_160}.

The model intensities obtained from Ring 1 appear to satisfactorily reproduce the input intensities at 24, 70, and 160 \mum, with a discrepancy of 
less than 10 percent. At longer wavelengths, the model tends to overestimates the input profiles and the discrepancy 
is more pronounced, of the order of 30 percent. This effect is likely related to the fact that dust associated with the cold untraced gas, which would emit at long wavelengths (e.g. $>$ 160 \mum), is mostly located outside Ring 1 (see Section \ref{sec:dustem}). Therefore its contribution is visibly missing when only Ring 1 is considered, and this produces the observed model overshooting.  Conversely, in the range 24 \mum $< \lambda <$ 160 \mum\ both dust and gas are properly accounted for  and the model is able to reproduce the input emission. We note that at 8 \mum\ the residuals are higher comparatively to other wavelengths and of the order of 40 percent. This is  
expected: at this wavelength the coarse assumption of the model of a one-to-one correlation between 
input intensity and column densities reveals its limitation, due to the degenerate PAH emission dependence on both radiation field amplitude  
and column density (Section \ref{sec:pearsons}).

The residuals obtained from our analysis are comparable to 
earlier results. For instance, from the longitude profiles of \citet{Paladini07}, in the region of overlap 
with this work, i.e. 29$^{\circ} <$ l $<$ 31$^{\circ}$, and between 60 \mum\ and 240 \mum, the residuals appear to be in the range 10 to 30 percent. 
\citet{Planck_Marshall11} provide residuals only at 1.4 GHz, 30 GHz and 857 GHz (350 \mum). No longitude 
profiles are given. At the common 350 \mum\ wavelength and for 29$^{\circ} <$ l $<$ 31$^{\circ}$, the residuals 
are of the order of 15 - 20 percent. Considering that both \citet{Paladini07} and \citet{Planck_Marshall11} work on the full sky at 
a 1$^{\circ}$ resolution, the residuals obtained for our 2$\times$2 square degree decomposition at 14.8 arcmin are in excellent agreement with these previous analyses. 

In summary, the test shows that Ring 1 accounts for approximately 70 to 90 percent of the total emission in the input maps 
at all wavelengths, while the combined Ring 2, 3 and 4 contributes only for the remaining 30 to 10 percent, therefore strongly corroborating 
our earlier statement that, despite the lack of meaningful convergence of the method in the other rings, the results of Ring 1 are robust. In 
Appendix \ref{app:model_test} we will return on this point, by further investigating possible biases introduced by Ring 2, 3 and 4.

\begin{figure*}
\centering
\includegraphics[width=16cm]{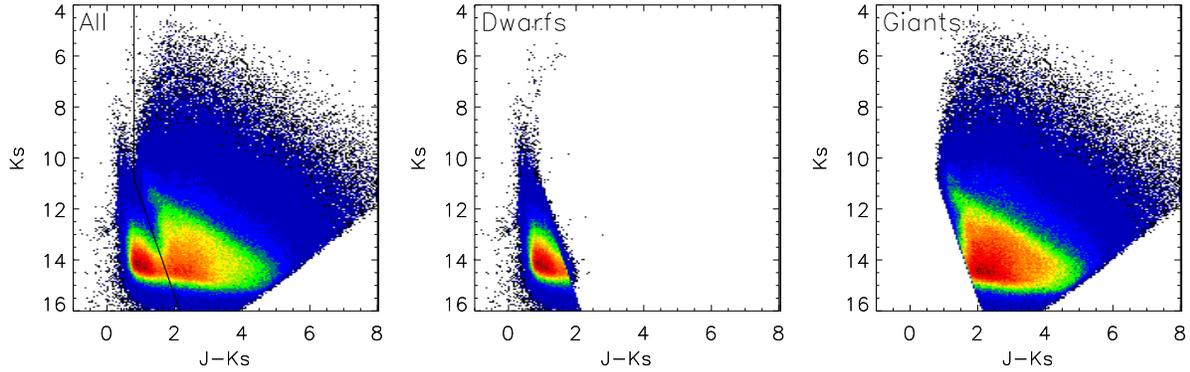}  
\caption{K$_{s}$ vs J-K$_{s}$ diagrams. The black line in the left panel corresponds to the cut applied to remove from the sample contaminating dwarfs. 
Identified dwarfs and giants are shown in the middle and right panel, respectively.}
\label{fig:color_magnitude}
\end{figure*}

\begin{figure*}
\centering
\includegraphics[width=15cm, height=10cm]{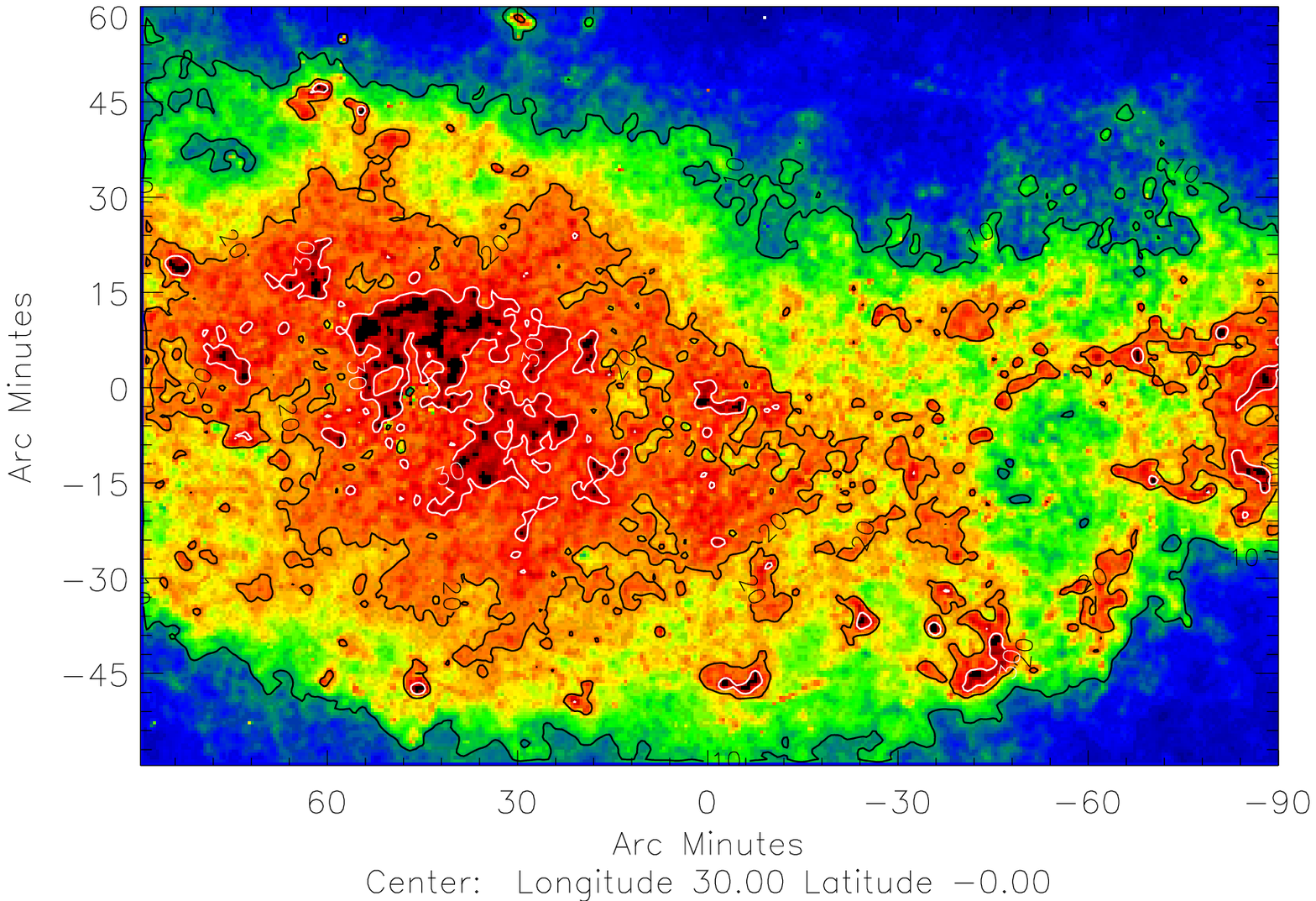}\\  
\caption{Extinction map obtained from giant stars in the UKIDSS, 2MASS and GLIMPSE catalogues. The resolution is 1 arcmin. 
Units are magnitude of \Av. The contours correspond to \Av\ of 10, 20 and 30 mag. 
Regions with the highest extinction (dark red) are associated with cold HI features.}
\label{fig:extinction_Laurent}
\end{figure*}

\subsection{Total column density and extinction maps}\label{sec:extinction}

An alternative method to recover the missing hydrogen column density is through extinction. We have attempted to generate an extinction map for SDPF1 using colour excess templates derived from observations 
of giant stars. For this purpose, we use UKIDSS \citep{Lawrence07} and 2MASS \citep{Skrutskie06} $JHK_{s}$ and GLIMPSE \citep{Churchwell09} 3.6 $\mu$m and 4.5 $\mu$m data. The first step is 
to minimize contamination from sources other than giants, in particular from dwarf stars. To this end, 
we first build colour-magnitude $J-K_{s}/K_{s}$ and $J-[3.6]/[3.6])$ diagrams using the photometric measurements provided in the UKIDSS/2MASS/GLIMPSE catalogues. 
Then we compare these with the predictions from the Besan\c{c}on Stellar Population Synthesis Model \citep{Robin03}. From this model, we derive the colour criteria 
to separate the dwarf from the giant stars, i.e.: $K_{s}\geq(J-K_{s})*3.8-7.8$ and $[3.6]\geq(J-[3.6])*3.2+7.4$ (see Figure \ref{fig:color_magnitude}). 
Applying these criteria, we select 45 percent of the sources in the original catalogues, in practice 4.4$\times10^{5}$ out of $\simeq10^{6}$ sources. 
The colour excess in  $H$ and $K_{s}$ bands measured for the selected sources is converted into extinction using the extinction law derived by \citet{Rieke85}. 
At longer wavelengths, the correct extinction law is still a matter of debate. Variations have been reported to occur 
from one molecular cloud to another, and even within the same cloud \citep[][and references therein]{Cambresy11}. Also, the extinction law 
appears to change with Galactocentric radius \citep{Zasowski09}. For this work, we adopt the extinction law derived by \citet{Cambresy11} in the Trifid Nebula. The extinction law values are: A$_{K_{s}}$/\Av=0.112, A$_{H}$/A$_{K_{s}}$=1.56, A$_{[3.6]}$/A$_{K_{s}}$=0.611 and A$_{[4.5]}$/A$_{K_{s}}$=0.500. 
The resulting \Av\ map is a combination of two maps: for visual extinctions $<10$ mag, it is generated using $H$ and $K_{s}$ bands,
while for \Av$>15$ mag it is obtained using the 3.6 $\mu$m  and 4.5 $\mu$m bands. In the range $10 \mathrm{\ mag}\leq$\Av$\leq15 \mathrm{\ mag}$, it is a linear combination of the two.  
The zero point (\Av=6 mag) is evaluated from the 2MASS data for $\vert l\vert\leq10^{\circ}$.

The final extinction map, with a resolution of 1 arcmin, is shown in Figure \ref{fig:extinction_Laurent}. 
This map is converted into total hydrogen column density using the relation \citep{Guver09} 

\begin{equation}
\mathrm{N(HI)} = 22.1\ A_{\mathrm{V}}  \qquad  [10^{20}\ \mathrm{atoms/cm}^{2}]
\end{equation}

\begin{figure}
\centering
\includegraphics[width=8cm]{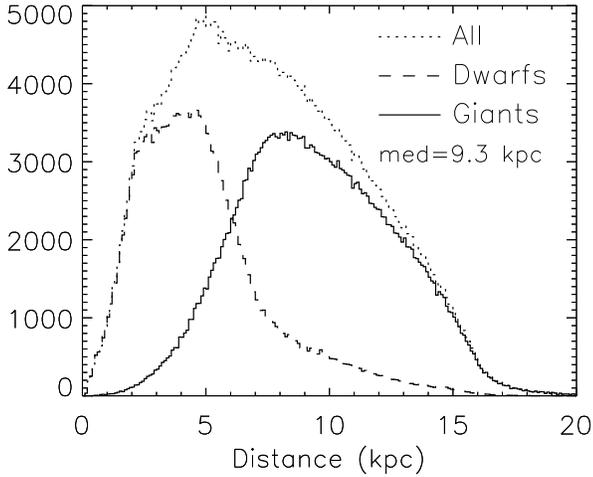}\\
\caption{Number of sources as a function of solar distance for both the selected giant stars and the dwarfs from the Besan\c{c}on model.} 
\label{fig:distance_extinction}
\end{figure}

\begin{figure}
\centering
\includegraphics[width=8.4cm]{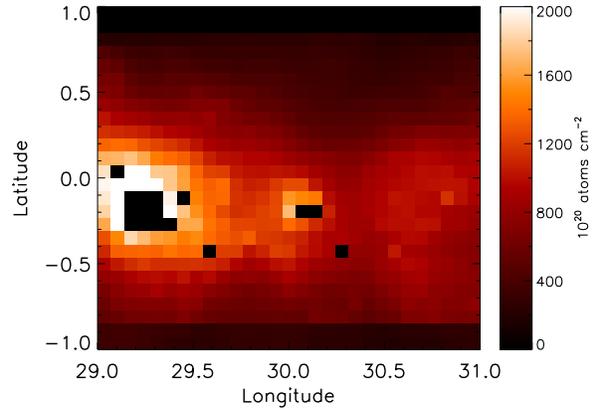}
\includegraphics[width=8.4cm]{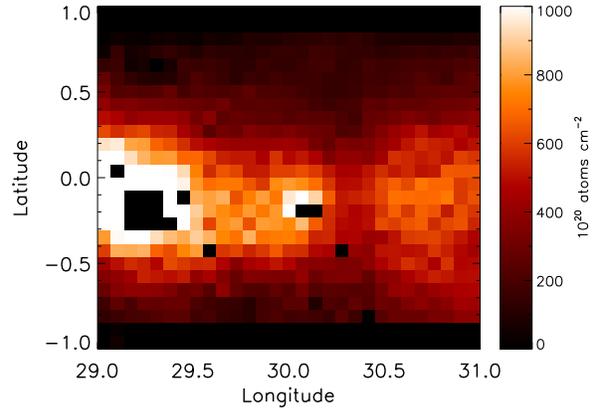}
\includegraphics[width=8.4cm]{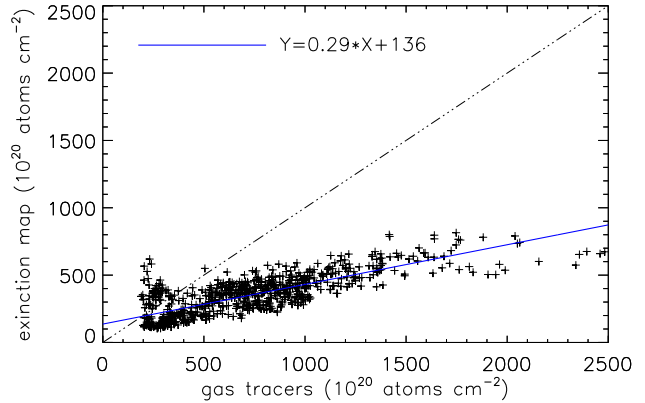}
\caption{Top panel: column density map obtained by summing up the contributions of the three gas phases in each each ring. 
Middle panel: column density map derived from
the extinction map generated using giant stars. Both maps are convolved to 14.8 arcmin. Bottom panel: pixel-to-pixel correlation plot of the two column density maps. 
The blue line shows the best-fit to the distribution. The black dot-dashed line denotes the y=x relation.}
\label{fig:extinction_laurent_column}
\end{figure}

We then compare the total column density map derived from extinction with the one derived from the gas tracers, obtained by summing 
up the individual contributions from the three gas phases and from each ring (see Figure \ref{fig:extinction_laurent_column}). From this comparison, we find that the extinction column density map, minus an offset of roughly 100 $\times$ 10$^{20}$ atoms cm$^{-2}$, is lower with respect to its gas tracers counterpart by $\sim$ 70 percent. We formulate the hypothesis 
that this discrepancy is mainly due to the fact that the gas tracers allow us to probe much larger distances with respect to the catalogued giants. 
To verify this scenario, we first note that Figure \ref{fig:radius_vs_intensity} shows that the bulk of material in SDPF1 is within $R <$ 8.5 kpc, corresponding for this LOS 
to 14.7 kpc, at the far distance. Therefore, the extinction map has to reach at least this distance in order to compare in column density 
with the gas tracers. Although we do not have an individual distance estimate for all our giants, 
we can use the Besan\c{c}on model to have a rough idea of their distance distribution. Figure \ref{fig:distance_extinction} illustrates 
the number of sources as a function of solar distance for the selected giants and dwarfs. Most of the sources appear to be located 
around 10 kpc, setting the approximate limit of our extinction map and showing that, at least with the available data, extinction is not yet a 
viable route for estimating accurate column densities towards the inner Galaxy.

We have also explored an alternative approach, based on the model of \citet{Marshall06} which allows building extinction maps in 3D. 
This model makes use of the combined 2MASS Point Source Catalog and Besan\c{c}on model to calculate the extinction at increasing solar distances. 
The maximum distance reached towards any given LOS depends on the completeness in J and K$_{s}$ bands in the 2MASS catalogue as well as on the effective column density. 
After generating an extinction map for SDPF1 following this prescription, we have compared the corresponding column density map with the one derived from the gas tracers.  
The comparison is not straightforward. In fact, although the \citet{Marshall06} extinction map has a resolution of 15 arcmin, which is comparable 
to our working resolution (14.8 arcmin), the data are not Nyquist sampled, as the pixel size is also set to 15 arcmin. Therefore, for consistency the spatial resolution of the 
gas tracers column density map is also downgraded to 15 arcmin using a 15 arcmin pixel size. With this procedure, we obtain a total of 50 pixels in each map.

\begin{figure}
\centering
\includegraphics[width=8.4cm]{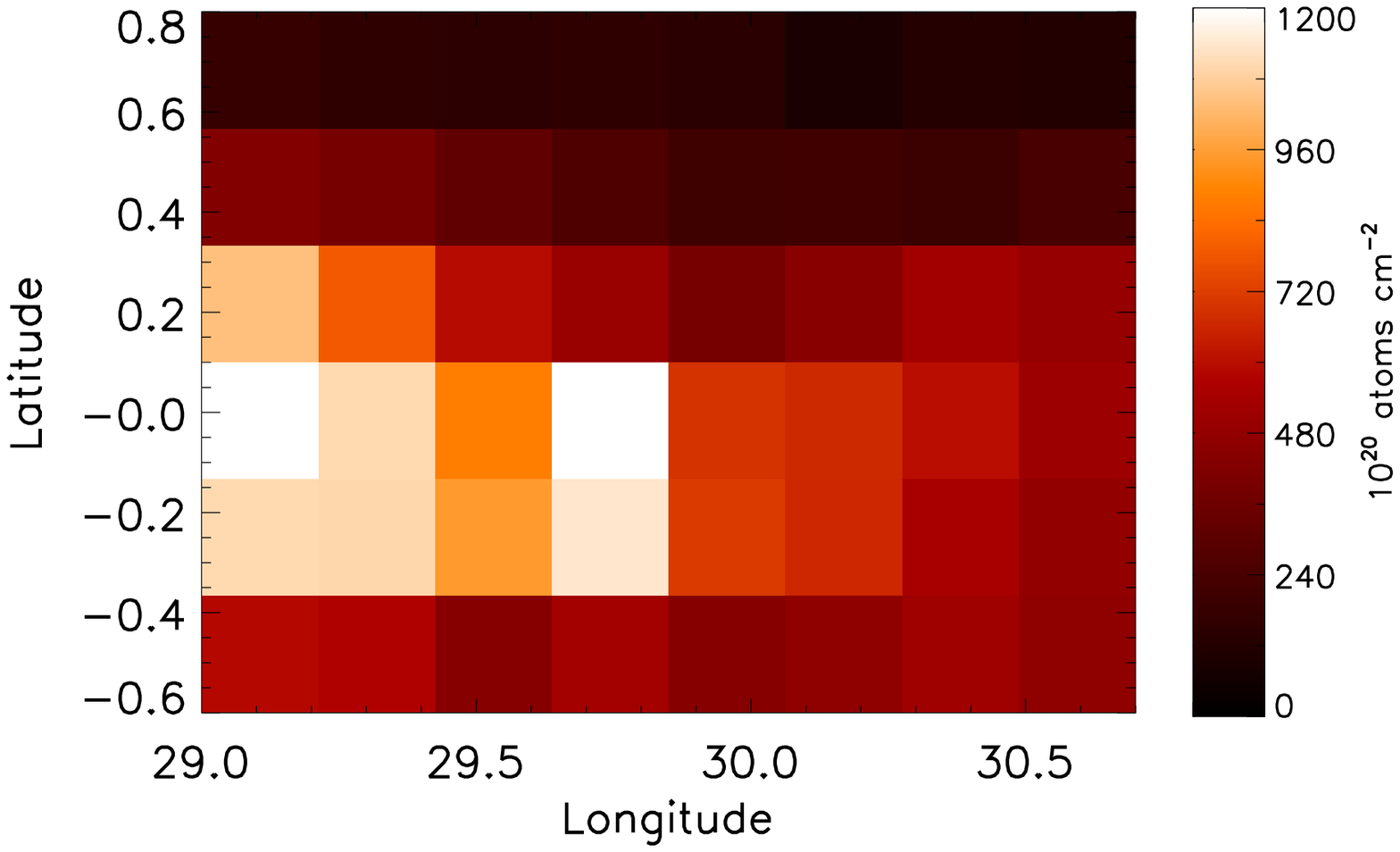}
\includegraphics[width=8.4cm]{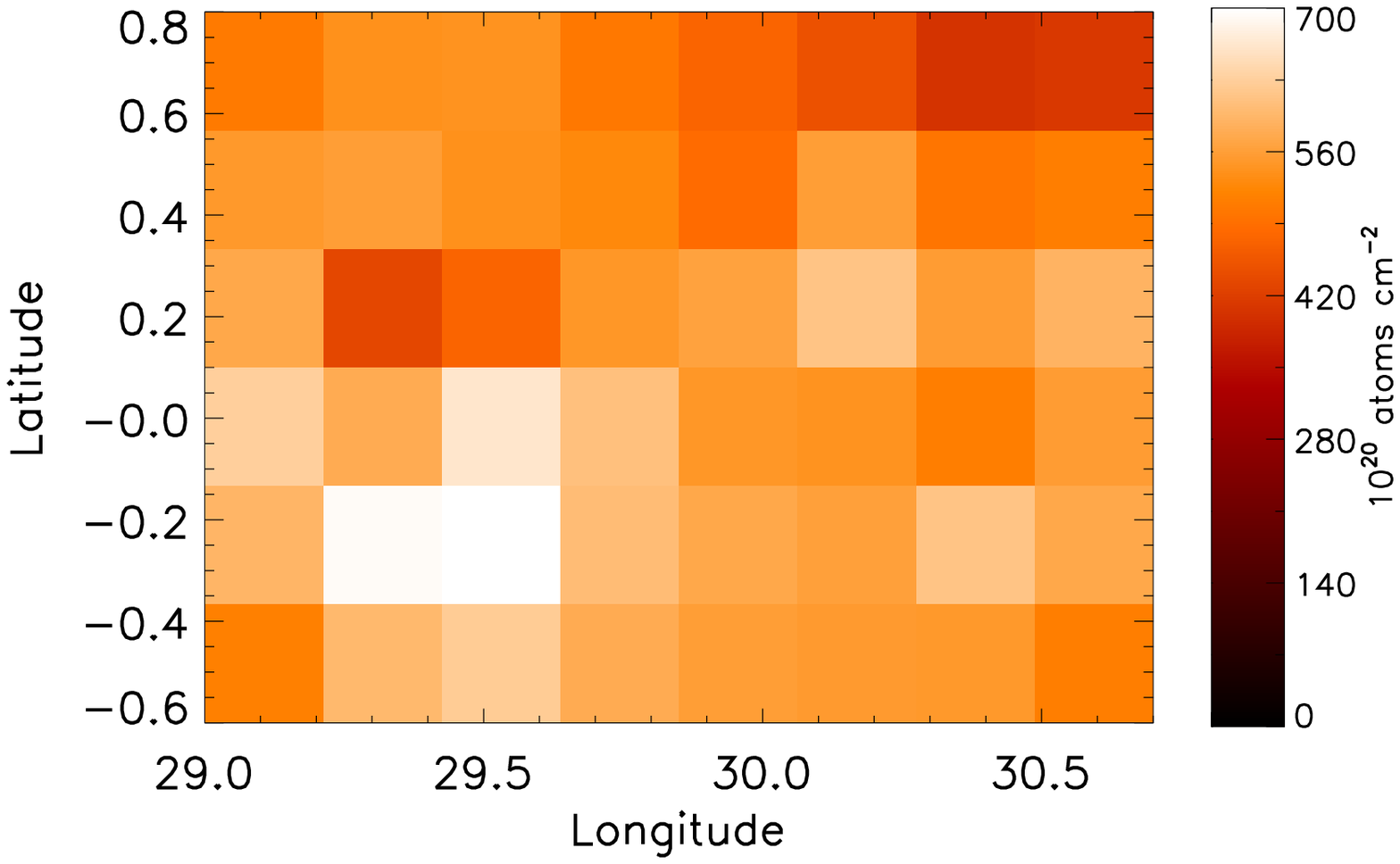}
\includegraphics[width=8.4cm]{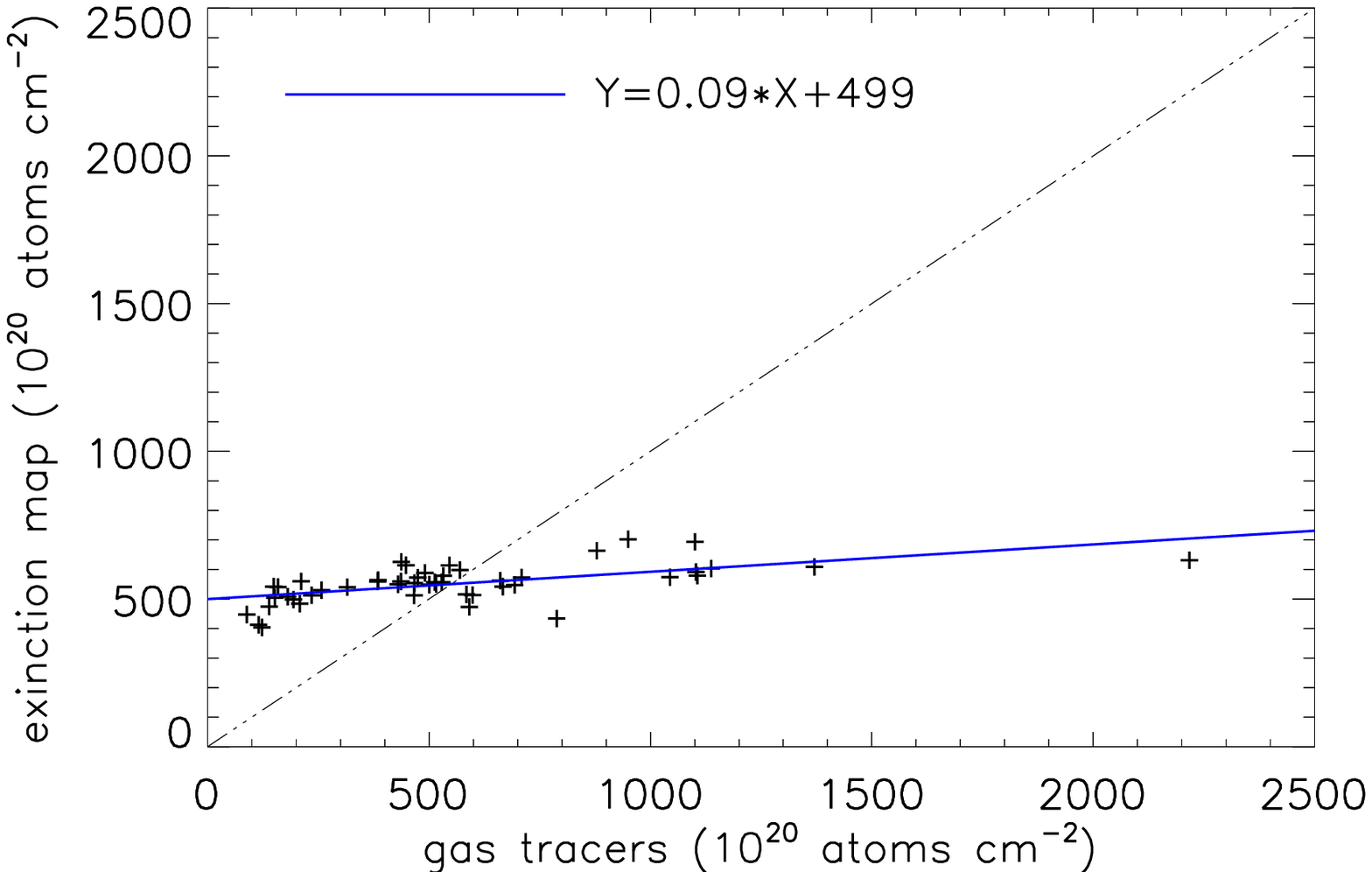}
\caption{Same as Figure \ref{fig:extinction_laurent_column} but for the extinction map obtained from the \citet{Marshall06} model (middle panel). Both the 
column density map from the gas tracers (top panel) and the extinction map are downgraded to 15 arcmin, using a pixel size of 15 arcmin.}
\label{fig:extinction_map}
\end{figure}

The analysis is limited to the Galactocentric ring containing the Scutum-Crux arm intersection, which is the only region accurately reconstructed by the \citet{Marshall06} model. Figure \ref{fig:extinction_map} shows 
the comparison of the column density estimated from the \citet{Marshall06} extinction map with the column density evaluated from 
the tracers of the gas phases. We notice, as for Figure \ref{fig:extinction_laurent_column}, that the two column density maps are separated by an offset which, in this case, is 
of the order of $\simeq$500 $\times$ 10$^{20}$ atoms cm$^{-2}$. Moreover, the column densities computed from the the 3D extinction map are even lower than 
those obtained from the previous extinction map, accounting only for $\sim$ 9 percent of the gas column densities.

\subsection{Testing the exclusion of the ionized phase from the inversion model}\label{sec:no_HII}

The Pearson's coefficients study described in Section \ref{sec:pearsons} show the existence, at every wavelength, of a high degree of correlation between the 
ionized gas column density and IR emission, in particular in Ring 1. 

In this section, we explore the consequences of performing inversion analysis without taking into account the ionized gas phase. 
For this purpose, we carry out a simple test, consisting in deriving the emissivity coefficients for Ring 1 ignoring 
the RRL data. We note that, if we do not make use of these data, we can afford to work at a higher resolution, hence with a 
larger number of pixels. By including only the atomic and molecular gas components, the pixel size is 
set by the \12CO data, i.e. 3 arcmin, and we obtain $\sim1200$ pixels in each map, almost twice the previous number. 

\begin{figure}
\centering
\includegraphics[width=8cm]{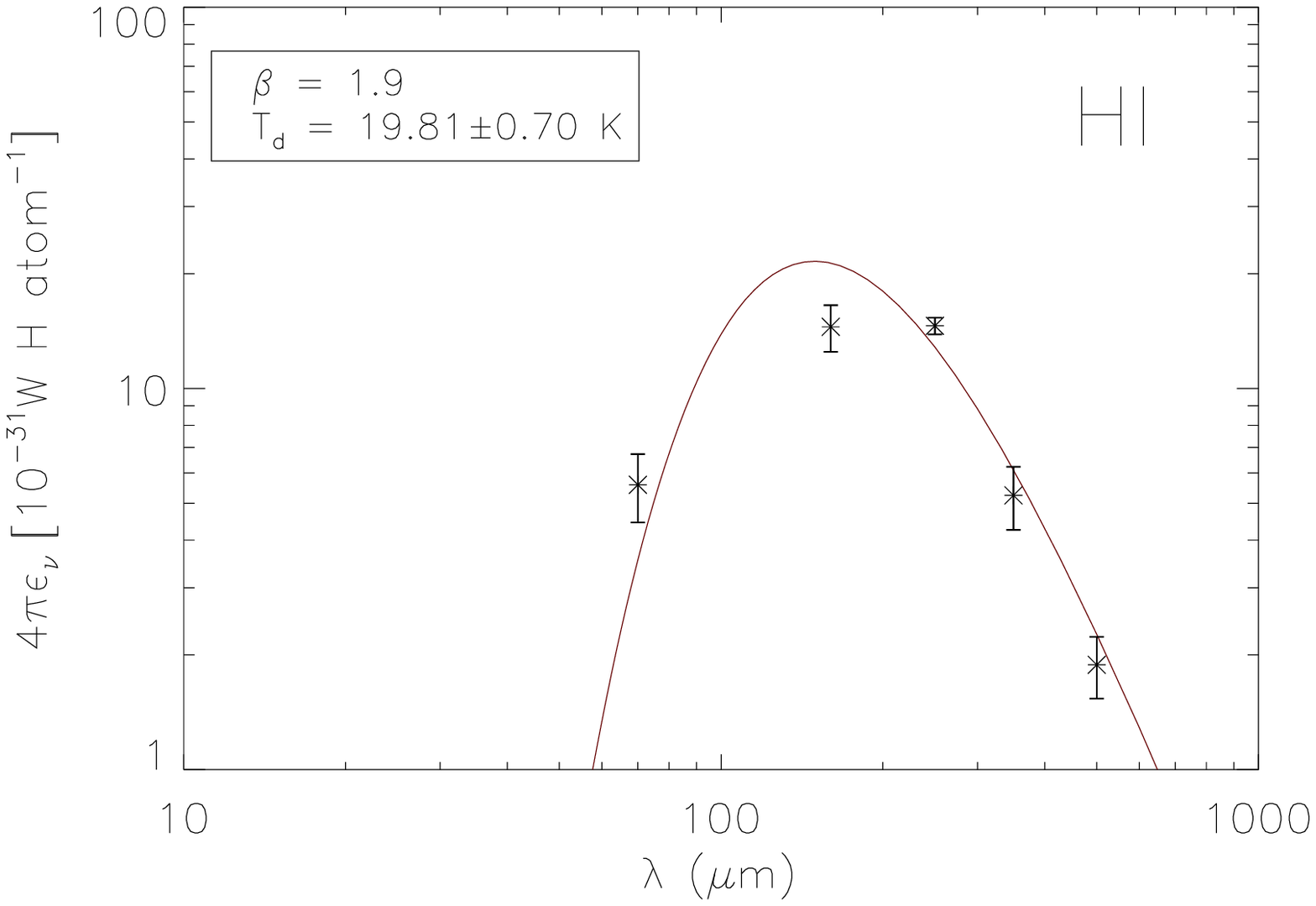}\\
\includegraphics[width=8cm]{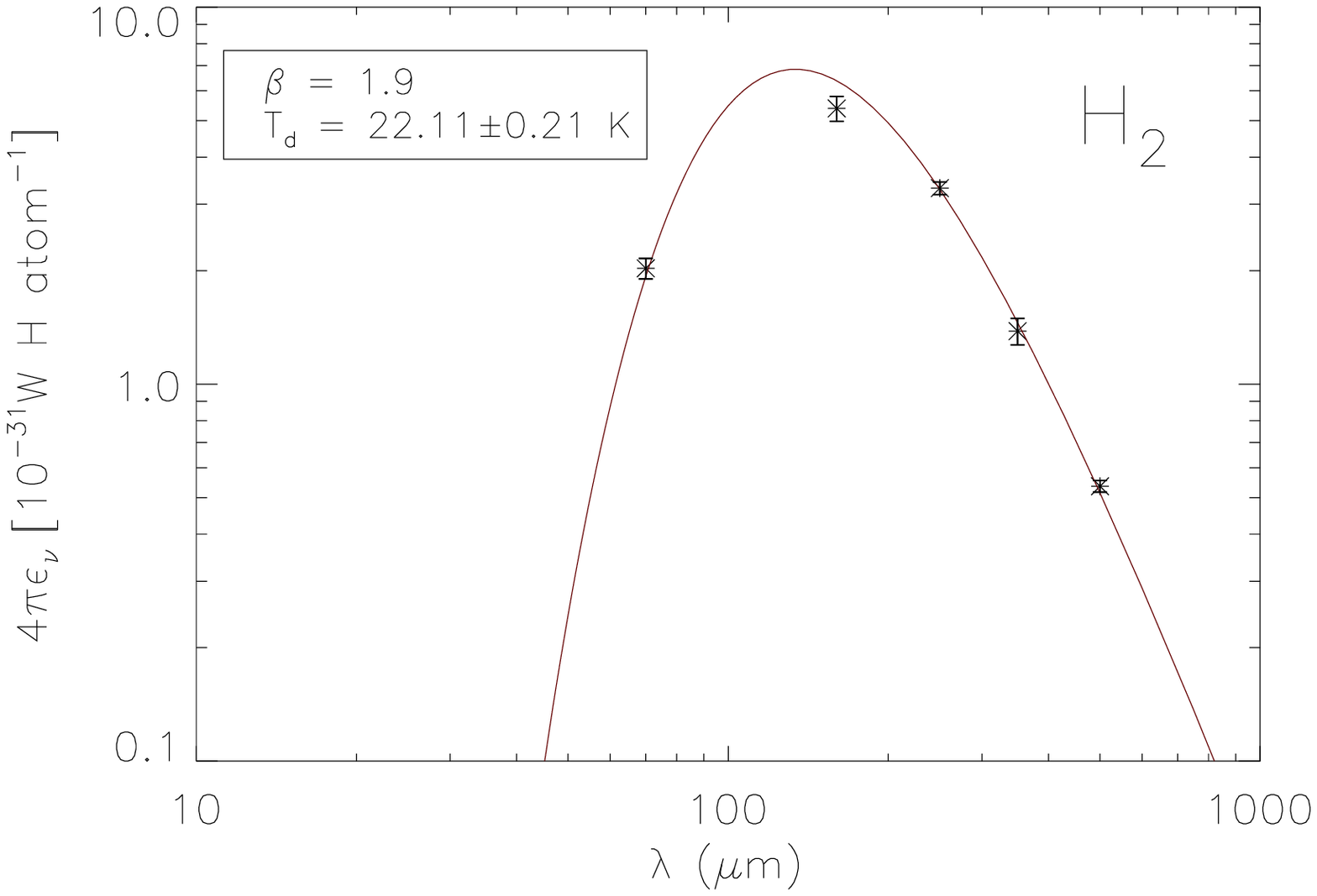}
\caption{Grey-body fits to the dust emissivities associated with the atomic (top panel) and molecular (bottom) gas phases. Results refer to Ring 1 when 
the ionized gas component (i.e. the RRL data) is not included in the inversion equation.}
\label{fig:no_RRLs}

\end{figure}

We now solve Equation \ref{eq:decomposition} (setting N(\HII)=0 in all rings and pixels) and analyse the recovered emissivities, focusing on the long 
wavelengths ($>$ 70 $\mu$m) which we can model with a simple grey-body. 
The fitted SEDs for both the atomic and molecular phases are shown in Figure \ref{fig:no_RRLs}. 
We fix the grey-body spectral emissivity index to $\beta=1.9$ to be consistent with the results obtained including also the RRLs in the analysis (see Section \ref{sec:dustem}). From the fit we obtain: T$_{d,HI}=19.81\pm0.70$ K and T$_{d,H_{2}}=22.11\pm0.21$ K. We note that the temperature of dust in the  
molecular phase is now higher compared to what we obtained in Section \ref{sec:dustem}, and comparable to our 
previous result for HII. This result can be explained once again in light of the Pearson's coefficients 
displayed in Table \ref{tab:scatterplot_columns_IR}: the correlation analysis shows that the second most correlated component, after 
the ionized gas, with the IR templates at all wavelengths, is the molecular gas phase. This, mathematically, translates 
into artificially boosted emissivities for dust associated with \H2, as the molecular phase compensates for the absence 
of the HII component.

\section{Conclusions}\label{sec:conclusions}

We have investigated dust properties in a 2$\times$2 square degrees 
Hi-GAL field (SDPF1) centred on (\textit{l,b})=(30$^{\circ}$,0$^{\circ}$), in the wavelength range 8 \um\ $\leq\lambda\leq500$ \um. 
For this purpose, we have used an inversion technique, first introduced by \citet{Bloemen86}, to decompose the observed integrated IR emission into 
individual contributions associated with dust in the atomic, molecular and ionized phase of the gas and located 
at different Galactocentric distances. We have used, for the first time in an inversion analysis, Radio Recombination Lines (RRLs) to trace the ionized gas. 
In addition, the decomposition into Galactocentric bins (or {\em{rings}}) is performed exploiting the natural boundaries of the 
structures (i.e. segments of spiral arms) as they appear in the gas data cubes. 

We have solved the inversion equation for all the decomposition rings (i.e. Ring 1, 2, 3 and 4), and obtained 
positive solutions only for Ring 1. A Pearson's coefficient and longitude profiles analyses reveal that Ring 1, which 
covers Galactocentric distances between 4.2 and 5.6 kpc and hosts the mini-starburst W43, dominates 
the total IR emission towards SDPF1. For this ring only, we have fitted with DustEM the emissivities retrieved by the inversion method. These fits 
allow estimating, for each phase of the gas, dust temperatures and abundances, as well as the intensity of the local radiation field 
normalized to the intensity of the radiation field in the solar neighbourhood. In particular we find, for the ionized gas phase with respect to the other gas phases, an indication of PAH depletion and an intensity of the local radiation field two times higher, which reflects into a higher dust temperature.

For the other rings (Ring 2, 3 and 4), the inversion equation either cannot be solved or returns 
negative emissivities. The Pearson's coefficients suggest a weak degree of correlation with the IR 
templates and, in a few cases,  even an anti-correlation. For Ring 2 and 3, this result might be ascribed to the presence   
of a large amount of \textit{untraced gas}, either associated with warm \H2 and/or cold HI. This hypothesis could find support in the fact that, in Ring 3,  
cold HI structures are indeed found. In this scenario, the column densities derived from the standard 
tracers would not be able to fully account for the observed IR emission, hence the assumption of the inversion model would break down  
and the resulting emissivities (e.g. negative) be unreliable. In Ring 4, which covers the outer Galaxy, the slight degree of anti-correlation with the input IR 
maps is probably indicative of the intrinsic low emissivity of the region, due to a combined drastic decrease of both hydrogen column density and intensity 
of the interstellar radiation field. 

We have investigated the role of extinction in evaluating total column densities along the LOS as an alternative method 
to gas tracers. For this purpose, we have attempted to build an extinction map for SDPF1  in two independent ways, i.e. using 
deep catalogues of giant stars and with a 3D-extinction model. Although both methods appear to be promising, they currently face 
the severe limitation of not being able to trace extinction beyond $\sim$ 10 -- 15 kpc from the Sun. 

Finally, we have explored the impact of neglecting the ionized gas phase in inversion analysis, as often done in the past. 
We have shown that, by not including this gas component, the temperature of dust associated with the molecular phase 
is artificially increased, due to its high degree of correlation with the input IR templates. 

We conclude with a general remark. In this work we have improved, with respect to previous inversion studies, the estimation of the HI, 
\H2 and HII column densities. However, we believe that this analysis has shown, above all, that local effects, such as departures from circular motion and 
the presence of cold HI structures (and, likely, warm \H2), 
become important when a 3D-inversion is performed in small sections of the Plane and with an angular resolution higher or comparable to 
the angular scale on which these effects dominate the total emission. Conversely, when the entire Galactic Plane is {\em{inverted}} at low angular resolution, the peculiarities
of each LOS are averaged out. Further developments of 3D-inversion models will have to take into account these limitations, both by including non-radial motions, 
i.e. the grand spiral design of the Galaxy, and by estimating total column densities accounting for the blending of cold and warm material along the LOS.

\section*{acknowledgements}
The authors want to thank Mark Calabretta and Lister Staveley-Smith for their work on the RRL data. AT is supported by an STFC grant to JBCA. CD acknowledges an STFC Advanced Fellowship and an EU Marie-Curie grant under the FP7. MIRA acknowledges the support by the European Research Council grant MISTIC (ERC-267934). \textit{Herschel} is an ESA space observatory with science instruments provided by European-led Principal Investigator consortia and with important participation from NASA.


\label{lastpage}

\bibliographystyle{mn2e}
\bibliography{bibliography.bib}


\appendix
\section{Deriving the \CO13 column density using both \12CO and \CO13 data}\label{sec:appendix}
Along each position (\textit{l,b}) in the sky, the $^{13}$CO column density is \citep{Duval10}:

\begin{equation}\label{eq:CO_column_density}
\frac{\textrm{N}(^{13}\textrm{CO})(l,b)}{\mathrm{cm}^{-2}}=2.6\times 10^{14}\int\frac{T_{ex}(l,b,\mathrm{v})\tau_{13}(l,b,\mathrm{v})}{1-e^{\frac{-5.3}{T_{ex}(l,b,\mathrm{v})}}}\frac{d\mathrm{v}}{\mathrm{km\ s}^{-1}}
\end{equation}
integrated over the $V_{\mathrm{LSR}}$ range corresponding to each ring. Both $T_{ex}$ and $\tau_{13}$ can be obtained starting from the detection equation \citep{Stahler05}:

\begin{equation}\label{eq:diluition_equation}
T_{B_{0}}=T_{0}[f(T_{ex})-f(T_{bg})][1-e^{-\tau_{0}}]
\end{equation}

Here, $T_{B_{0}}$ is the brightness temperature of each pixel at a given frequency which is related to the observed antenna temperature through the beam efficiency 
and beam dilution corrections. $\tau_{0}$ is the optical thickness in the same position.  Assuming the radiation behind the clouds is due only to 
the Cosmic Microwave Background, $T_{bg}=2.7$ K. $T_{0}$ is the equivalent temperature of the transition: $T_{0}=h\nu_{0}/K_{B}$. The function $f(T)$ is:

\begin{equation}
f(T)=[\exp(T_{0}/T)-1]^{-1}
\end{equation}

Inverting Equation \ref{eq:diluition_equation} and fixing $T_{0}=5.29$ K for the $^{13}$CO $J=1\rightarrow0$ transition \citep{Pineda10}, one can obtain 
the expression for the $^{13}$CO optical depth at each position \citep{Duval10}:

\begin{equation}
\tau^{13}(l,b,\mathrm{v})=-\ln\bigg[1-\frac{T_{B_{0}}(l,b,\mathrm{v})}{5.29}\big([e^{5.29/T^{13}_{ex}(l,b,\mathrm{v})}-1]^{-1}-0.16\big)^{-1}\bigg]
\end{equation}

$T^{13}_{ex}$ is estimated starting from the $^{12}$CO $J=1\rightarrow0$ transition: since in the colder region of the molecular clouds this line is optically thick, 
its population can be considered in LTE. It is in general a good approximation to consider 
the lower levels of $^{13}$CO transitions in LTE within the same regions. Since the collisional and radiative transition rates per $^{13}$CO molecule are very close to those of $^{12}$CO, 
LTE implies that the relative population in the lowest ($J=1\rightarrow0$) level for the two isotopes is the same \citep{Stahler05}. Then:

\begin{equation}
T^{13}_{ex}=T^{12}_{ex}
\end{equation}

Equation \ref{eq:diluition_equation} can now be solved in the optically thick condition $\tau>>1$ for $^{12}$CO, obtaining:

\begin{equation}
T^{12}_{ex}(l,b,\mathrm{v})=T^{13}_{ex}(l,b,\mathrm{v})=\frac{5.53}{\ln\big(1+\frac{5.53}{T^{12}(l,b,\mathrm{v})+0.837}\big)}
\end{equation}

Once $T^{13}_{ex}$ and $\tau^{13}$ are computed, from Equation \ref{eq:CO_column_density} we estimate $\textrm{N}^{13}(\textrm{CO})$.

\section{Cross-correlation among different rings}\label{app:model_test}
In this Section we show that, when the cross correlation between rings is minimal and the standard tracers account for the total column density in a given ring, the model converges to stable results in that ring, regardless of what happens in the other rings.

In order to validate our results for Ring 1 we run two different series of tests, modifying both the ring configuration and the gas column densities. The first test demonstrates that the dust emission in Ring 1 is dominant with respect to the dust emission in the other rings; the second test shows that the results obtained for Ring 1 are independent from column density variations in the other rings. We use dust temperatures as an indicator of the quality of the result obtained from each individual test. From each set of emissivities, we estimate the dust temperature associated with each gas phase and ring configuration with a simple grey-body model. This model is less sophisticated than DustEM, given that it assumes that all the IR emission comes from dust in thermal equilibrium with the RF, however, it allows us to run a large number of tests using only the Hi-GAL wavelengths, for which dust emission is dominated by the BGs contribution. At 70 \mum, we neglect the additional contributions from VSGs and PAHs.

In performing the tests, we use as a reference the dust temperatures obtained for Ring 1 for each of the gas phases by adopting the ring configuration described in Section \ref{sec:ring_selection}, a spectral emissivity index  $\beta=1.9$ and a grey-body model: $T_{d,\ \mathrm{HI}}=19.3\pm1.0$ K, $T_{d,\ \mathrm{H_{2}}}=19.5\pm0.9$ K and $T_{d,\ \mathrm{HII}}=23.3\pm0.3$ K.

\begin{table*}
\begin{center}
\begin{tabular}{c|c|c}
\hline
\hline
\textit{Gas phase} & configuration A & configuration B\\   
& T (K) & T (K)\\
\hline
\\
HI & 12.8 $\pm$ 1.0 & 21.5 $\pm$ 0.6\\
\\
H$_{2}$ & 20.1 $\pm$ 0.4 & 20.4 $\pm$ 0.7\\
\\
HII & 23.6 $\pm$ 0.7 & 23.5 $\pm$ 0.5\\
\hline 
\end{tabular} 
\end{center}
\caption{Temperature, and corresponding uncertainties, of dust associated with the three gas phases in configuration A (1 ring integrated along the LOS) and configuration B (2 rings), Ring 1.}
\label{tab:temperature_1Ring}
\end{table*}

\begin{table*}
\begin{center}
\begin{tabular}{c|c|c|c|c}
\hline
\hline
\textit{Modified} & \textit{Modified} & HI & \H2 & HII \\
\textit{Ring} & \textit{gas phase} & T(K) & T(K) & T(K)\\
\hline
\\
Ring 1 & HI & 19.6 $\pm$ 1.4 & 18.2 $\pm$ 0.9 & 23.2 $\pm$ 0.3\\
\\
Ring 2 & HI & 20.3 $\pm$ 0.7 & 18.8 $\pm$ 0.9 & 23.5 $\pm$ 0.5\\
\\
Ring 3 & HI & 19.6 $\pm$ 0.8 & 18.9 $\pm$ 0.9 & 23.5 $\pm$ 0.3\\
\\
Ring 1 & \H2 & 19.6 $\pm$ 0.9 & 13.8 $\pm$ 3.6 & 23.4 $\pm$ 0.4\\
\\
Ring 2 & \H2 & 19.7 $\pm$ 1.1 & 19.0 $\pm$ 0.8 & 23.7 $\pm$ 0.3\\
\\
Ring 3 & \H2 & 20.1 $\pm$ 0.6 & 19.4 $\pm$ 1.3 & 23.6 $\pm$ 0.4\\
\\
Ring 1 & HII & 20.7 $\pm$ 1.1 & 19.7 $\pm$ 0.7 & 23.2 $\pm$ 0.6\\
\\
Ring 2 & HII & 19.3 $\pm$ 0.6 & 19.1 $\pm$ 1.1 & 23.8 $\pm$ 0.3\\
\\
Ring 3 & HII & 19.9 $\pm$ 0.6 & 18.1 $\pm$ 1.1 & 23.8 $\pm$ 0.4\\
\\
\hline 
\end{tabular} 
\end{center}
\caption{Temperature, and corresponding uncertainties, of dust associated with the three gas phases in the 9 test cases described in Appendix \ref{app:model_test}.}
\label{tab:temperature_altered_ring_gasphase}
\end{table*}

We run the following two sets of tests:

\begin{itemize}
\item[1.] the first type of tests aims at checking the robustness of the results as a function of the adopted ring configuration. We create two different configurations, A and B. In configuration A we consider the extreme case of having only one single ring, obtained by the integration of all the emission along the LOS. In configuration B, we have instead two rings: Ring 1, spanning the same range of Galactocentric radii as in the standard configuration ($4.5\leq\mathrm{R}\leq5.6$ kpc) and Ring 2, given by the sum of the default Ring 2, 3 and 4 ($5.6\leq\mathrm{R}\leq16.0$ kpc). In configuration A the code converges, giving positive emissivities for all the three components. In configuration B the code converges in Ring 1 but not in Ring 2. Therefore, the majority of the emission arises from the region delimited by $4.5\leq\mathrm{R}\leq5.6$ kpc, allowing the convergence of the code independently from the column density underestimation in Ring 2-3-4. However, the dust temperatures are strongly influenced by the blending of gas and dust along uncorrelated LOS in both configurations. The temperature of dust associated with the three different components for Ring 1 are in Table \ref{tab:temperature_1Ring}. The dust emissivities associated with the HI gas phase in configuration A have very high uncertainties at all wavelengths, and the resulting temperature is very low compared to the 4-ring configuration result. At the same time, the \H2 dust temperature is higher than its counterpart in the 4-ring configuration. These results are consequence of the ring cross-correlation. The cold regions are mostly located in a well defined portion of the sky (the region $7.4\leq\mathrm{R}\leq8.5$ kpc, see Section \ref{sec:missing_column}), but in these tests they are blended together with the warm gas. If the cold regions are not properly isolated (i.e. the rings are not separated as in Section \ref{sec:ring_selection}), the blending makes the dust in the rings which include these features colder. To compensate for that, similarly to what happens if we ignore the ionized gas in the inversion (the warmest component, see Section \ref{sec:no_HII}), the code makes artificially warmer the other rings. In addition, since the majority of the HII gas is in Ring 1, with a minor contribution from the other rings, the emissivities and the temperature of dust associated with this phase are very similar to the 4-ring configuration.

The same effect is evident in configuration B, where both HI and \H2 are warmer than their counterparts in the 4-ring configuration. The blending of different HI components for $5.6\leq\mathrm{R}\leq 16.0$ kpc (and of \H2 components for $5.6\leq\mathrm{R}\leq 8.5$ kpc) prevents the convergence of the code in Ring 2 and makes artificially warmer the temperature of dust associated with HI and \H2 in Ring 1.

On the contrary, if the rings are well-defined into physically independent regions, the propagation of errors across rings is minimal, as demonstrated with the second series of tests.

\item[2.] In these tests we fixed the 4-ring configuration and we simulate the effect of a poor column density estimation.  We run 9 different configurations: in each configuration we add, for each Ring 1 to 3 and phase of the gas, a synthetic gas column density to the gas column densities evaluated in Section \ref{sec:dataset}. 

We generate unbiased synthetic column density maps for each gas component and ring as a random Gaussian distribution with a variance equal to the dispersion of each original map. Independently from the configuration, the model always converges in Ring 1 and does not converge in Ring 2-3-4. The dust temperatures for each gas phase in Ring 1, for the 9 different configurations, are in Table \ref{tab:temperature_altered_ring_gasphase}. With the exception of the altered gas phase and ring, the dust temperatures are the same, within the errors, of the reference values. Since the correlation among gas phases and rings is minimal in this configuration, the error in the gas column density estimation for a given ring and phase of the gas translates into a poor estimation of the emissivity associated with that ring and that phase of the gas, but it does not propagate to the other rings and gas phases. 

\end{itemize}

In conclusion, the results for Ring 1 are robust and are not affected, within the inversion model errors, by even severe gas column density underestimations occurring in Ring 2, 3 and 4.


\section{Masking the W43 region}\label{sec:testing_II}
\begin{figure*}
\centering
\includegraphics[width=14cm]{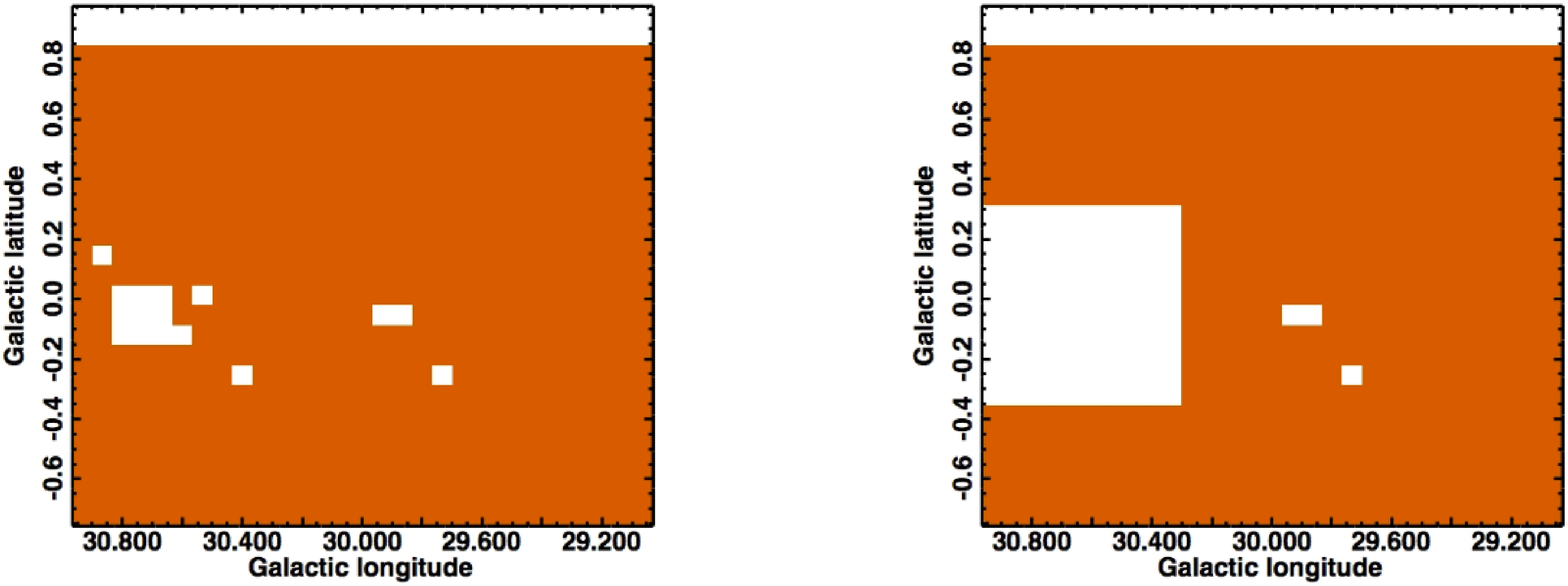}\\
\caption{\textit{Left}: mask used in the main run at the final working resolution of 14.8 arcmin and a pixel size of 4 arcmin. The white pixels are the flagged pixels and they are mainly in correspondence of the strong continuum HI emission observed along bright HII regions LOS, as discussed in Section \ref{sec:atomic_column}. \textit{Right}: the mask used in the test described in Section \ref{sec:testing_II}. It is a square with a side of 0.7$^{\circ}$ and it fully encompass the W43 region.}
\label{fig:W43_mask_comparison}

\end{figure*}

In this Section we demonstrate that the emissivities obtained in Section \ref{sec:dustem} are representative of SDPF1 as a whole and are not biased by specific features present in the field, such as the bright 
HII region W43. In Figure \ref{fig:W43_mask_comparison} we show the comparison between the default mask used in the inversion (left panel) and a new mask which entirely covers W43 (right panel). This new mask was generated for the purpose of investigating the robustness of the results described in the paper. The pixel size of these masks is 4 arcmin. The masked pixels are highlighted in white in the figure. In the default mask these pixels correspond to HI 21 cm line emission associated with strong continuum emission, in correspondence of bright HII regions, as discussed in Section \ref{sec:atomic_column}. The W43 region is already partially masked. The extended mask (Figure \ref{fig:W43_mask_comparison}, right panel) covers a squared region with a side of $\simeq0.7^{\circ}$ centred on W43. This mask reduces the number of good pixels from 680 to 593, a number that still assures the convergence of the code.

\begin{figure*}
\centering
\includegraphics[width=9cm]{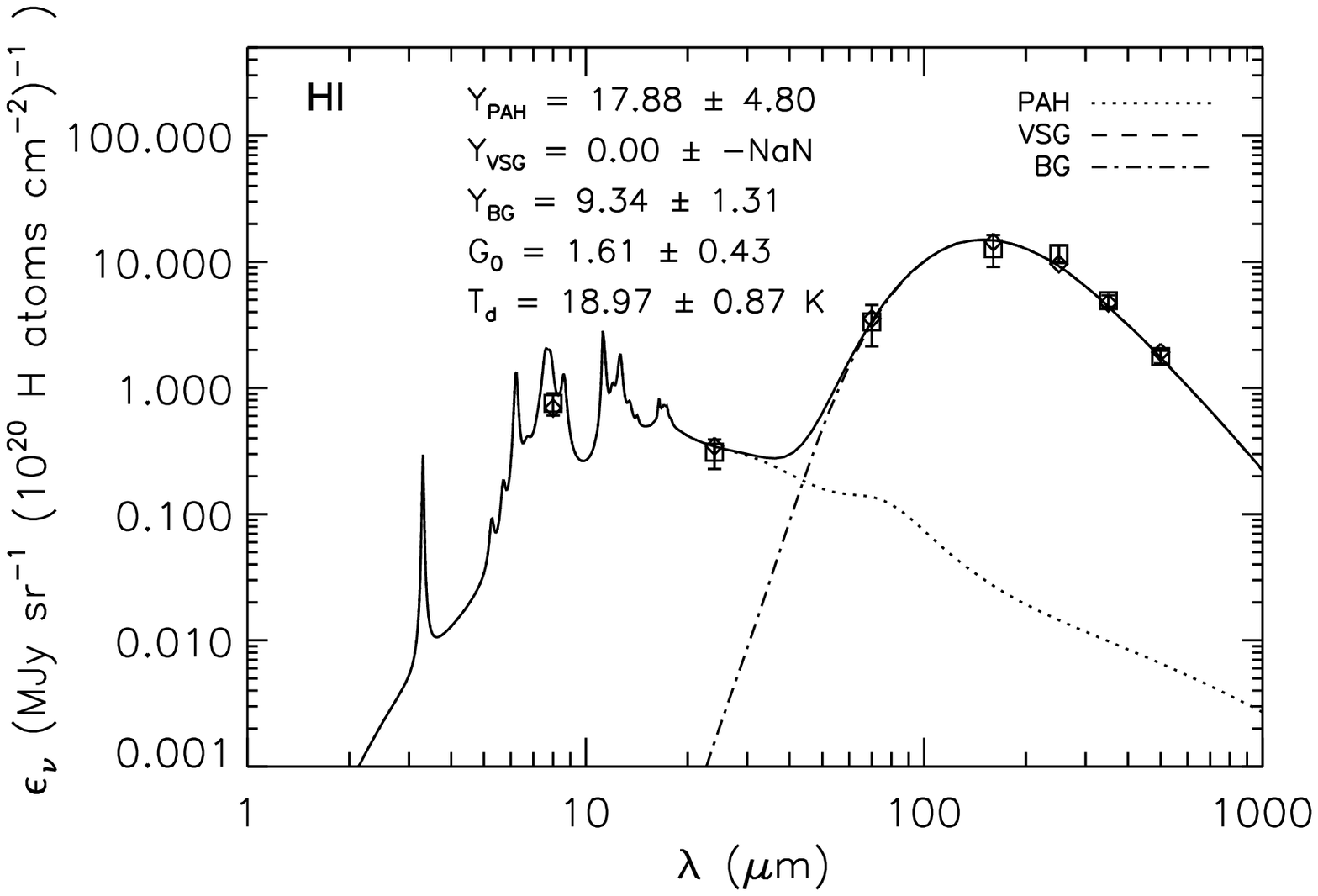}\includegraphics[width=9cm]{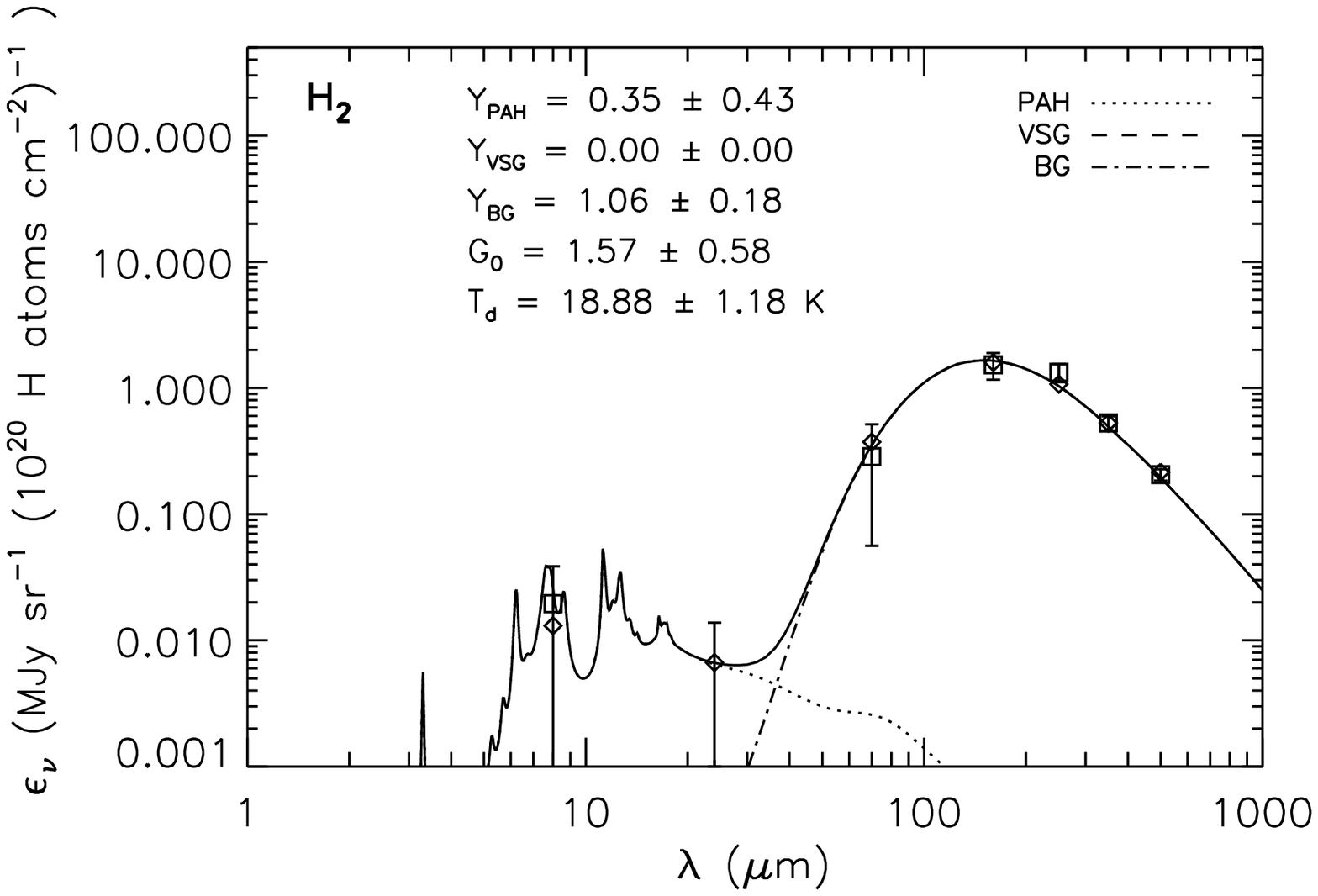}\\
\includegraphics[width=9cm]{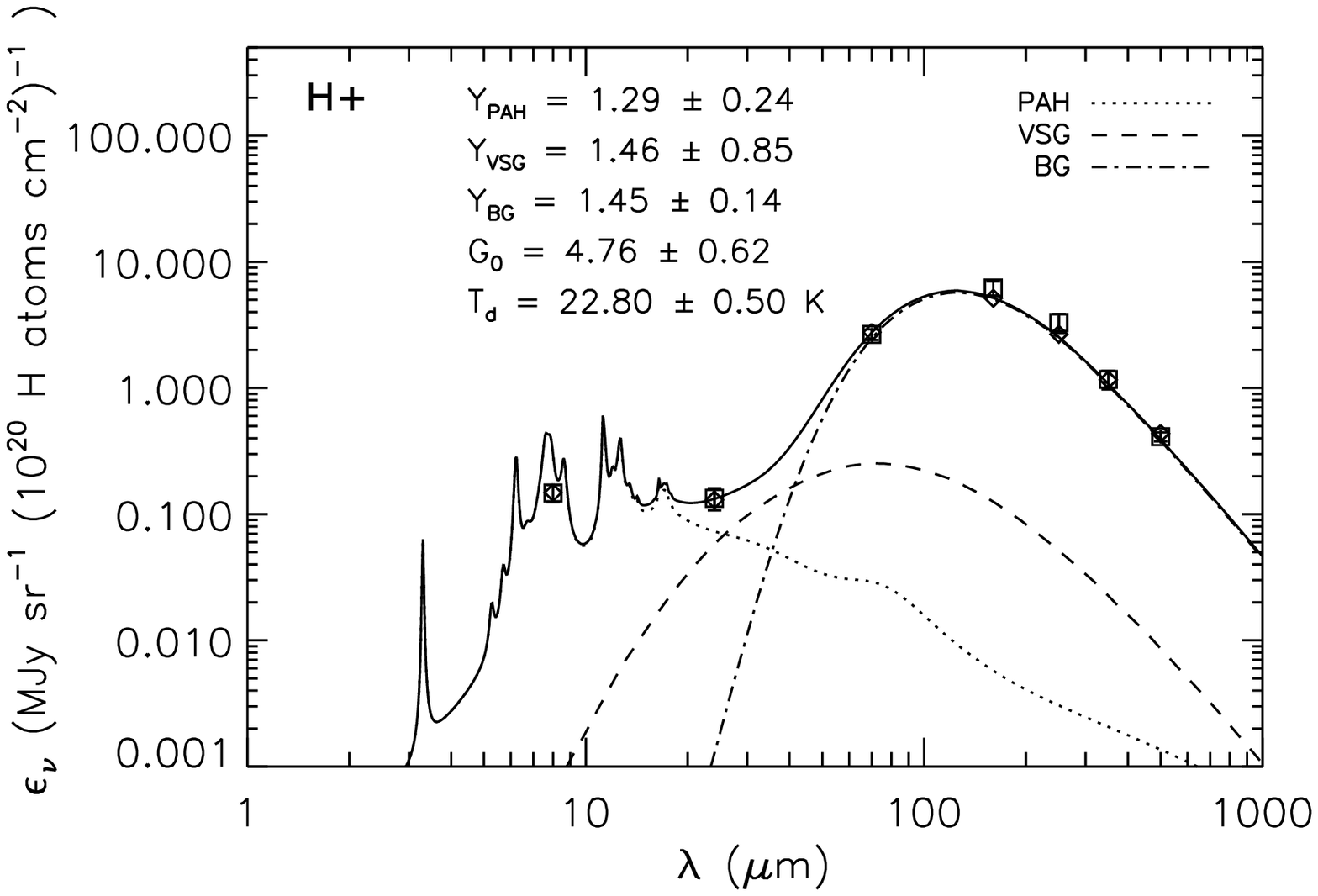}\\
\caption{Dustem fit of the emissivities in Table \ref{tab:emissivities_W43_mask} using the W43 extended mask for the three gas phases in Ring 1. The dust temperatures and dust grain abundances are in agreement within the errors with the main results described in Section \ref{sec:dustem} using the default mask showed in Figure \ref{fig:W43_mask_comparison}.}
\label{fig:dustem_W43_masked}

\end{figure*}

The emissivities associated with Ring 2-3-4 are still negative or poorly constrained, as showed in Table \ref{tab:emissivities_W43_mask}. Also the dust temperature and dust grain abundances for all the three gas phases in Ring 1 are not significantly affected by the extended mask. The DustEM fit obtained with this new set of emissivities are in Figure \ref{fig:dustem_W43_masked}. They are in agreement within the errors with the main results described in Section \ref{sec:dustem}. 

This test evidences the unbiased nature of our results and stresses that the main limitation in recovering the dust emissivities in SDPF1 relies on our inability to accurately reproduce the gas content along each LOS across the field .

\begin{landscape}
\begin{table}
\begin{center}
\begin{tabular}{c|c|c|c|c|c|c|c|c|c|c}
\hline
\hline
\textit{Band} & \multicolumn{10}{c}{$\epsilon_{\nu}$}\\
($\mu$m) & \multicolumn{10}{c}{[ MJy sr$^{-1}$ (10$^{20}$ H atoms cm$^{-2}$)$^{-1}$ ]}\\
\hline
&  \multicolumn{4}{c}{HI} & \multicolumn{3}{c}{\H2} &  \multicolumn{3}{c}{HII} \\
& \textit{Ring 1} & \textit{Ring 2} & \textit{Ring 3} & \textit{Ring 4} & \textit{Ring 1} & \textit{Ring 2} & \textit{Ring 3} & \textit{Ring 1} & \textit{Ring 2} & \textit{Ring 3} \\
\hline
\\
8 & 0.757 $\pm$ 0.152 & 0.022 $\pm$ 0.153 & -1.598 $\pm$ 0.558 & 0.022 $\pm$ 0.093 & 0.020 $\pm$ 0.019 & 0.079 $\pm$ 0.029 & 0.106 $\pm$ 0.156 & 0.147 $\pm$ 0.023 & 0.024 $\pm$ 0.043 & -0.011 $\pm$ 0.226\\
\\
24 & 0.309 $\pm$ 0.081 & 0.158 $\pm$ 0.094 & -0.782 $\pm$ 0.466 & 0.090 $\pm$ 0.066 & -0.001 $\pm$ 0.014 & -0.025 $\pm$ 0.038 & 0.048 $\pm$ 0.044 & 0.134 $\pm$ 0.027 & 0.156 $\pm$ 0.032 & 0.117 $\pm$ 0.147 \\ 
\\
70 & 3.346 $\pm$ 1.209 & 2.435 $\pm$ 1.153 & -8.288 $\pm$ 6.892 & -3.837 $\pm$ 1.216 & 0.286 $\pm$ 0.230 & -0.853 $\pm$ 0.684 & 2.008 $\pm$ 0.582 & 2.659 $\pm$ 0.254 & 2.720 $\pm$ 0.899 & 3.822 $\pm$2.283\\ 
\\
160 & 12.699 $\pm$ 3.598 & 4.796 $\pm$ 2.359 & -14.850 $\pm$ 13.388 & -16.719 $\pm$ 4.120 & 1.528 $\pm$ 0.363 & 1.909 $\pm$ 2.041 & 0.926 $\pm$ 2.046 & 6.197 $\pm$ 0.793 & 4.207 $\pm$ 2.264 & 8.744 $\pm$ 5.098 \\ 
\\
250 & 11.626 $\pm$ 1.826 & 2.726 $\pm$ 2.942 & -23.089 $\pm$ 9.234 & -7.821 $\pm$ 1.859 & 1.331 $\pm$ 0.218 & 2.137 $\pm$ 0.913 & -2.851 $\pm$ 1.626 & 3.299 $\pm$ 0.568 & 0.289 $\pm$ 0.626 & 3.975 $\pm$3.674 \\
\\
350 & 4.916 $\pm$ 0.514 & 1.174 $\pm$ 1.050 & -6.294 $\pm$ 1.286 & -3.067 $\pm$ 0.961 & 0.528 $\pm$ 0.069 & 0.633 $\pm$ 0.472 & -1.072 $\pm$ 0.553 & 1.170 $\pm$ 0.198 & 0.348 $\pm$ 0.553 & 2.119 $\pm$ 0.714 \\
\\
500 & 1.771 $\pm$ 0.203 & 0.299 $\pm$ 0.488 & -2.807 $\pm$ 0.824 & -0.825 $\pm$ 0.236 & 0.206 $\pm$ 0.022 & 0.239 $\pm$ 0.177 & -0.431 $\pm$ 0.187 & 0.411 $\pm$ 0.033 & 0.074 $\pm$ 0.184 & 0.933 $\pm$ 0.522 \\ 
\hline 
\end{tabular} 
\end{center}
\caption{Emissivities, and corresponding uncertainties, of dust associated with the three gas phases in Ring 1-2-3-4 evaluated applying to the data the extended W43 mask showed in Figure \ref{fig:W43_mask_comparison}, right panel. The overall behaviour does not differ sensibly from the main results and in particular the emissivities associated with Ring 2-3-4 are still negative or poorly constrained by the model.}
\label{tab:emissivities_W43_mask}
\end{table}
\end{landscape}


\section{Input and column density maps}\label{sec:column_density_maps}



\begin{figure*} 
\centering
\includegraphics[width=11.5cm]{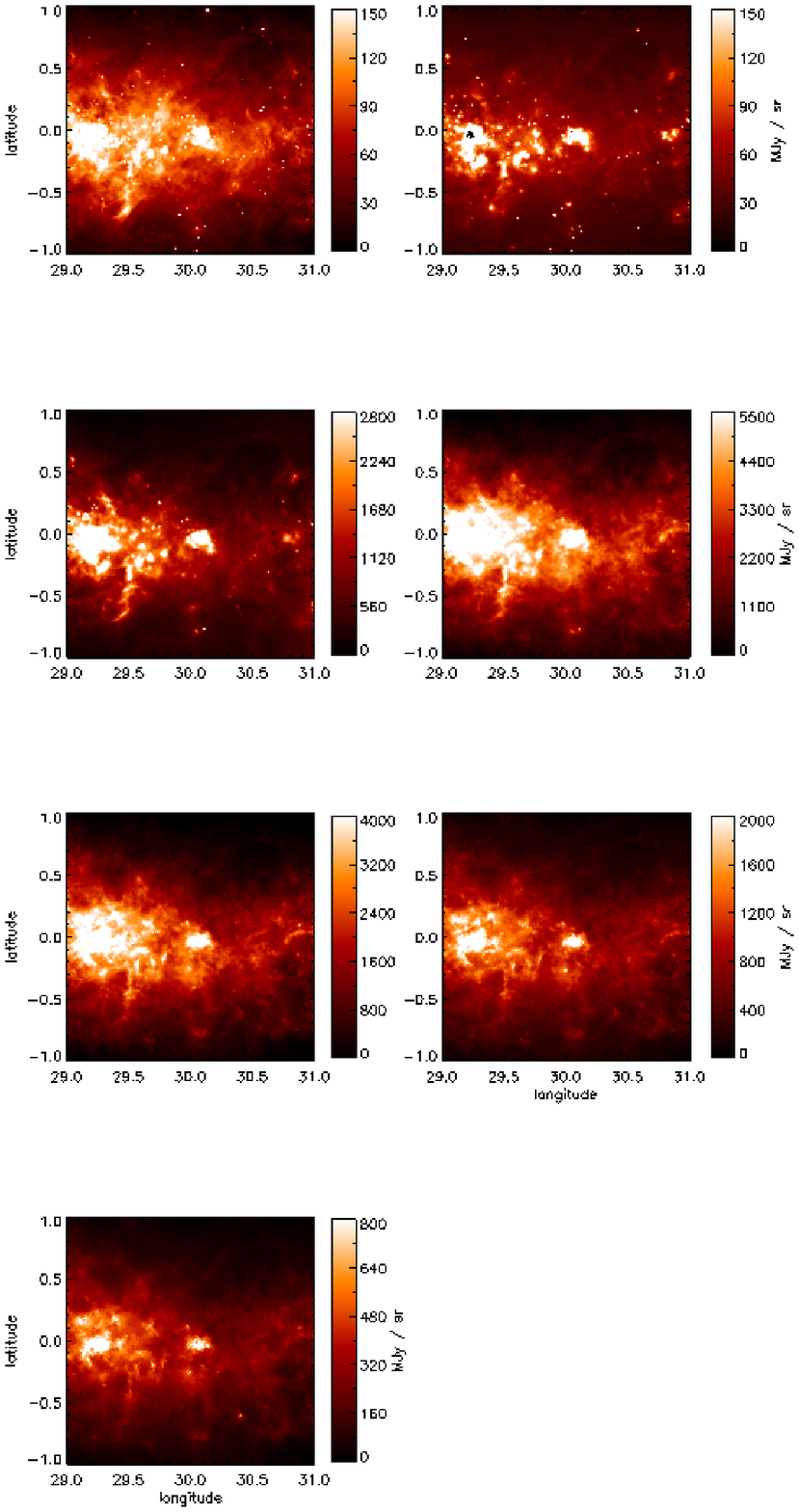}
\caption{IR maps used as input in Equation \ref{eq:decomposition}. From the top to the bottom, left to right: IRAC 8 \mum, MIPS 24 \mum, PACS 70 and 160 \mum, SPIRE 250, 350 and 500 \mum. All maps have been calibrated in MJy / sr and point-source subtracted as described in Section \ref{sec:IR_data}.}
\label{fig:IR_maps}
\end{figure*}

\begin{figure*} 
\centering
\includegraphics[width=12cm,angle=90]{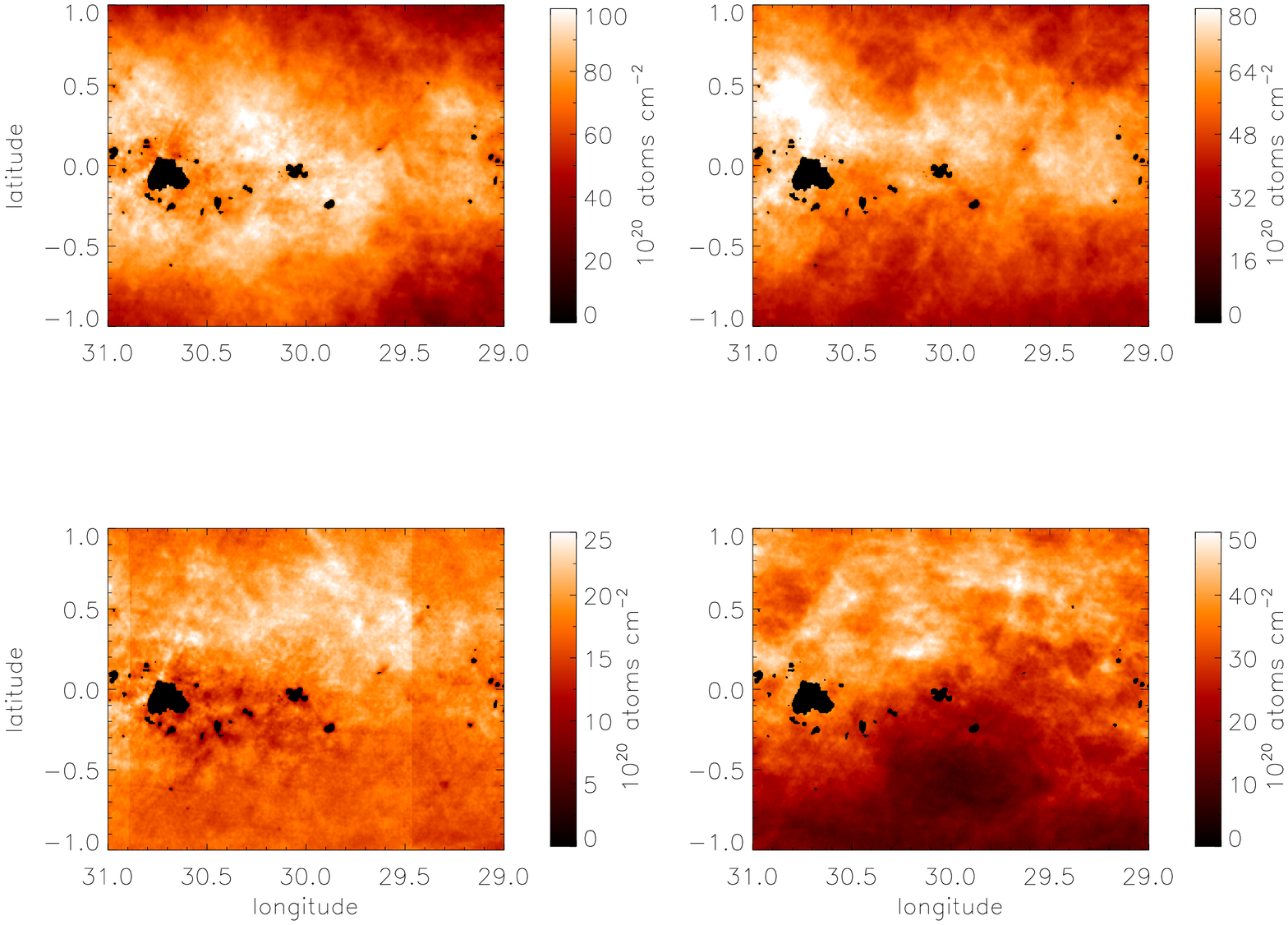}
\caption{HI column density maps in Ring 1 (upper left), Ring 2 (upper right), Ring 3 (lower left) and Ring 4 (lower right) at native VGPS angular resolution (1 arcmin). The black pixels are the masked pixels in correspondence of strong continuum emission (see Section \ref{sec:atomic_column}). The steps in Ring 3 around \textit{l}=29.5$^{\circ}$ and \textit{l}=30.8$^{\circ}$ are at the longitudes in which the emission is split among Ring 2 and Ring 3, and it is of $\simeq 1\times 10^{20}\  \mathrm{H\ atoms / cm}^{2}$. The same steps are indeed present in Ring 2, but they are not visible in the map.}
\label{fig:HI_column_density}
\end{figure*}

\begin{figure*} 
\centering
\includegraphics[width=12cm,angle=90]{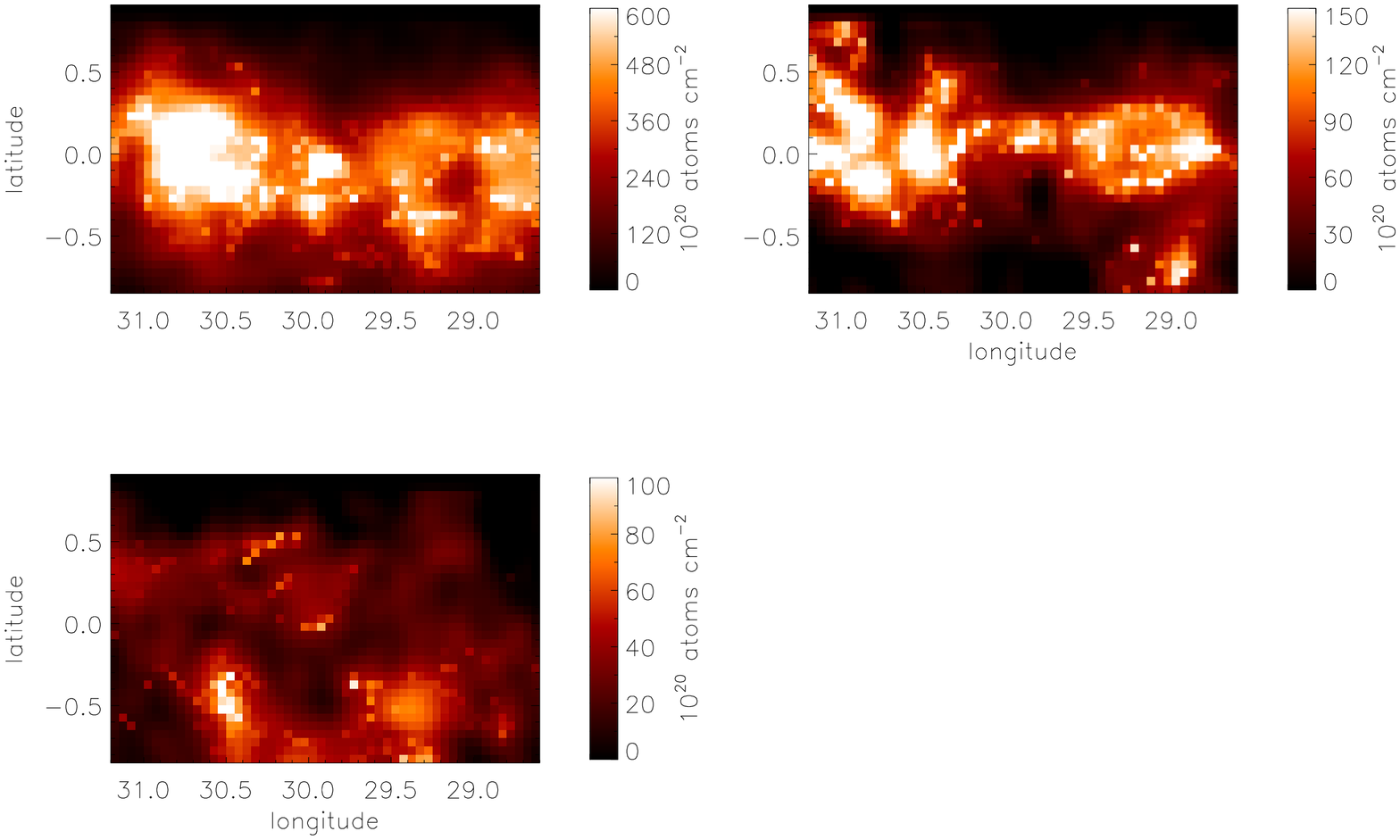}
\caption{\H2 column density maps in Ring 1 (upper left), Ring 2 (upper right) and Ring 3 (lower left). The bright pixels are in correspondence of high column density regions where both \12CO and \CO13 emission have been observed. The resolution is 9 arcmin with a pixel scale of 3 arcmin, obtained combining \12CO and \CO13 datasets (see Section \ref{sec:H2_column}).}
\label{fig:H2_column_density}
\end{figure*}

\begin{figure*}
\centering
\includegraphics[width=12cm,angle=90]{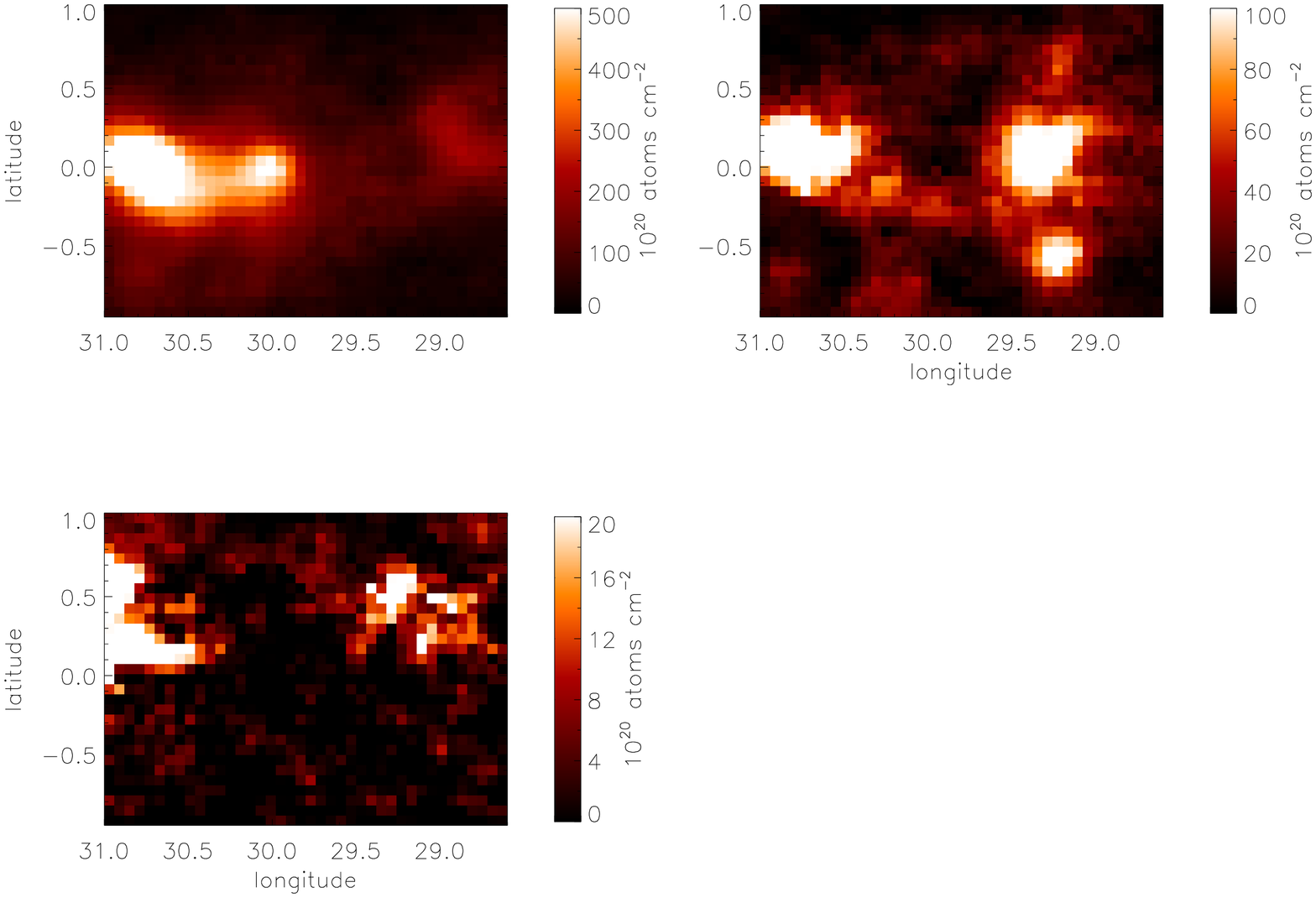}
\caption{HII column density maps in Ring 1 (upper left), Ring 2 (upper right) and Ring 3 (lower left). The majority of the ionized gas is in Ring 1, in correspondence of W43 complex. The map resolution is 14.8 arcmin.}
\label{fig:HII_column_density}
\end{figure*}

\end{document}